\documentclass[a4paper,11pt]{article}
\pdfoutput=1
\usepackage{a4wide}
\usepackage{ascii}
\usepackage[T1]{fontenc}
\usepackage{amsmath,amssymb,amsthm,mathtools,array,booktabs}
\usepackage{stmaryrd}
\usepackage{graphicx,hyperref}
\usepackage{xcolor}
\usepackage{cancel}
\usepackage{tikz}
\usetikzlibrary{cd}
\usepackage{amscd}
\usepackage{latexsym,cite}
\usepackage{subcaption}
\usepackage[boxsize=0.8em]{ytableau}
\usepackage{pictexwd,dcpic}
\usepackage{youngtab}

\Yboxdim{5pt}

\input xy
\xyoption{all}

\hypersetup{
  unicode,
  bookmarksnumbered,
  linktoc = all,
  pdfborderstyle = {/S/U/W 0.5}
}

\newcommand{\bbC}{\mathbb{C}}

\newcommand{\bbI}{\mathbb{I}}

\newcommand{\bbP}{\mathbb{P}}
\newcommand{\bbR}{\mathbb{R}}
\newcommand{\bbZ}{\mathbb{Z}}

\newcommand{\bbW}{\mathbb{W}}

\newcommand{\calB}{\mathcal{B}}
\newcommand{\calC}{\mathcal{C}}

\newcommand{\calE}{\mathcal{E}}

\newcommand{\calH}{\mathcal{H}}

\newcommand{\calL}{\mathcal{L}}
\newcommand{\calM}{\mathcal{M}}
\newcommand{\calN}{\mathcal{N}}
\newcommand{\calO}{\mathcal{O}}

\newcommand{\calS}{\mathcal{S}}

\newcommand{\calV}{\mathcal{V}}
\newcommand{\calW}{\mathcal{W}}

\renewcommand{\Re}{\operatorname{Re}}
\renewcommand{\Im}{\operatorname{Im}}

\DeclareMathOperator{\Ch}{ch}

\DeclareMathOperator{\Id}{Id}

\DeclareMathOperator{\Sym}{Sym}

\newtheorem{thm}{\color{blue} Theorem} \numberwithin{thm}{section}

\newtheorem{rmk}{\color{blue} Remark}  \numberwithin{rmk}{subsection}

\newtheorem{eg}{\color{blue} Example}  \numberwithin{eg}{subsection}

\newtheorem{conj}{\color{blue} Conjecture} \numberwithin{conj}{section}

\newtheorem{prop}{\color{blue} Proposition} \numberwithin{prop}{section}

 \numberwithin{lem}{section}

\newtheorem{cor}{\color{blue} Corollary} \numberwithin{cor}{section}

\newtheorem{defi}{\color{blue} Definition} \numberwithin{defi}{section}

\begin{document}
\thispagestyle{empty}
\begin{flushright}
preprint
\end{flushright}
\vspace{1cm}
\begin{center}
{\LARGE\bf Monodromy of Calabi-Yau threefold flops via grade restriction rule and their quantum K\"ahler moduli} 
\end{center}
\vspace{8mm}
\begin{center}
Ban Lin\footnote{{\tt banlin@kias.re.kr}}$^{,\alpha,\gamma}$
\\
Mauricio Romo\footnote{{\tt mromoj@simis.cn}}$^{,\beta,\gamma}$
\end{center}
\vspace{6mm}
\begin{center}
$^\alpha$Korea Institute for Advanced Study, Seoul, Republic of Korea, 02455
\\
$^\beta$Center for Mathematics and Interdisciplinary Sciences, Fudan University, Shanghai, China, 200433
\\
$^\gamma$Shanghai Institute for Mathematics and Interdisciplinary Sciences, Shanghai, China, 200433
\end{center}
\vspace{15mm}

\begin{abstract}
\noindent
We present exact expressions, based on the grade restriction rule and window categories, for monodromies associated to certain Calabi-Yau threefold flops. We show a general formula for the monodromy action on the lattice of B-brane charges, based on the hemisphere partition function for abelian and nonabelian gauged linear sigma models. We exploit the explicit form of the discriminant in the quantum K\"ahler moduli to further refine the form of the monodromies, in several examples, using their relation to the fundamental group of nested torus links. 
\end{abstract}
\newpage
\setcounter{tocdepth}{3}
\tableofcontents
\setcounter{footnote}{0}

\section{Introduction}

This work is a follow up to \cite{Lin:2024fpz} on the study of autoequivalences induced by certain Calabi-Yau (CY) threefold flops. More precisely we consider a three dimensional CY complete intersection (CYCI):
\begin{equation}
X_{\zeta_{+}}\subset \mathbb{P}^{n}\times \vec{\bbP},\qquad \vec{\bbP}:=Gr(k_{1},m_{1})\times\cdots\times Gr(k_{r},m_{r})
\end{equation}
defined in \eqref{Xplus}, which can be realized as the geometric phase of a gauge linear sigma model (GLSM), described in sect. \ref{sec:flopsGLSM}, with gauge group $G=U(1)_{0}\times U(k_{1})\times\ldots U(k_{r})$. Such GLSM have $r+1$ FI-theta parameters, denoted as
\begin{equation}
t_{\alpha}=\zeta_{\alpha}-i\theta_{\alpha},\qquad\alpha=0,\ldots, r
\end{equation}
they are coordinates in the stringy K\"ahler moduli $\calM_{K}$ of such a model. We will focus on two phases in $\calM_{K}$, namely:
\begin{equation}
\zeta_{\pm}:=\{\zeta_{0}\gg \pm1,\zeta_{1}\gg1,\ldots\zeta_{r}\gg1\}
\end{equation}
then, the $\zeta_{+}$ RG flows to a nonlinear sigma model with target space the aforementioned CICY $X_{\zeta_{+}}$. We use the grade restriction rule \cite{Herbst:2008jq,Hori:2013ika} to find an appropriate set of generators for $D^b\mathrm{Coh}(X_{\zeta_{+}})$. More precisely, we describe explicitly objects (B-branes) $(\widehat{\calE}^{(0)}_{-},\calB^{(0)}_{D_{\alpha}},\calB^{(0)}_{{\alpha}},\calB^{(0)}_{\mathrm{pt}})$ in the window subcategory $\mathbb{W}(0)\subset MF_{G}(W)$, described in \eqref{windowabdef}, which RG flow, in the $\zeta_{+}$ phase, to what we call the Doran-Morgan basis of generators: $(\calO_{X_{\zeta_{+}}},\calO_{D_{\alpha}},\calO_{C_{\alpha}},\calO_{P})$. We denote the vector of central charges of $(\widehat{\calE}^{(0)}_{-},\calB^{(0)}_{D_{\alpha}},\calB^{(0)}_{{\alpha}},\calB^{(0)}_{\mathrm{pt}})$ as $\Pi$:
\begin{equation}
\Pi(t):=\left(Z_{\widehat{\calE}^{(0)}_{-}}(t),Z_{\calB^{(0)}_{D_{\alpha}}}(t),Z_{\calB^{(0)}_{{\alpha}}}(t),Z_{\calB^{(0)}_{\mathrm{pt}}}(t)\right)
\end{equation}
where $Z_{\calB}$ are defined via the hemisphere partition function \cite{Hori:2013ika} in \eqref{eqn:ZBt}. On the other hand, the central charge of the objects $(\widehat{\calE}^{(0)}_{-},\calB^{(0)}_{D_{\alpha}},\calB^{(0)}_{{\alpha}},\calB^{(0)}_{\mathrm{pt}})$, which we also call A-periods, can be independently defined in purely geometric terms as we review in sect. \ref{sec:Aperiods}. If we denote $\Pi_{+}:=\pi_{\zeta_{+}}\circ \Pi$, the RG flow of $\Pi$ to the $\zeta_{+}$ phase,it is straightforward to prove prop. \ref{PropClassicalA}, which states
\begin{prop}
$\Pi_{+}$ coincides with the A-periods of $(\widehat{\calE}^{(0)}_{-},\calB^{(0)}_{D_{\alpha}},\calB^{(0)}_{{\alpha}},\calB^{(0)}_{\mathrm{pt}})$, in the zero-instanton sector (as defined in sect. \ref{sec:Aperiods}, rmk. \ref{zeroinstrmk}).
\end{prop}
In principle, a proof of prop. \ref{PropClassicalA} for the full A-period should be also straightforward, using prop. \ref{propresgrass} but will require full knowledge of the geometric A-period for $X_{\zeta_{+}}$, which is only available for abelian GLSMs \cite{Aleshkin:2023hbu} (and for some nonabelian cases, relevant for our context \cite{Priddis:2024qeb}). However, just with prop. \ref{propresgrass} at hand we can prove our main theorem \ref{thm:TPN}
\begin{thm}
Under the monodromy $M_{0}$ (defined on sect. \ref{sec:wincatMon}), we have 
\begin{equation}
    \Pi_{+}(t)|_{M_{0}(\mathcal{B}^{(0)})}=\Pi_{+}(t)\cdot ^t\mathrm{M}_{0},\qquad \mathrm{M}_{0}:=T_N\cdot L_{K_{Y}}
\end{equation}
where 
\begin{equation}
   T_N:=(T_{C_0})^N=\begin{pmatrix}
       1 & & & 
       \\
        & \ \ \delta_{\alpha\beta} & &
       \\
        & -N\delta_{\alpha\beta,00}\   & \delta_{\alpha\beta} &
       \\
       & & & \ \ \  1
   \end{pmatrix},
\end{equation}
corresponds to a matrix of $N$ times of the spherical twist of $C_0$ acting trivially on all components of $\Pi_{+}(t)$, except $Z_{\mathcal{O}_{D_{0}}}(t)\rightarrow Z_{\mathcal{O}_{D_{0}}}(t)-NZ_{\mathcal{O}_{C_{0}}}(t)$. The coefficient $N$ is defined as:
\begin{equation}
    N:=\int_{Y}(A_2(\vec{n})^2-A_1(\vec{n})A_3(\vec{n})),
\end{equation}
and \footnote{$\delta_{\mu\alpha}$ and $c_{\mu\mu\alpha}$ must be read as column vectors of rank $r+1$. Likewise, $b_{\mu\beta}$ and $\delta_{\mu\beta}$ must be read as rank $r+1$ row vectors.}
\begin{equation}
   L_{K_{Y}}:= \prod_{\alpha=1}^r L_{\alpha}^{-n^{(\alpha)}},\qquad L_\mu=\begin{pmatrix}
       \ 1 &  & & 
        \\
        \  \delta_{\mu\alpha} & \ \delta_{\alpha\beta} &  & 
        \\
        c_{\mu\mu\alpha} & c_{\mu\alpha\beta} & \delta_{\alpha\beta} &
        \\
       \ 0 & \ b_{\mu\beta} & \delta_{\mu\beta} & \ 1
   \end{pmatrix},\quad  b_{\alpha\beta}:=\frac{c_{\alpha\alpha\beta}-c_{\alpha\beta\beta}}{2}
   \end{equation}
\end{thm}
where $X_{\zeta_{+}}$ is written as the zeroes of a section of a vector bundle $\calN\rightarrow Y$ over a Fano variety $Y\subset \bbP^n\times \vec\bbP$ (defined in sect. \ref{sec:globalflps}) and the characteristic classes $A_{k}(\vec n)$ are defined in appendix \ref{sec:appcharclass}. We develop several refinements of theorem \ref{thm:TPN} along this work, although we do not present proofs for them in full generality only in families of examples. These refinements are listed below:
\begin{enumerate}
    \item In theorem \ref{thm:TPN} and in most of the sections of the paper, we work under the assumption that the bundle $\calN\rightarrow Y$ is a sum of line bundles, however, as we point out in rmk. \ref{rmkN}, it is straightforward to generalize theorem \ref{thm:TPN} for $\calN$ given by a sum of vector bundles when those bundles satisfy some technical conditions. We put this generalization into practice in the examples associated to nonabelian GLSMs, in sect. \ref{sec:NAexamples}.
    \item In all the examples studied in sect. \ref{sec:examples} we refine the monodromy $M_{0}$ by writing it as products of matrices associated to spherical twists. For this, we use the exact form of the discriminant $\Delta$: the intersection $\mathfrak{L}=\Delta\cap S^{3}$, with $S^{3}$ centered at $\Delta\cap \{\exp(-t_{\alpha})=0:\alpha\neq0\}$, is a nested torus link (see appendix \ref{sec:appknot}) in the case $\mathrm{dim}\calM_{K}=2$, and we can associate the spherical twists $T_{\calO_{X}}$, $T_{\calS_{X}}$ and twists $L_{\alpha}$, to different generators of $\pi_{1}(S^3\setminus \mathfrak{L})$, depending on the case. For the cases $\mathrm{dim}\calM_{K}\geq2$, and under certain conditions for $\Delta$, we propose a conjecture \ref{conjecturenested} that allows us to apply a recursive formula, also based on fundamental groups of nested torus links.
    \item In appendix \ref{sec:weighted}, we present one example where we allow $\vec{\bbP}$ to be a weighted projective space showing a possible a generalization beyond smooth $Y$.
\end{enumerate}

The paper is organized as follows. 
In sect. \ref{sec:flopsGLSM} we construct the GLSM that realizes $X_{\zeta_{+}}$ as one of its geometric phases and its flop $X_{\zeta_{-}}$. This construction generalizes some of the ones realized in \cite{Jockers:2012zr,Brodie:2021toe}. We comment on several aspects of its quantum K\"ahler moduli $\calM_{K}$, as well as purely geometric aspects of the flop, in sect. \ref{sec:globalflps}, for instance the existence of an extremal transition to another smooth CY threefold $X^{\natural}$. In sect. \ref{section3} after reviewing the construction of B-branes for GLSMs, following mostly \cite{Herbst:2008jq,Hori:2013ika,hori2019notes} we construct a set of B-branes in $MF_{G}(W)$ that RG-flow to holomorphic cycle in $X_{\zeta_{+}}$, and then in sect. \ref{sec:wincatMon} we describe the monodromy associated to the flop, in terms of window subcategories of  $MF_{G}(W)$, using grade restriction rules, we find the action of this monodromy that we call $M_{0}$ into our choice of generators of a given window category. In sect. \ref{sec:Aperiods} we review the hemisphere partition function $Z_\calB(t)$ for B-branes on GLSMs and the geometric A-periods $Z_{\calE}$ for objects $\calE\in D^{b}\mathrm{Coh}(X)$. We also present explicit formulae for both and show they agree in the zero-instanton sector. Then, we proceed to show our main result, theorem \ref{thm:TPN}. In sect. \ref{sec:examples} we present several examples, abelian and nonabelian, all based in different flops constructed from a given $X^{\natural}$. In each example we relate their discriminant $\Delta$ to nested torus links, whose properties we collect in appendix \ref{sec:appknot}.

\section*{Acknowledgements}

We thank W. Donovan, J. Knapp, S. Lee, E. Scheidegger, Y. Wen, Z. Yu, H. Zou and Y. Zhou for helpful and enlightening discussions. BL is supported by the KIAS Individual Grant (PG100701) at Korea Institute for Advanced Study. MR thanks Wuhan University, Beijing Institute of Mathematical Sciences and Applications (BIMSA), St. Petersburg University, Instituto de Matem\'atica Pura e Aplicada (IMPA) and Aquam\'atica for hospitality while part of this work has been performed. BL also thanks the hospitality of Wuhan University, BIMSA, Fudan University, Seoul National University and Ulsan National Institute of Science and Technology and for allowing  him to present these results. In particular BL thanks Shanghai Institute for Mathematics and Interdisciplinary Sciences (SIMIS) for hosting him during the Fall semester, 2024, when most of these results were developed.

\section{GLSMs for certain global flops}\label{sec:flopsGLSM}
In this section, we introduce a family of GLSMs that implements a global flop. Let us first consider the gauge group
\begin{equation}
G=U(1)\times U(k_{1})\times\cdots\times U(k_{r})
\end{equation}
and the representation $\rho:G\rightarrow GL(V)$, where $V$ is a complex vector space of rank
\begin{equation}
\mathrm{rk}V=n+1+\sum_{\alpha=1}^{r}m_{\alpha}k_{\alpha}+K,
\end{equation}
with $m_{\alpha}\in \mathbb{Z}_{\geq k_{\alpha}}$ and $K\in\mathbb{N}_{>n}$, determining the matter superfields (chiral superfields). Specifically, we consider $\rho$ given by
\begin{eqnarray}
\mathbb{C}(1)^{\oplus(n+1)}\oplus\mathbf{k}_{1}^{\oplus m_{1}}\oplus\cdots \oplus \mathbf{k}_{r}^{\oplus m_{r}}\bigoplus_{I=1}^{n+1}(\mathbb{C}(-1)\otimes \mathcal{R}_{I})\bigoplus_{I=n+2}^{K}\mathcal{R}_{I} \label{eqn:chiralfields}
\end{eqnarray}
where $\mathbb{C}(q)$ with $q\in\mathbb{Z}$ denotes an irreducible representation of $U(1)\subset G$ of weight $q$, $\mathbf{k}_{\alpha}$ denotes the fundamental representation of $U(k_{\alpha})\subset G$, $\alpha=1,\ldots,r$, and $\mathcal{R}_{I}$ denotes the rank one representation:
\begin{eqnarray}
\mathcal{R}_{I}:=\mathrm{det}^{-n^{(1)}_{I}}\otimes\cdots \otimes \mathrm{det}^{-n^{(r)}_{I}},
\end{eqnarray}
with $n^{(\alpha)}_{I}\in\mathbb{N}$ and $\mathrm{det}^{-n^{(\alpha)}_{I}}$ denoting the $-n^{(\alpha)}_{I}$th power of the determinant representation of $U(k_{\alpha})\subset G$. We denote the coordinates in $V$ as
\begin{eqnarray}
x^{(\alpha)}&&\text{ \ \ coordinates on \ \ }\mathbf{k}_{\alpha}^{\oplus m_{\alpha}},\qquad \alpha=1,\ldots,r\nonumber\\
p_{I}&&\text{ \ \ coordinates on \ \ }\mathcal{R}_{I},\qquad I=1,\ldots, K\nonumber\\
y &&\text{ \ \ coordinates on \ \ } \mathbb{C}(1)^{\oplus(n+1)},
\end{eqnarray}
The coordinates $x^{(\alpha)}$ will be eventually identified with Stiefel coordinates of the Grassmannian $G(k_{\alpha},m_{\alpha})$, hence we can think of them as $m_{\alpha}\times k_{\alpha}$ matrices, upon fixing a basis for the vector space $\mathbf{k}_{\alpha}^{\oplus m_{\alpha}}$, then we can construct Pl\"ucker coordinates $B^{(\alpha)}(x)$ by considering the $k_{\alpha}\times k_{\alpha}$ minors of $x^{(\alpha)}$. The coordinates $B^{(\alpha)}(x)$ are $SU(k_{\alpha})$ invariant and transform on the $\mathrm{det}$ representation of $U(k_{\alpha})$. Then, given a (homogeneous) polynomial $f(x,y)$ transforming on the representation $\mathbb{C}(d_{0})\otimes \mathrm{det}^{d_{1}}\otimes\cdots\otimes \mathrm{det}^{d_{r}}$ we denote its degree by $(d_{0},\ldots,d_{r})$. Then we can define the $G$-invariant superpotential $W$ to specify the matter interaction:
\begin{equation}\label{superpot}
W:=\sum_{I=1}^K p_{I}F_{I}(x,y),
\end{equation}
where
\begin{eqnarray}
F_{I}(x,y)&&\text{ \ \ homogeneous of degree \ \ }(1, n^{(1)}_{I},\ldots,n^{(r)}_{I}),\qquad I=1,\ldots,n+1\nonumber\\
F_{I}(x)&&\text{ \ \ homogeneous of degree \ \ }(0, n^{(1)}_{I},\ldots,n^{(r)}_{I}),\qquad I=n+2,\ldots,K.
\end{eqnarray}
In summary, the matter content of GLSM is indicated in the following table:
\begin{equation}
    \begin{array}{c|cccccc}
         & x^{(1)} & \cdots & x^{(r)} & y_{1,\cdots,n+1} & p_{I=1,\cdots,n+1} & p_{I=n+2,\cdots,K}
         \\\hline
       U(1)  & 0 & \cdots & 0 & 1 & -1 & 0 
       \\
       U(k_1) & \mathbf{k}^{m_1}_1 & \cdots & 0 & 0 & \mathrm{det}^{-n^{(1)}_I} & \mathrm{det}^{-n^{(1)}_I}
       \\
       \vdots &  & \ddots & & \vdots & \vdots & \vdots
       \\
       U(k_r) & 0 & \cdots & \mathbf{k}_r^{m_r} & 0 & \mathrm{det}^{-n^{(r)}_I} & \mathrm{det}^{-n^{(r)}_I}
       \\\hline
       U(1)_R & \varepsilon_1 & \cdots & \varepsilon_r & \varepsilon_0 & \epsilon_{I} & \widetilde \epsilon_{I}
    \end{array}
\end{equation}
where the vector R-charges are denoted $\varepsilon_0,\varepsilon_\alpha,\epsilon_I,\widetilde \epsilon_I\in[0,2)$ and they are subjected to the constraints (since $W$ must have total vector R-charge $2$):
\begin{equation} 
    \begin{array}{ll}
     \epsilon_I=2-\varepsilon_0-2\sum_{\alpha=1}^rn_I^{(\alpha)}\varepsilon_\alpha,   &  \quad I=1,\cdots,n+1,
         \\
    \widetilde \epsilon_I= 2-2\sum_{\alpha=1}^rn_I^{(\alpha)}\varepsilon_\alpha,     &  \quad I=n+2,\cdots,K.
    \end{array}
   \label{eqn:Rcharge}
\end{equation}
To guarantee that our model is anomaly free, we impose the CY condition:
\begin{eqnarray}
\sum_{I=1}^{K}n^{(\alpha)}_{I}=m_{\alpha},\qquad \alpha=1,\ldots,r.
\end{eqnarray}
This guaranties the representation $\rho$ factors through $SL(V)$. The twisted superpotential potential, specifying the interaction of the twisted chiral field, is characterized by the FI-theta parameters that can be written as
$t=\zeta-i\theta$, where \cite{hori2019notes}
  \begin{eqnarray}
  t\in \left(\frac{\mathfrak{t}^{\vee}_{\mathbb{C}}}{2\pi i
\mathrm{P}}\right)^{W_{G}}\cong\frac{\mathfrak{z}^{\vee}_{\mathbb{C}}}{2\pi i
\mathrm{P}^{W_{G}}},
  \end{eqnarray}
  here $\mathrm{P}$ denotes the weight lattice, $W_{G}$ the Weyl subgroup of
$G$,
$\mathfrak{t}$ the Cartan subalgebra of $\mathfrak{g}=\mathrm{Lie}(G)$ and
$\mathfrak{z}=\mathrm{Lie}(Z(G))$. In the present case we can choose a basis where the FI-theta parameters for the $U(1)$ and $U(k_{i})$ subgroups of $G$, can be denoted $t_0$ and $t_\alpha$, $\alpha=1,\ldots,r$, respectively. This GLSM, by construction have a classical Higgs phase on the regime $\zeta_{0},\ldots,\zeta_{k}\gg 1$, given by
\begin{equation}
X_{\zeta_{+}}:=\mu^{-1}(\zeta)/G\cap \{dW^{-1}(0)\}
\end{equation}
where $\mu:V\rightarrow \mathfrak{g}^{\vee}$ denotes the moment map, associated to $\rho$, on the
vector space $V$ (whose coordinates are $(y,x,p)$). For generic polynomials $F_{I}(x,y)$, $X_{\zeta_{+}}$ is given by the smooth complete intersection
\begin{equation}\label{Xplus}
X_{\zeta_{+}}=\bigcap_{I=1}^{K}\{F_{I}(x,y)=0\}\subset \mathbb{P}^{n}\times G(k_{1},m_{1})\times\cdots\times G(k_{r},m_{r})
\end{equation}
The central charge of this GLSM and its central charge $\hat{c}$, coincide with $d=\mathrm{dim}X_{\zeta_{+}}$:
\begin{equation}
\hat{c}=d=n+\sum_{\alpha=1}^{r}k_{\alpha}(m_{\alpha}-k_{\alpha})-K
\end{equation}
We will use the following notation for the complete intersection CY \cite{Green:1986ck,Candelas:1987kf}(CICY) $X_{\zeta_{+}}$:
\begin{equation}
\begin{aligned}
    X_{\zeta_{+}}=\left[ \begin{array}{c|cccccc}
      \bbP^n   &  1 & \cdots & 1 & 0 & \cdots & 0
    \\
       \vec{\bbP}  & \vec{n}_1 & \cdots & \vec{n}_{n+1} & \vec{n}_{n+2} & \cdots & \vec{n}_K
    \end{array}\right],
    \label{eqn: CICYI}
\end{aligned}
\end{equation}
where $\vec{\bbP}:=G(k_{1},m_{1})\times\cdots\times G(k_{r},m_{r})$. For the abelian case: $k_{\alpha}=1$ for all $\alpha$, these configurations have been studied in \cite{Candelas:1987kf,Green:1988wa,Brodie:2021toe} as examples of global flops and termed \emph{splitting configurations}. The phase $-\zeta_{0},\zeta_{1},\ldots,\zeta_{k}\gg 1$ will also be relevant in our analysis, so we proceed to describe it here. The analysis goes likewise. It is also a pure Higgs phase described by a sigma model with target space $X_{\zeta_{-}}$ which is a complete intersection:
\begin{eqnarray}
X_{\zeta_{-}}&=&\bigcap_{I=1}^{n+1}\{F'_{I}(x,p)=0\}\bigcap_{I=n+2}^{K}\{F_{I}(x)=0\}\subset\mathbb{P}_{-} \nonumber\\
\mathbb{P}_{-}&:=&\mathbb{P}\left(\bigoplus_{I=1}^{n+1}\mathcal{O}(-n^{(1)}_{I},\ldots,-n^{(r)}_{I})\right)\rightarrow G(k_{1},m_{1})\times\cdots\times G(k_{r},m_{r})
\end{eqnarray}
where $\mathcal{O}(-n^{(\alpha)}_{I})$ denotes the line bundle $(\mathrm{det}\calS_{\alpha})^{\otimes n^{(\alpha)}_{I}}\rightarrow G(k_{\alpha},m_{\alpha})$ with $\calS_{\alpha}$ the tautological bundle of rank $k_{\alpha}$ and,
\begin{eqnarray}
F'_{I}(x,p):=\frac{\partial}{\partial y_{I}} \sum_{J=1}^{n+1}p_{J}F_{J}(x,y)
\end{eqnarray}

\textbf{Comment on vector R-charges on $\zeta_{\pm}$ phases:} From \eqref{eqn:Rcharge}, the vector R-charge for matter fields, in the phases $\zeta_{\pm}$ \eqref{pmphases},  can be assigned as
\begin{equation}
    \begin{array}{cl}
        (\varepsilon_\alpha,\ \varepsilon_0,\ \epsilon_I,\ \widetilde\epsilon_I)=(0,0,2,2) &  \text{for } \zeta_+ \text{ \ phase}
         \\
        (\varepsilon_\alpha,\ \varepsilon_0,\ \epsilon_I,\ \widetilde\epsilon_I)=(0,2,0,2) & \text{for }\zeta_- \text{ \ phase}
    \end{array}\label{eqn:RInt}
\end{equation}
These R-charges are not strictly inside the interval $(0,2)$, however, in this limit, physical correlators such as the hemisphere partition function are well defined, so we will work with R-charges \eqref{eqn:RInt}, whenever we need an explicit assignment.

\subsection{Effective twisted potential and discriminant}

The effective twisted potential is the potential for the scalar fields $\sigma$ in the vector multiplet \cite{Witten:1993yc,Morrison:1994fr}. In a generic region of the Coulomb branch, i.e. where the eigenvalues of the VEV $\langle\sigma\rangle\in \mathfrak{t}_{\mathbb{C}}:=\mathrm{Lie}(T_{G})\otimes \mathbb{C}$ are distinct and their magnitude is large (compared with the energy scale). Upon integration of the charged chirals, the effective twisted potential is given by \cite{hori2019notes,Morrison:1994fr}
\begin{equation}\label{genericWeff}
\widetilde{W}_{\mathrm{eff}}(\sigma)=-t(\sigma)+\pi i\sum_{\alpha>0} \alpha(\sigma)-\sum_{\sf{a}}Q_{\sf{a}}(\sigma)(\log(Q_{\sf{a}}(\sigma)/\Lambda)-1)
\end{equation}
where $\alpha>0$ denotes the set of positive roots of $G$, the sum $\sum_{\sf{a}}$ is over the set of all weights of $\rho$ and $\Lambda$ is a UV energy cut-off scale\footnote{Note, in the equation \eqref{genericWeff}, $\sigma$ has units of energy, (in natural units).}. Then, the generic component of the discriminant loci $\Delta\subset \exp(\mathfrak{z}^{\vee}_{\mathbb{C}})$ in the $t$-space can be computed by the system of algebraic equations:
\begin{equation}\label{genericWeffeqs}
\exp\left(\frac{\partial\widetilde{W}_{\mathrm{eff}}(\sigma)}{\partial \sigma_{a}}\right)=1,\qquad \sigma\in\mathfrak{t}_{\mathbb{C}}/W_{G},
\end{equation}
upon some choice of a basis for the vector space $\mathfrak{t}_{\mathbb{C}}$. Due to the condition on our GLSM models to be nonanomalous, the eqs. \eqref{genericWeffeqs} depends only on $t$, not in $\sigma$, however, it is well known that, there exists several models where $\Delta$ gets contributions from regions were the large VEV's $\langle \sigma\rangle$ are not generic \cite{Morrison:1994fr}. These are known as mixed Coulomb-Higgs branches. So, more precisely if we denote by $\Delta_{\mathrm{gen}}$ the solutions to \eqref{genericWeffeqs}, we have
\begin{equation}
\Delta_{\mathrm{gen}}\subseteq \Delta.
\end{equation}
Only for $G$ abelian, there exist an algorithmic way to compute these mixed components (and in some cases, they may not exist). For general $G$, we only have some explicit families of examples. We will be interested only on the phase crossing between $\zeta_{0}\ll-1$ and $\zeta_{0}\gg 1$, while keeping $\zeta_{\alpha}\gg 1$ for $\alpha=1,\ldots, r$. Therefore, for our computations, besides $\Delta_{\mathrm{gen}}$, the only relevant mixed Coulomb-Higgs branch are the ones that break the subgroup $U(1)\subset G$. In the case $G$ abelian, where $\vec{\bbP}:=\mathbb{P}^{m_{1}-1}\times\cdots\times \mathbb{P}^{m_{r}-1}$, we have that such a mixed branch can only exist if for each $I=n+2,\ldots,K$, there always exist at least one $\alpha=1,\ldots,r$ such that $n^{(\alpha)}_{I}$ does not vanish. Otherwise, if there exists $i_{*}\in\{1,\ldots,r\}$ such that $n^{(\alpha_{*})}_{I}=0$ for all $I=n+2,\ldots,K$, we cannot break just the subgroup $U(1)\subset G$. If we do, then $\zeta_{\alpha_{*}}$ can only take values in $\mathbb{R}_{\geq 0}$ giving us an invalid configuration \cite{Morrison:1994fr}. Therefore, we need to consider the mixed Coulomb-Higgs branch arising when we break at least $U(1)\times U(k_{\alpha_{*}})\subset G$ (with $k_{\alpha_{*}}=1$). In the latter case, we still need to check if $\zeta_{\alpha_{*}}\gg 1$ is compatible with the equations derived from the effective twisted potential in this sector.\\

\subsubsection{Comment on $\Delta$ for $G=U(1)\times U(1)$}\label{sec:comment}

The case $G=U(1)\times U(k_{1})$ is particularly interesting because of the relation between monodromy and the fundamental group of links in $S^{3}$ \cite{Ceresole:1993nz,Aspinwall:2001zq,Cota:2019cjx,Lin:2024fpz}. The equations for $\Delta_{\mathrm{gen}}$ are quite involved to analyze, for a generic nonanomalous model in the case $k_{1}>1$. Here we will content to give some general argument for the equations defining $\Delta_{\mathrm{gen}}$ for the case $k_{1}=1$. Upon a choice of basis for $\mathfrak{t}_{\mathbb{C}}\cong \mathbb{C}^{2}$, we can write \eqref{genericWeffeqs} as:
\begin{equation}\label{disc}
z_{\alpha}=\prod_{\sf{a}}(Q_{\sf{a}}^1\sigma_{1}+Q_{\sf{a}}^0\sigma_{0})^{Q_{\sf{a}}^\alpha}, \qquad \alpha=0,1,
\end{equation}
where $z_{\alpha}:=\exp(-t_{\alpha})$ are the exponentiated FI-theta parameters for $U(1)$ and $U(k_{1}=1)$, respectively. Imputing charges $Q^{\alpha}_{\sf{a}}$, determined by \eqref{eqn: CICYI}, one can reduce \eqref{disc} to:
\begin{equation}
    z_0=\prod_{I=1}^{n+1}(-n^{(1)}_I\tau-1)^{-1},\quad z_1=\tau^{m_{1}}\prod_{I=1}^K(-n^{(1)}_{I}\tau-1)^{-n^{(1)}_I}.
\end{equation}
where $\sum_{I=1}^{K}n^{(1)}_I=m_{1}$ and $\tau:=\sigma_{1}/\sigma_{0}$. We can solve for the parameter $\tau$ using the method of resultants and obtain a single polynomial equation (independent of $\tau$) $f(z_0,z_1)=0$ defining $\Delta_{\mathrm{gen}}$. For generic values of $n^{(1)}_I$, obtaining an explicit expression for $f(z_0,z_1)=0$ does not give something very illuminating. Instead, since $f(z_0,z_1)=0$ is independent of $\tau$, we propose to study it by Taylor expanding $(z_{0},z_{1})$ around $\tau=0$ and obtain an approximate expression for $f(z_0,z_1)$. Moreover, we change coordinates from $(z_{0},z_{1})$ to $(u_{0},u_{1})$, centered at the intersection $\{f=0\}\cap\{z_{1}=0\}$. At $\tau=0$, $\{f=0\}\cap\{z_{1}=0\}=((-1)^{n+1},0)$. Then, a first order approximation for the coordinates $(u_{0},u_{1})$ is given by:
\begin{equation}
    \begin{aligned}
        {u}_0:=&z_0-(-1)^{n+1}
        \\
        =&-\tau(-1)^{n+1}\sum_{I=1}^{n+1}n^{(1)}_{I}+\calO(\tau^2)
        \\
        {u}_1:=&z_1-0
        \\
        =&(- {\tau})^{m_{1}}+\calO(\tau^{m_{1}+1})
        \\
        =&\frac{{u}_0^{m_{1}}}{(-1)^{n+1}\sum_{I=1}^{n+1}n^{(1)}_I}+\calO(u_0^{m_{1}+1})
    \end{aligned}\label{eqn:Puiseux}
\end{equation}
which, together with the component $\{u_{1}=0\}$, parameterizes a $\mathfrak{L}^{(1)}_{m_{1}}$ link (a simple nested torus link, defined in Appendix \ref{sec:appknot}) up to higher terms, upon intersection with a 3-sphere $S^{3}$, of small radius, centered at $(u_{0},u_{1})=(0,0)$. When $f$ is irreducible, \eqref{eqn:Puiseux} is nothing but the Puiseux expansion of the hypersurface singularity $f_{m_{1}}(u_{0})=u_{1}$ where $f_{m_{1}}(u_{0})$ is a polynomial of degree $m_{1}$, namely:
\begin{equation}
    u_0=\tau,\quad u_1=a \tau^{m_1}+\mathcal{O}(\tau^{m_{1}-1}).
\end{equation}
Then, together with the component $\{u_1=0\}$, the expansion gives the simple nested torus link $\mathfrak{L}^{(1)}_{m_{1}}$. However, if $f$ is reducible: $f=f_1\cdots f_g$, with $g$ irreducible components, the Puiseux expansion for each component: 
\begin{equation}
    f_i(u_0,u_1)=0:\quad u_0=\tau^{p_i},\quad u_1=a_i\tau^{q_i}+\mathcal{O}(\tau^{q_{i}-1}),\qquad i=1,\ldots,g.
\end{equation}
If the leading coefficients of each component satisfy the strict inequalities $|a_1|<\cdots<|a_g|$, then $\{f=0\}\cup\{ u_1=0\}$, upon intersection with $S^{3}$, as before, a more general nested torus link with each component a $(p_i,q_i)$-torus knot. These nested torus links are studied in detail in \cite{Argyres:2019kpy}. If $p_{i}=1$ for all $i=1,\dots,g$, then we obtain a simple torus link $\mathfrak{L}^{(g)}_{q_{1},\ldots,q_{g}}$.

\subsection{Models for global flops for CY threefolds}\label{sec:globalflps}

The abelian models studied on this papers where studied previously in \cite{Brodie:2021toe}. In this subsection we review some aspects of their construction and generalize them to a broader family of models that can be fit into GLSMs. We will concentrate in the main actors of this these paper, namely CY threefolds, so, we fix from now on:
 \begin{equation}
\hat{c}=d=3
    \end{equation}
However, we do not study all the class of these models at one on the present work, we just content with the case of $\vec{\mathbb{P}}$ being a product of Grassmannians. Let us start by recalling the definition of a simple flop of CY threefolds:
\begin{defi}
Let $X_+$ and $X_-$ be smooth CY threefolds and $\pi_\pm$ are small contractions (i.e. isomorphisms in codimension $2$) of $N$ curves $\{ C_\pm^{(i)} \}\subset X_\pm$, $i=1,\ldots,N$ to singular points $\{ P^{(i)} \}\in X_{\mathrm{sing}}$:
\begin{equation}
\begin{aligned}
    X_+\xrightarrow{\pi_+} X_{\mathrm{sing}}\xleftarrow{\pi_-} X_-
\end{aligned}
\end{equation}
Then the birational map 
\begin{equation}
   (\pi_-)^{-1}\circ \pi_+:\ X_+\dashrightarrow X_-
\end{equation}
is a flop if there exists divisors $D_\pm\subset X_\pm$ such that $(\pi_-)^{-1}\circ \pi_+(D_+)=D_-$ and  for all $i$, $D_+\cdot C_+^{(i)}>0$ and $D_-\cdot C_-^{(i)}<0$.
\end{defi}
Let us start reviewing the splitting configuration first i.e. $k_1=\cdots=k_r=1$. Then, the phases $X_{\zeta_\pm}$ are equipped with small contractions to the singular variety $X_{\mathrm{sing}}$:
\begin{equation}
    X_{\pm}\xrightarrow{\pi_\pm} X_{\mathrm{sing}}\subset\vec\bbP.
\end{equation}
Denote $F_I(x,y)=F^j_I(x)y_j$, $F(x)$ the $(n+1)\times (n+1)$ matrix whose entries are $F^j_I(x)$ and $F_{\hat I}^{\hat j}$ the sub-matrix of $F$ obtained by deleting the $I$th row and $j$th column. Then, $X_{\mathrm{sing}}$ is given by\footnote{Conventionally, $\vec\bbP[\vec n,\vec n_{n+2},\cdots,\vec n_K]$ denotes always a smooth configuration (e.g. the $X^\natural$ later in this paper) but not a singular one like here.}
\begin{equation}
      X_{\mathrm{sing}}=\vec\bbP[ (\det F=0)  ,\vec{n}_{n+2},\cdots,\vec{n}_K]=\vec\bbP[ \vec{n}  ,\vec{n}_{n+2},\cdots,\vec{n}_K], \qquad \vec{n}:=\sum_{I=1}^{n+1}\vec{n}_{I}.
    \end{equation}
The singular loci in $X_{\mathrm{sing}}$ is given by the points:
 \begin{equation}\label{singpts}
        \vec\bbP[ (\operatorname{rank}F<n)  ,\vec{n}_{n+2},\cdots,\vec{n}_K].
    \end{equation}
For threefolds, it is straightforward to see, using for example the elementary results found in \cite{Jockers:2012zr}, that \eqref{singpts} are indeed codimension $3$ when $\operatorname{rank}F=n-1$ and the exceptional divisors corresponds to 
\begin{equation}\label{exceptdiv}
        \{y\in\mathbb{P}^{n}:y\in \mathrm{Ker}F\}\subset \mathbb{P}^{n}.
    \end{equation}
Moreover, in the generic case, we cannot have points $x\in\vec{\mathbb{P}}$ such that $\operatorname{rank}F<n-1$, because their codimension will be $>3$. 
There are two natural classes of divisors on $X_{+}$ given by
    \begin{itemize}
        \item[1.] $H_{(i)}=\{ y_i=0 \}\cap X_+$ that intersects the exceptional divisors \eqref{exceptdiv} at a point and contracts to $X_{\mathrm{sing}}$ as
        \begin{equation}
        \pi_+(H_{(i)})=\vec\bbP[ (\operatorname{rank}F^{\hat i}<n)  ,\vec{n}_{n+2},\cdots,\vec{n}_K].
    \end{equation}

        \item[2.] $\calH_{{(I)}}=\{{\det F_{\hat I}^{\hat j}}/{y_j}=0\}\cap X_+$, an effective divisor on $X_+$ which is well defined on $X_+$\footnote{Indeed one can show that the divisors $D_{I,j}:=\{{\det F_{\hat I}^{\hat j}}/{y_j}=0\}\cap X_+$ and $D_{I,i}:=\{{\det F_{\hat I}^{\hat i}}/{y_i}=0\}\cap X_+$ are linearly equivalent. For this simply note that $\det F_{\hat I}^{\hat j}=\varepsilon_{j}\mathrm{adj}(F)^{I}_{j}$ i.e., up to a sign $\varepsilon_{j}\in\{\pm1\}$, $\det F_{\hat I}^{\hat j}$ corresponds to the $(I,j)$ component of the adjugate matrix $\mathrm{adj}(F)$ of $F$. Moreover, if $\operatorname{rank}F=n$, the $\operatorname{rank}\mathrm{adj}(F)=1$ and we can write $\mathrm{adj}(F)^{I}_{j}=v^{I}u_{j}$, where $v,u$ are vectors and in particular, $u$ satisfies $F_{I}^{k}u_{k}=0$ for all $I$. Therefore, when intersected with $X_{+}$, we can write, up to a proportionality constant (that we can absorb in $v$) $\mathrm{adj}(F)^{I}_{j}=v^{I}y_{j}$. Hence, on $X_+$, $\varepsilon_{j}D_{I,j}-\varepsilon_{i}D_{I,i}=0$. If $\operatorname{rank}F<n$ this identity is authomatically satisfied. Therefore $D_{I,j}$ $D_{I,i}$ are linearly equivalent as divisors in $X_{+}$.}. The exceptional divisors are contained in  $\calH_{{(I)}}$ and contracts to $X_{\mathrm{sing}}$ as
        \begin{equation}
            \pi_+(\calH_{(I)})=\vec\bbP[ (\operatorname{rank}F_{\hat I}<n)  ,\vec{n}_{n+2},\cdots,\vec{n}_K].
        \end{equation}
    \end{itemize}
    
    Similarly, there are two other class of divisors $H_{(I)}'$ and $\calH_{(i)}'$ on $X_-$ given by exchanging $y_i\leftrightarrow p_I$ and $F\leftrightarrow \;^tF$. Then 
    
\begin{prop}\label{propbrodie}\cite{Brodie:2021toe}
    Denote the basis of divisor classes that is dual to the curve classes $(C_0,C_1,\cdots, C_r)$ of $\bbP^n\times\vec\bbP$ (and equivalently, that of $\bbP(\oplus_I\mathcal{R}_I)\rightarrow\vec\bbP$) as
    \[
    (D_0,\vec D)=(D_0,D_1,\cdots,D_r).
    \]
     Let $\vec n=\vec n_1+\cdots + \vec n_{n+1}$, then the above classes are given by\footnote{A class $(a,\vec b)$ means $aD_{0}+\vec{b}\cdot \vec{D}$.}
    \begin{equation}
        \begin{aligned}
            X_+:\quad& [H_{(i)}]=(1,\vec 0),\quad [\calH_{(I)}]=(-1,\vec n-\vec n_I)
            \\
            X_-:\quad& [H'_{(I)}]=(1,-\vec n_I),\quad [\calH'_{(i)}]=(-1,\vec n).
        \end{aligned}
    \end{equation}
    Moreover,
    \begin{equation}
        \pi_+(H_{(i)})=\pi_-(\calH_{(i)}'),\quad \pi_+(\calH_{(I)})=\pi_-(H_{(I)}'),
    \end{equation}
    and from their classes is clear that they satisfy the intersection conditions. Thus $X_+$ and $X_-$ form a flop.
\end{prop}

\begin{eg}
    Consider $n=1$, the defining polynomial is in a $2\times2$ matrix as
    \begin{equation}
        F=\begin{pmatrix}
            F^{1}_{1}(x) & F^{2}_{1}(x) \\ F^{1}_{2}(x) & F^{2}_{2}(x)
        \end{pmatrix}
    \end{equation}
   then, the singular points, upon contraction, are $P:=\{ F=0 \}\subset X_{\mathrm{sing}}=\{\det F=0\}\subset Y$, where $Y:=\vec\bbP[\vec n_{3},\cdots,\vec n_K]$, then
    \begin{equation}
    \begin{aligned}
        X_+=&\{ F\cdot y =0\}\subset Y_+:=\bbP^1\times Y
        \\
        X_-=&\{ p\cdot F=0  \}\subset Y_-:=\bbP(\oplus_I\mathcal R_I)\rightarrow Y
    \end{aligned}
    \end{equation}
where, by abuse of notation, we denoted $\mathcal{R}_{I}=\mathcal{O}(-\vec{n}_{I})$, $I=1,2$. The preimage of $P$ corresponds to the curves $C_+=\bbP^1=\pi_+^{-1}(P)$ (resp. $C_-=\bbP(\oplus_I\mathcal{R}_I))=\pi_-^{-1}(P)$). Consider the divisor 
    \begin{equation}
        H_+=\{y_1=0\}\cap X_+=\{ F^{2}_{1}=F^{2}_{2}=y_1=0 \}\subset Y_+.
    \end{equation}
    Thus $H_{+}\cdot C_+=1>0$ and $\pi_-^{-1}\circ\pi_+(H_{+})$ is given by 
    \begin{equation}
        \calH_-=\{ F^{2}_{1}=F^{2}_{2}=p_1F_{1}^{1}+p_2F_{2}^{1}=0\}\subset Y_-,
    \end{equation}
    which contains $C_-$ entirely. Thus $H_-\cdot C_-=-1<0$.
    
\end{eg}

The previous family of examples, realized as abelian GLSMs, have a natural generalization. Consider a smooth, projective Fano variety $Y$ and a nef partition of $-K_{Y}$ given by
\begin{equation}
   -K_Y= \calL_1+\cdots+\calL_{n+1} \label{eqn:nefp}
\end{equation}
where $\mathcal{L}_I$, $I=1,\ldots, n+1$ denotes a collection of line bundles (or divisors) on $Y$. Denote by $\mathcal{V}$ the trivial rank $n+1$ bundle over $Y$ and $\mathcal{N}:=\oplus_{I=1}^{n+1}\mathcal{L}_{I}$. Given a generic section 
\begin{equation}
F\in\Gamma(Y,\mathrm{Hom}(\calV,\calN)),
\end{equation}
we can define the smooth CY threefolds $X_{\pm}$ as
\begin{equation}\label{gencase}
\begin{aligned}
    X_+=&\{ F\cdot y=0 \}\subset Y_+,\quad Y_+=\bbP(\calV)\rightarrow Y,
    \\
    X_-=&\{ p\cdot F=0 \}\subset Y_-,\quad Y_-=\bbP(\calN^\vee)\rightarrow Y,
\end{aligned}
\end{equation}
where $y$ and $p$ denote coordinates on the fibers of $Y_{+}$ and $Y_{-}$, respectively. The proof of proposition \ref{propbrodie} in \cite{Brodie:2021toe} can be straightforwardly generalized to this case, and so we can state

\begin{prop}\label{propbrodiegeneral} The CY threefolds $X_{\pm}$ defined in \eqref{gencase} with $\pi_{\pm}:X_{\pm}\rightarrow X_{\mathrm{sing}}$, where $X_{\mathrm{sing}}$ is defined in \eqref{gensing} form a flop.
\end{prop}

\begin{cor}
    The phases $X_{\zeta_\pm}$ for $\vec\bbP=Gr(k_1,m_1)\times\cdots Gr(k_r,m_r)$ form a flop.
\end{cor}
\begin{rmk}\label{rmkbundles}
    One can propose more general nonabelian GLSMs realizing cases where the nef partition of $-K_{Y}$ is replaced by a vector bundle $\mathcal{N}$ satisfying $c_1(\calN)=-c_1(K_Y)$. This include, for example, the PAX models proposed in \cite{Jockers:2012zr}. 
\end{rmk}
In the general case \eqref{gencase}, we have 
\begin{equation}\label{gensing}
X_{\mathrm{sing}}=\{\operatorname{rank}F<n+1\}\subset Y,
\end{equation}
$X_{\text{sing}}$ admits a deformation to a smooth CY $X^\natural\subset Y$. This corresponds to an extremal transition, where the singular points are deformed to 3-spheres $S^{3}$ \cite{Candelas:1987kf,Green:1988wa,Candelas:1989js}:
\begin{equation}\label{extrtransition}
    \vcenter{\xymatrix{
    X_+ \ar[dr]^{\pi_+}&  &   
    \\
      &  X_{\text{sing}}  \ar[r]^{
      \text{def}
      } & X^\natural
    \\
    X_- \ar[ur]_{\pi_-}&  &
    }}
\end{equation}
As a consequence, the Euler number between conifold transition changes by
\begin{equation}
    \chi(X_\pm)-\chi(X^\natural)=N(\chi(\bbP^1)-\chi(S^3))=2N.
\end{equation}
where $N$ corresponds to the number of singular points in $X_{\mathrm{sing}}$ and can be computed the Giambelli-Thom-Porteous formula in co-dimension four \cite{fulton2013intersection, Jockers:2012zr}:
\begin{equation}
    [\text{Sing}(F)]=\left\vert\begin{array}{cc}
        c_2(\calN) & c_3(\calN)  
         \\
        c_1(\calN) & c_2(\calN)
    \end{array}\right\vert= N[P]_Y.
\end{equation}
where $[P]_Y$ is the cohomology class of a point in $Y$, normalized as $\int_Y[P]_Y=1$ and then, $N$ is exactly the number of singular points in $X_{\text{sing}}$. This number will  show up in many of our monodromy computations.

\section{Universal monodromy from window categories} \label{section3}

An important characteristic of the GLSMs under consideration is the existence of a vector R-symmetry $U(1)_{V}$. The action on the fields is given by a, not necessarily faithful, representation $R:U(1)_{V}\rightarrow GL(V)$ (hence, the weights can be real) under which the superpotential $W:V\rightarrow \mathbb{C},\phi\mapsto W(\phi)$ (where $\phi$ denote the coordinates of $V$) satisfies 
\begin{equation}
   W(R(\lambda)\cdot \phi)=\lambda^{2}W(\phi),\quad\text{for all \ } \lambda\in U(1)_V.
\end{equation}
It is clear that at least one representation $R$ exists for our class of superpotentials. In general, the weights of $R$ cannot be fixed in a GLSM, since it is subjected to ambiguities. At an IR fixed point, it is expected that the weights of $R$ are fixed, under RG flow, to values in the interval $(0,2)$ \cite{Lerche:1989uy}. These values can be different, for different phases (or chambers) in the stringy K\"ahler moduli $\mathcal{M}_{K}$ (defined below).\\

\begin{figure}[h]
\centering
\includegraphics[width=.6\textwidth]{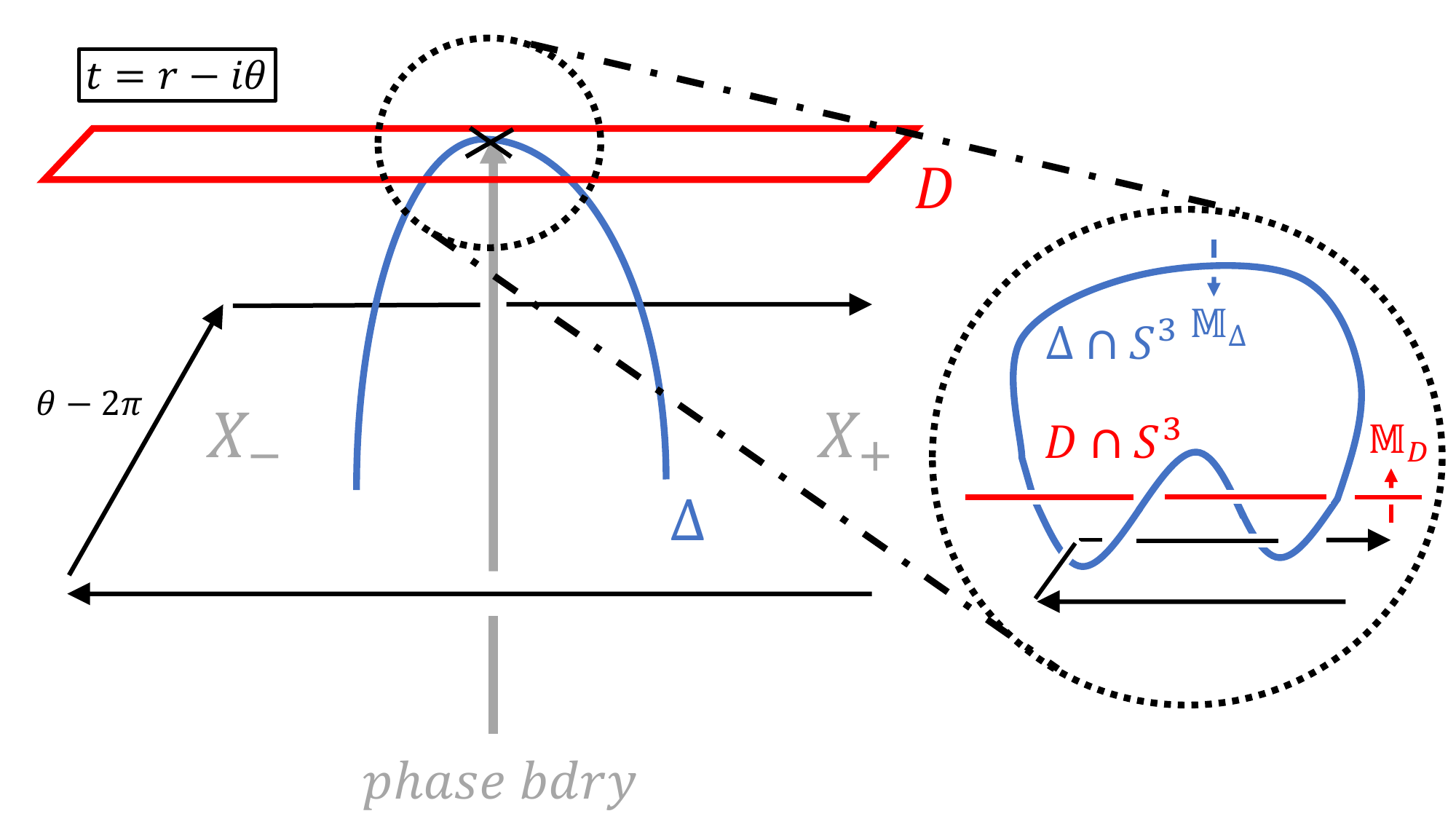}
\caption{Here the monodromy $M$ is illustrated by black arrows and $D:=\{z_{1}=0\}$. We write $\mathbb{M}_{\Delta}$ and $\mathbb{M}_{D}$ for the monodromies around the components  $S^{3}\cap D$ and $S^{3}\cap \Delta$, respectively.}\label{fig:windowshift2}
\end{figure}

B-type i.e. $\mathbf{2}_{B}\subset (2,2)$ supersymmetry preserving boundary conditions on GLSMs equipped with a superpotential $W$ are characterized by $G$-equivariant matrix factorizations of $W$, given by the triple $\calB:=(\mathbf{T},\rho_{M},R_{M})$\cite{Herbst:2008jq,Hori:2013ika} where $\mathbf{T}\in\mathrm{End}^{\mathrm{odd}}(M)$, is an odd endomorphism of the finite, $\mathbb{Z}_{2}$-graded, free $\mathrm{Sym} V^{\vee}$-module (it is customary to denote $M=M_{0}\oplus M_{1}$ for the even and odd factors of $M$). This triplet $\calB$ satisfies:\cite{Herbst:2008jq,Hori:2013ika}
\begin{equation}
\begin{aligned}
    \mathbf{T}^2&=W\cdot\Id_{M},
    \\
    \rho_M^{-1}(g) \mathbf{T}(\rho(g)\cdot\phi)\rho_M(g)&= \mathbf{T}(\phi),\quad \text{for all \ }g\in G
    \\
    R_{M}(\lambda) \mathbf{T}(R(\lambda)\cdot\phi)R_{M}^{-1}(\lambda)&=\lambda \mathbf{T}(\phi),\quad\text{for all \ } \lambda\in U(1)_V.
\end{aligned}\label{eqn:MF}
\end{equation}
We will denote the category where B-branes belong, as $MF_{G}(W)$ (see \cite{ballard2019variation} for more details, in a more general case). B-type boundary conditions on GLSMs need more data than just $\calB$. The boundary conditions for the vector multiplet must also be specified. The complete analysis of B-type boundary conditions in GLSMs, leads to the Grade Restriction Rule (GRR) \cite{Herbst:2008jq} and the definition of window categories (which are subcategories of $MF_{G}(W)$ \cite{ballard2019variation,halpern2015derived,segal2011equivalences}). We will not need to recall in detail the definition of the vector multiplet boundary condition \cite{Herbst:2008jq,Hori:2013ika} since we will be concerned with a very specific class of monodromies. The stringy K\"ahler moduli space $\mathcal{M}_{K}$ of a GLSM is the space spanned by the (exponentiated) FI-theta parameters $\exp t$. This space takes generically the form $(\mathbb{C}^{*})^{\mathrm{rk}\mathfrak{z}}\setminus \Delta$ and the GLSM provides a natural compactification of it \cite{Morrison:1994fr}. It is well known that $\mathcal{M}_{K}$ is subdivided into open chambers termed \emph{phases} \cite{Witten:1993yc}. For a nonanomalous GLSM, each point of $\mathcal{M}_{K}$ determines a SCFT (by RG flow to the IR fixed point), and B-type boundary conditions of a $\mathcal{N}=(2,2)$ SCFT form a triangulated category (more precisely a $A_{\infty}$ category, see for example \cite{Aspinwall:2009isa} for a review). If we denote this latter category $\mathcal{C}_{c}$ for $t$ in a given chamber $c$ of $\mathcal{M}_{K}$, we always have a projection functor:
\begin{equation}\label{projfunctor}
\pi_{c}:MF_{G}(W)\rightarrow \mathcal{C}_{c}.
\end{equation}
When a chamber $c$ corresponds to a geometric phase, where the IR fixed point corresponds to the fixed point of a sigma model with target space $X$, we have
\begin{equation}
\mathcal{C}_{c}\cong D(X):=D^{b}\mathrm{Coh}(X),\qquad c \text{ \ is geometric}.
\end{equation}
Given a point in the covering $\widetilde{\mathcal{M}}_{K}\rightarrow\mathcal{M}_{K}$, defined simply by the logarithmic map i.e. the coordinates $t\in\widetilde{\mathcal{M}}_{K}$ are just defined by 'unwrapping' the $\theta$ coordinate, allowing it to take values in $\mathbb{R}$. A point in $\widetilde{\mathcal{M}}_{K}$ determines a subcategory of $\mathbb{W}\subset MF_{G}(W)$ called a window category. In the following, we will describe it for the cases of interest. Consider the model of Sect. \ref{sec:flopsGLSM} and the phases:
\begin{equation}\label{pmphases}
\zeta_{\pm}:=(\pm\zeta_{0}\gg 1,\zeta_{1}\gg 1,\ldots,\zeta_{r}\gg 1)
\end{equation}
then, the window subcategory defined by a straight line running between phases $\zeta_{\pm}$ is defined by restricting the weights of $\rho_{M}$ to the band\footnote{It is called a 'band'\cite{Herbst:2008jq} because the rest of the weights are unrestricted.}:
\begin{equation}\label{abGRR}
-\frac{n+1}{2}<q^{0}+\frac{\theta^{0}}{2\pi}<\frac{n+1}{2}
\end{equation}
where $q^{0}$ denotes the weight of the $U(1)\subset G$ subgroup. Then the window subcategories, became, explicitly:
\begin{equation}\label{windowabdef}
\mathbb{W}(l):=\{\mathcal{B}\in MF_{G}(W):\text{weights of \ } \rho_{M}\text{ \ satisfy \eqref{abGRR} }\},\qquad
l:=\left\lfloor{\frac{\theta^{0}}{2\pi}}\right\rfloor.
\end{equation}
Using the notation of Sect. \ref{sec:comment}, when $\mathrm{dim}( \calM_K)=2$, the monodromy giving rise to the window categories $\mathbb{W}(l)$ corresponds to loops running parallel to $\{z_{1}=0\}\cap S^{3}$, surrounding the point $(z_{0},z_{1})=((-1)^{n+1},0)$ (in the abelian case). Therefore we expect that the autoequivalence $M\in \mathrm{Aut}(D(X_{+}))$ corresponding to this monodromy can be decomposed in a composition of other elements of $\mathrm{Aut}(D(X_{+}))$ consistent with the relations of $\pi_1(S^3\setminus\mathfrak{L})$ where $\mathfrak{L}=S^{3}\cap (\{z_{1}=0\}\cup \Delta)$. This is sketched in figure~\ref{fig:windowshift2}.

\subsection{B-branes for the Doran-Morgan basis}

In Sect. \ref{sec:Aperiods} we will concentrate on the action of the monodromy on A-periods, for which is very convenient to work on the Doran-Morgan basis in \cite{Doran:2005gu, Almkvist:2005qoo}, that is described in detail in  Sect. \ref{sec:Aperiods}. For the geometric category $D(X)$ working on this basis translates on a choice of generators corresponding to the structure sheaves of point, curves and divisors as well as the structure sheaf $\mathcal{O}_{X}$. In this section, we will proceed to construct the corresponding objects in $MF_{G}(W)$, that project to $D(X)$ as \eqref{projfunctor} as well as their grade restrictions to different window categories $\mathbb{W}(l)$. The Chan-Paton space $M$ of an object $\mathcal{B}\in\mathbb{W}(l)$ must be a direct sum of the following subclass of representations of $G$:
\begin{equation}
    \calW(-n+l,\vec \lambda),\ \calW(-n+1+l,\vec \lambda),\ \cdots,\ \calW(-1+l,\vec \lambda),\ \calW(l,\vec \lambda),\quad \text{for any }\vec\lambda. \label{eqn:W}
\end{equation}
Where $\vec\lambda=(\lambda_{1},\ldots,\lambda_{r})$ is a collection of (generalized) Young diagrams characterizing an irreducible representation $R_{\vec\lambda}$ of $U(k_{1})\times \cdots\times U(k_{r})$ (see for example \cite{Taylor} for a review), and
\begin{equation}
    \calW(a,\vec \lambda):=\mathbb{C}(a)\otimes R_{\vec\lambda}
\end{equation}
however of particular interest for us will be when $R_{\vec\lambda}$ corresponds to a tensor product of determinant representations, so we will use a special notation for it:
\begin{equation}\label{notationdet}
    \calW(a,\mathbf{d}):=\mathbb{C}(a)\bigotimes_{i=1}^r \mathrm{det}^{d_{i}},\quad \mathbf{d}\in\mathbb{Z}^{r}.
\end{equation}
For convenience we write the superpotential \eqref{superpot} as a sum of two potentials:
\begin{equation}
W=W_{1}+W_{2}\qquad W_{1}:=\sum_{I=1}^{n+1}p_{I}F_{I}(y,x),\quad W_{2}:=\sum_{I=n+2}^{K}p_{I}F_{I}(x),
\end{equation}

\section*{Empty branes and structure sheaf branes}

We consider the following object $\mathcal{B}_{\pm}=(\mathbf{T}_{\pm},\rho_{M_{\pm}},R_{M_{\pm}})\in MF_{G}(W_{1})$  where
\begin{eqnarray}
\mathbf{T}_{+}&:=&\sum_{I=1}^{n+1}\left(y_{I}\eta^{I}+\sum_{J=1}^{n+1}p_{J}\frac{\partial}{\partial y_{I}}F_{J}(y,x)\bar{\eta}_{I}\right)\nonumber\\
\mathbf{T}_{-}&:=&\sum_{I=1}^{n+1}\left(p_{I}\eta^{I}+F_{I}(y,x)\bar{\eta}_{I}\right),
\end{eqnarray}
where $\eta^{I},\bar{\eta}_{I}$ are Clifford matrices satisfying $\{\eta^{I},\bar{\eta}_{J}\}=\delta^{I}_{J}\mathrm{id}$. The spaces $M_{\pm}$ are constructed by choosing a Clifford vacuum $|\Omega\rangle$ satisfying ${\eta} |\Omega\rangle=0$. We choose $|\Omega\rangle$ on the trivial representation of $G$ and with zero R-charge. Then (using the notation \eqref{notationdet})
\begin{eqnarray}
M_{+}&=&\bigoplus_{j=0}^{n+1}\wedge^{j}\mathcal{V}_{+},\qquad \mathcal{V}_{+}:=\mathbb{C}(1)^{\oplus n+1}\nonumber\\
M_{-}&=&\bigoplus_{j=0}^{n+1}\wedge^{j}\mathcal{V}_{-},\qquad \mathcal{V}_{-}:=\bigoplus_{I=1}^{n+1}\mathcal{W}(-1,(-n_{I}^{(1)},\ldots,-n_{I}^{(r)})).
\end{eqnarray}
Given any object $\widetilde{\mathcal{B}}\in MF_{G}(W_{2})$ and using the following definition:
\begin{defi}(Definition 1.2 in \cite{Yoshino_1998}). The Thom-Sebastiani sum of matrix factorizations corresponds to the operation:
    \begin{equation}
        \otimes:\ \mathrm{MF}(W_1)\times \mathrm{MF}(W_2)\rightarrow \mathrm{MF}(W_1+W_2).
    \end{equation}
In particular, the Chan-Paton space of the image is the (graded) tensor product of the Chan-Paton space of each of its factors.
\end{defi}
We can then show that
$\mathcal{B}_{\pm}\otimes \widetilde{\mathcal{B}}\in MF_{G}(W)$ satisfies:
\begin{eqnarray}
\pi_{\zeta_{\pm}}(\mathcal{B}_{\pm}\otimes \widetilde{\mathcal{B}})\text{ \ is null homotopic in\ }D(X_{\zeta_{\pm}}).
\end{eqnarray}
we will consider the special case $\widetilde{\mathcal{B}}_{\calO}=(\widetilde{\mathbf{T}}_{\calO},\rho_{\widetilde{M}_{\calO}},R_{\widetilde{M}_{\calO}})$ with
\begin{eqnarray}
\widetilde{\mathbf{T}}_{\calO}:=\sum_{I=n+2}^{K}\left(F_{I}(x)\eta^{I}+p_{I}\bar{\eta}_{I}\right)
\end{eqnarray}
then
\begin{eqnarray}
\mathcal{E}_{\pm}(q,\vec\lambda):=\mathcal{B}_{\pm}(q,\vec\lambda)\otimes \widetilde{\mathcal{B}}_{\calO}
\end{eqnarray}
where $(q,\vec\lambda)$ denotes the choice of representation of the Clifford vacuum on $\mathcal{B}_{\pm}$. For convenience we write $\widehat{\mathcal{E}}_{\pm}(q,\vec\lambda)\in MF_{G}(W)$ for exactly the same objects defined above but with the role of $\bar{\eta}$ and $\eta$ exchanged for the $\mathcal{B}_{\pm}$ component, respectively. It can be argued then that:
\begin{eqnarray}
\pi_{\zeta_{+}}(\widehat{\mathcal{E}}_{-}(q,\vec\lambda))\simeq \mathcal{O}_{X_{\zeta_{+}}}(q,\vec\lambda)\in D(X_{\zeta_{+}}),\nonumber\\
\pi_{\zeta_{-}}(\widehat{\mathcal{E}}_{+}(q,\vec\lambda))\simeq \mathcal{O}_{X_{\zeta_{-}}}(-q,\vec\lambda)\in D(X_{\zeta_{-}}),
\end{eqnarray}

\section*{B-branes for divisor classes}

Consider the matrix factorization $\mathcal{B}_{\alpha}$ of $W\equiv 0$ given by
\begin{eqnarray}
l^{(\alpha)}(y,x)\eta
\end{eqnarray}
where $l^{(\alpha)}(y,x)$, $\alpha=0,\ldots, r$ is a collection of homogeneous linear functions of the coordinates of $V$, satisfying: $l^{(0)}(y,x)=l^{(0)}(y)$ and $l^{(i)}(x)=l^{(i)}(B^{(i)}(x))$ is a linear function of the Pl\"ucker coordinates $B^{(i)}(x)$ only. Then
\begin{eqnarray}
\mathcal{B}_{D_{\alpha}}:=\widehat{\mathcal{E}}_{-}(0,0)\otimes\mathcal{B}_{\alpha}
\end{eqnarray}
and we have that
\begin{eqnarray}
\pi_{\zeta_{+}}(\mathcal{B}_{D_{\alpha}})=\mathcal{O}_{D_{\alpha}}\in D(X_{\zeta_{+}})
\end{eqnarray}
where $D_{\alpha}=\{l^{(\alpha)}(y,x)=0\}\cap X_{\zeta_{+}}$.

\section*{B-branes for curve and point classes}

Our objective is to construct objects in $MF_{G}(W)$ that are mapped to structure sheaves of points and curves under $\pi_{\zeta_{+}}$. It is easier to start from the fact that, since we are dealing with a CICY, we can simply restrict to study curves and points in the ambient space $\mathbb{P}^{n}\times \vec{\mathbb{P}}$. For this consider the following collection of sections:
\begin{eqnarray}
H(x,y)&\in&\Gamma\left(\mathbb{P}^{n}\times \vec{\mathbb{P}},\ \mathcal{O}^{n}_{\mathbb{P}^{n}}(1)\bigoplus_{\alpha=1}^{r}(\mathcal{S}^{\vee(\alpha)})^{m_{\alpha}-k_{\alpha}}\right)\nonumber
\\
H_{\mu}(x,y)&\in&\Gamma\left(\mathbb{P}^{n}\times \vec{\mathbb{P}},\ \mathcal{O}^{n}_{\mathbb{P}^{n}}(1)\bigoplus_{\alpha=1}^{r}(\mathcal{S}^{\vee(\alpha)})^{m_{\alpha}-k_{\alpha}-\delta_{\alpha,\mu}}\bigoplus(\det\calS^{\vee(\mu)})^{\oplus(k_\mu-1)}\right)\nonumber
\\
H_{0}(x,y)&\in&\Gamma\left(\mathbb{P}^{n}\times \vec{\mathbb{P}},\ \mathcal{O}^{n-1}_{\mathbb{P}^{n}}(1)\bigoplus_{\alpha=1}^{r}(\mathcal{S}^{\vee(\alpha)})^{m_{\alpha}-k_{\alpha}}\right),
\end{eqnarray}\label{pointcurvesects}
with $\mathcal{S}^{\vee(\alpha)}$ and $\det\calS^{\vee (\alpha)}$ denoting the dual of the rank $k_{\alpha}$ tautological sheaf and tautological line bundle over $G(k_{\alpha},m_{\alpha})$ (pulled back to $\mathbb{P}^{n}\times \vec{\mathbb{P}}$). The sections are chosen such that $\{H=0\}$, $\{H_{\mu}=0\}$ and $\{H_{0}=0\}$ define a point and curves contained in $\mathbb{P}^{n}\times \vec{\mathbb{P}}$, $G(k_{\mu},m_{\mu})\subset \mathbb{P}^{n}\times \vec{\mathbb{P}}$ and $\mathbb{P}^{n}\subset \mathbb{P}^{n}\times \vec{\mathbb{P}}$, respectively, and moreover, all contained in $X_{\zeta_{+}}$. Then is straightforward to lift these sections to representations of $G$, namely
\begin{eqnarray}\label{repsH}
H(x,y)&\rightarrow& \mathbb{C}^{n}(1)\bigoplus_{\alpha=1}^{r}\mathbf{k}_{\alpha}^{m_{\alpha}-k_{\alpha}}\nonumber
\\
H_{\mu}(x,y)&\rightarrow& \mathbb{C}^{n}(1)\bigoplus({\det}_{\mu})^{\oplus (k_\mu-1)}\bigoplus_{\alpha=1}^{r}\mathbf{k}_{\alpha}^{m_{\alpha}-k_{\alpha}-\delta_{\alpha,\mu}}
\nonumber
\\
H_{0}(x,y)&\rightarrow &\mathbb{C}^{n-1}(1)\bigoplus_{\alpha=1}^{r}\mathbf{k}_{\alpha}^{m_{\alpha}-k_{\alpha}}.
\end{eqnarray}
Moreover, because of Hilbert's Nullstallensatz, we can factorize the superpotential in the form $W=H\cdot L$, where $H$ can be either $H(x,y)$, $H_{0}(x,y)$ or $H_{\mu}(x,y)$ and $L$ is a polynomial in $(p,x,y)$ variables that transforms on the dual representation of the one pointed out in \eqref{repsH}, respectively. Then, we have the following matrix factorizations:
\begin{eqnarray}\label{mfpointcurve}
\mathbf{T}_{\mathrm{pt}}&:=& \left(H(x,y)\cdot \eta+L(p,x,y)\cdot\bar{\eta}\right)\nonumber
\\
\mathbf{T}_{C_0}&:=& \left(H_{0}(x,y)\cdot \eta+L_{0}(p,x,y)\cdot\bar{\eta}\right)\nonumber
\\
\mathbf{T}_{C_\mu}&:=& \left(H_{\mu}(x,y)\cdot \eta+L_{\mu}(p,x,y)\cdot\bar{\eta}\right),\qquad \mu=1,\ldots, r\nonumber
\\
\end{eqnarray}
where, again $\eta$ must be chosen to transform in the representation of $G$ dual to \eqref{repsH}. The Clifford
vacuum on each case of \eqref{mfpointcurve} is chosen to have trivial charges, then this completely determines an object of $MF_{G}(W)$ which we denote $\mathcal{B}_{\mathrm{pt}}$, $\mathcal{B}_{C_0}$ and $\mathcal{B}_{C_\mu}$, respectively. They satisfy:
\begin{eqnarray}
\pi_{\zeta_{+}}(\mathcal{B}_{\mathrm{pt}})=\mathcal{O}_{\mathrm{pt}},\qquad \pi_{\zeta_{+}}(\mathcal{B}_{\mathrm{C_0}})=\mathcal{O}_{C_{0}},\qquad \pi_{\zeta_{+}}(\mathcal{B}_{C_\mu})=\mathcal{O}_{C_\mu}.
\end{eqnarray}

\subsection{Monodromy functor from window categories}\label{sec:wincatMon}

Now we have all set to compute the monodromy functor associated to a loop, surrounding the phase boundary corresponding to the geometric phases $X_{\zeta_{+}}$ and $X_{\zeta_{-}}$ in $\mathcal{M}_{K}$ with base point near $\zeta_{+}$ i.e., near the limit point of the geometric chamber corresponding to $X_{\zeta_{+}}$. To be more precise, we take a loop that keeps $z_{i}=\exp(-t_{i})=\varepsilon_{i}=\mathrm{const.}$, $i=1,\ldots,r$ fixed at $|\varepsilon_{i}|\ll 1$ for all $i$ and we only vary $z_{0}$ on a loop surrounding the points
\begin{eqnarray}
\Delta\cap\bigcap_{i=1}^{r}\{ z_{i}=\varepsilon_{i}\}\subset \mathcal{M}_{K} 
\end{eqnarray}
once. This loop is sketched, for the case 
$\mathrm{dim}\mathcal{M}_{K}=2$ in figure
\ref{fig:windowshift2}. We can then fix $\theta_{0}$ so that the corresponding window categories \eqref{windowabdef} for the loop corresponds to the following sequence of equivalences:
\begin{eqnarray}
F_{\mathcal{E}_{-}}&:&D(X_{\zeta_{+}})\cong \mathbb{W}(0)\rightarrow\mathbb{W}(-1),\nonumber\\
F_{\mathcal{E}_{+}}&:&\mathbb{W}(-1)\rightarrow\mathbb{W}(0)\cong D(X_{\zeta_{+}})
\end{eqnarray}
where $F_{\mathcal{E}_{\pm}}$ corresponds to taking cones with (twists of) $\mathcal{E}_{\pm}\in MF_{G}(W)$. This will result on the monodromy functor:
\begin{eqnarray}
M_{0}:=F_{\mathcal{E}_{+}}\circ F_{\mathcal{E}_{-}}\in\mathrm{Aut}(\mathbb{W}(0))\cong \mathrm{Aut}(D(X_{\zeta_{+}})).
\end{eqnarray}
We face the problem that not all the objects $\mathcal{B}_{\mathrm{pt}},\mathcal{B}_{D_{\alpha}},\mathcal{B}_{\mu},\widehat{\calE}_{-}\in MF_{G}(W)$ are in $\mathbb{W}(0)$ i.e. we need to grade restrict them to $\mathbb{W}(0)$. This is possible by just taking multiple cones with $\mathcal{E}_{+}$ and their twists. We will illustrate this procedure and the derivation of the autoequivalence $M_{0}$, as it acts on $\mathcal{B}_{\mathrm{pt}},\mathcal{B}_{D_{\alpha}},\mathcal{B}_{\mu}$, $\widehat{\calE}_{-}$ in the following.

\section*{Structure sheaf}

The object $\widehat{\mathcal{E}}_{-}$ satisfies $\pi_{\zeta_{+}}(\widehat{\mathcal{E}}_{-})=\mathcal{O}_{X_{\zeta_{+}}}$. However 
\begin{eqnarray}
\widehat{\mathcal{E}}_{-}\notin \mathbb{W}(0).
\end{eqnarray}
In order to find a quasi-isomorphic object that belongs to $\mathbb{W}(0)$, we need to take cones with twists of $\mathcal{E}_{+}$, since $\pi_{\zeta_{+}}(\mathcal{E}_{+})\simeq 0$ it guarantees that the resulting B-brane will be quasi-isomorphic. The object  $\widehat{\mathcal{E}}_{-}$ is written as a tensor product of two matrix factorizations $\widehat{\mathcal{E}}_{-}=\widehat{\mathbf{T}}_{-}\otimes \widetilde{\mathbf{T}}_{\mathcal{O}}$. Therefore, the Chan-Paton vector space of $\widehat{\mathcal{E}}_{-}$ takes the form of a tensor product $M_{-}\otimes \widetilde{M}_{\mathcal{O}}$. The representation $\widetilde{M}_{\mathcal{O}}$, as a representation of $U(1)\subset G$, is trivial and the only factor in $M_{-}$ that does not belong to $\mathbb{W}(0)$ is $\mathcal{W}(-n-1,-\vec n)$. Indeed, $\widehat{\mathbf{T}}_{-}$ takes the form:
\begin{equation}
\begin{aligned}
\widehat{\mathbf{T}}_{-}:\xymatrix{
   \calW(-n-1,-\vec n) \ar@<2pt>[r]^-F & \ar@<2pt>[l]^-p \wedge^{n} \mathcal{V}_{-} \ar@<2pt>[r]^-F & \ar@<2pt>[l]^-p \wedge^{n-1}\mathcal{V}_{-}\cdots 
 \ar@<2pt>[r]^-F & \ar@<2pt>[l]^-p \mathcal{V}_{-}\ar@<2pt>[r]^-F & \ar@<2pt>[l]^-p \calW(0,0),
}
\end{aligned}
\end{equation}
with the wedge products $ \wedge^{k} \mathcal{V}_{-}$ corresponding to multiple copies of the representation $\mathbb{C}(-k)$ of $U(1)\subset G$. We can remove the factor $\mathcal{W}(-n-1,-\vec n)$ by taking a cone with $\mathcal{E}_{+}(0,-\vec{n})=\mathbf{T}_{+}(0,-\vec{n})\otimes \widetilde{\mathbf{T}}_{\mathcal{O}}$ by noting first that, $\mathbf{T}_{+}(0,-\vec{n})$ takes the form
\begin{equation}
\mathbf{T}_{+}(0,-\vec{n}):\xymatrix
{\calW(-n-1,-\vec n) \ar[r]^-y& \ar@<2pt>[l]^-{p\partial_{y}F}\calW(-n,-\vec n)^{\oplus (n+1)} \ar[r]^-y &\ar@<2pt>[l]^-{p\partial_{y}F}\cdots \ar[r]^-y  &\ar@<2pt>[l]^-{p\partial_{y}F}\calW(0,-\vec n).
}
\end{equation}
Since the term we want to replace in $\widehat{\mathcal{E}}_{-}$ is $ \calW(-n-1,-\vec n) \otimes \widetilde{M}_{\mathcal{O}}$ we need to extend to $MF_{G}(W_{1}+W_{2})$ the cone $\mathrm{Cone}(\varphi:\mathbf{T}_{+}(0,-\vec{n})\rightarrow \widehat{\mathbf{T}}_{-})$ given by:
\begin{equation}
\xymatrix
{
&\calW(-n-1,-\vec n) \ar@<2pt>[r] &\ar@<2pt>[l]  \cdots \ar@<2pt>[r]  &\ar@<2pt>[l]\mathcal{V}_{-} \ar@<2pt>[r]  &\ar@<2pt>[l]\calW(0,0)
\\
\calW(-n-1,-\vec n)\ar@<2pt>[r]\ar[ru]^{\Id} & \ar@<2pt>[l] \calW(-n,-\vec n)^{\oplus (n+1)} \ar@<2pt>[r]\ar[ru]^{\varphi} & \ar@<2pt>[l] \cdots \ar@<2pt>[r]\ar[ru]^{\varphi}  & \ar@<2pt>[l] \calW(0,-\vec n)\ar[ru]^{\varphi}
}
\end{equation}
where the maps simply indicated as $\varphi$ can be uniquely determined using the techniques in \cite{Herbst:2008jq}, but they are not important for us. The important fact for us is that, on one hand, we can extend the morphism $\varphi$ to a morphism $\mathcal{E}_{+}(0,\vec{n})\rightarrow \widehat{\mathcal{E}}_{-}$ by using the results in \cite{Yoshino_1998}. And on the other hand we can remove the subcomplexes of the form 
\begin{equation}
\calW(-n-1,-\vec n)\otimes\widetilde{M}_{\mathcal{O}}\xrightarrow{\mathrm{Id}}\calW(-n-1,-\vec n)\otimes\widetilde{M}_{\mathcal{O}}
\end{equation}
at the price that we will have to add more arrows in order to have an object of $MF_{G}(W)$ \cite{Herbst:2008jq}, but these additional arrows are irrelevant for us. In summary, we completed the grade restriction of $\widehat{\mathcal{E}}_{-}$:
\begin{equation}
\widehat{\mathcal{E}}_{-}^{(0)}:=
\mathrm{Cone}(\mathcal{E}_{+}(0,\vec{n})\rightarrow \widehat{\mathcal{E}}_{-})\in\mathbb{W}(0).
\end{equation}
Next, we need to take cones with twists of $\mathcal{E}_{-}$ in order to map the objects from $\mathbb{W}(0)$ to $\mathbb{W}(-1)$ by a quasi-isomorphism when projected by $\pi_{\zeta_{-}}$. The factors of $\widehat{\mathcal{E}}_{-}^{(0)}$ whose weights do not belong to $\mathbb{W}(-1)$ are given by $\calW(0,0)\otimes\widetilde{M}_{\mathcal{O}}$ and $\calW(0,-\vec n)\otimes\widetilde{M}_{\mathcal{O}}$. They can be removed by taking cones with $\widehat{\mathcal{E}}_{-}(0,-\vec{n})[1]$ and $\widehat{\mathcal{E}}_{-}$, respectively. This will give a quasi-isomorphic object under $\pi_{\zeta_{-}}$, since $\pi_{\zeta_{-}}(\widehat{\mathcal{E}}_{-})\simeq 0$. A straightforward manipulation of complexes gives:
\begin{equation}\label{complexesOX}
F_{\mathcal{E}_{-}}(\widehat{\mathcal{E}}_{-}^{(0)})=\mathrm{Cone}(\widehat{\mathcal{E}}_{-}\rightarrow\mathrm{Cone}(\widehat{\mathcal{E}}_{-}(0,-\vec{n})[1]\rightarrow \widehat{\mathcal{E}}_{-}^{(0)}))\in\mathbb{W}(-1)
\end{equation}
it is important to remark, that, as an object in $MF_{G}(W)$:
\begin{equation}\label{strshfquasi}
F_{\mathcal{E}_{-}}(\widehat{\mathcal{E}}_{-}^{(0)})\simeq \widehat{\mathcal{E}}_{-}(0,-\vec{n})[2]
\end{equation}
where the quasi-isomorphism in \eqref{strshfquasi} must be understood as in $MF_{G}(W)$, not in $\mathbb{W}(-1)$. The functor $F_{\mathcal{E}_{+}}$ will not affect the projection of the resulting object under $\pi_{\zeta_{+}}$, neither its central charge, when evaluated in the $X_{\zeta_{+}}$, therefore we can ignore its details and state:
\begin{equation}
M_{0}(\widehat{\mathcal{E}}_{-}^{(0)})=F_{\mathcal{E}_{+}}\circ F_{\mathcal{E}_{-}}(\widehat{\mathcal{E}}_{-}^{(0)})\xmapsto{\pi_{\zeta_{+}}}\mathcal{O}_{X_{\zeta_{+}}}[2]\otimes (K_Y|_X) \label{OXtwistK}
\end{equation}
which is the result we will need in the following.

\section*{Divisor classes}

The B-branes $\mathcal{B}_{D_{\alpha}}\in MF_{G}(W)$, $\alpha=1,\ldots, r$ RG flow to divisors supported in $\vec{\mathbb{P}}$, restricted to $X_{\zeta_{+}}$ i.e. $\pi_{\zeta_{+}}(\mathcal{O}_{D_{\alpha}})$.  The matrix factorization for divisors is written as a tensor product $\mathcal{B}_{D_{\alpha}}=\widehat{\mathcal{E}}_{-}\otimes \mathcal{B}_{\alpha}$, where $\mathcal{B}_{\alpha}\in MF_{G}(0)$ is given by
\begin{equation}
\mathcal{W}(0,-\mathbf{e}_{\alpha})\rightarrow\mathcal{W}(0,0).
\end{equation}
Therefore, since $\mathcal{B}_{\alpha}$ is trivial w.r.t $U(1)\subset G$,  we can use an analogous argument than the one above, to show that
\begin{equation}\label{strshfquasi}
F_{\mathcal{E}_{-}}(\mathcal{B}^{(0)}_{D_{\alpha}}
)\simeq \mathcal{B}_{D_{\alpha}}(0,-\vec{n})[2],\qquad M_{0}(\widehat{\mathcal{E}}_{-}^{(0)})=F_{\mathcal{E}_{+}}\circ F_{\mathcal{E}_{-}}(\mathcal{B}^{(0)}_{D_{\alpha}})\xrightarrow{\pi_{\zeta_{+}}}\mathcal{O}_{D_{\alpha}}[2]\otimes K_{Y}|_X 
\end{equation}
for $i=1,\ldots,r$. The only object that requires a more careful treatment is $\mathcal{B}_{D_{0}}$. First, note that $\mathbf{T}_{D_{0}}=\widehat{\mathbf{T}}_{0}\otimes\widetilde{\mathbf{T}}_{\mathcal{O}}$, where 
\begin{equation}\label{hatT0}
\widehat{\mathbf{T}}_{0}:=\mathrm{Cone}(\widehat{\mathcal{E}}_{-}\otimes\mathcal{W}(-1,0)\rightarrow \widehat{\mathcal{E}}_{-})
\end{equation}
and the morphism in \eqref{hatT0} is given by $l^{(0)}(y)$, explicitly:
\begin{equation}
\widehat{\mathbf{T}}_{0}:\xymatrix
{
&\calW(-n-1,-\vec n) \ar@<2pt>[r] &\ar@<2pt>[l]  \wedge^{n}\mathcal{V}_{-} \ar@<2pt>[r]  &\ar@<2pt>[l]\cdots \ar@<2pt>[r]  &\ar@<2pt>[l]\underline{\calW(0,0)}
\\
\calW(-n-2,-\vec n)\ar@<2pt>[r]\ar[ru]^{l^{(0)}} & \ar@<2pt>[l]\wedge^{n}\mathcal{V}_{-}\otimes\mathcal{W}(-1,0) \ar@<2pt>[r]\ar[ru]^{l^{(0)}} & \ar@<2pt>[l] \cdots \ar@<2pt>[r]\ar[ru]^{l^{(0)}}  & \ar@<2pt>[l] \calW(-1,0)\ar[ru]^{l^{(0)}}
}
\end{equation}
where the underlined factor is the chosen Clifford vacuum. In order grade restrict $\mathcal{B}_{D_{0}}$ to $\mathbb{W}(0)$ we need to remove the factors $\mathcal{W}(-n-1,-\vec{n})\otimes \widetilde{M}_{\mathcal{O}}$, $\mathcal{W}(-n-2,-\vec{n})\otimes \widetilde{M}_{\mathcal{O}}$ and $\wedge^{n}\mathcal{V}_{-}\otimes\mathcal{W}(-1,0)\otimes \widetilde{M}_{\mathcal{O}}$ by taking cones with twists of $\mathcal{E}_{+}$. It is straightforward to see that this can be done by
\begin{equation}
\mathcal{B}_{D_{0}}^{(0)}:=\mathrm{Cone}\left(\mathcal{E}_{+}(0,-\vec{n})\oplus\bigoplus_{I=1}^{n+1}\mathcal{E}_{+}(0,-\vec{n}+\vec{n}_{I})\oplus \mathcal{E}_{+}(-1,-\vec{n})[1]\oplus \mathcal{E}_{+}^{\oplus n+1}(0,-\vec{n})[1]\rightarrow\mathcal{B}_{D_{0}}\right)
\end{equation}
then, we need to take cones with $\widehat{\mathcal{E}}_{-}$ (and twists of) grade restrict $\mathcal{B}_{D_{0}}^{(0)}$ to $\mathbb{W}(-1)$. The factors on $\mathcal{B}_{D_{0}}^{(0)}$ that do not belong to $\mathbb{W}(-1)$ are
\[
\mathcal{W}(0,0)\otimes \widetilde{M}_{\mathcal{O}},\quad  \mathcal{W}(0,-\vec{n})\otimes \widetilde{M}_{\mathcal{O}},\quad  \bigoplus_{I=1}^{n+1}\mathcal{W}(0,-\vec{n}+\vec{n}_{I})\otimes \widetilde{M}_{\mathcal{O}}\quad \text{and}\quad \mathcal{W}^{\oplus n+1}(0,-\vec{n})\otimes \widetilde{M}_{\mathcal{O}}. 
\]
This results in the following:
\begin{equation}\label{FEminusB0}
F_{\mathcal{E_{-}}}(\mathcal{B}_{D_{0}}^{(0)})
=
\mathrm{Cone}
\left(\widehat{\mathcal{E}}_{-}\oplus \widehat{\mathcal{E}}_{-}(0,-\vec{n})[1]\oplus \bigoplus_{I=1}^{n+1}\widehat{\mathcal{E}}_{-}(0,-\vec{n}+\vec{n}_{I})[1]\oplus \widehat{\mathcal{E}}^{\oplus n+1}_{-}(0,-\vec{n})[2]\rightarrow\mathcal{B}^{(0)}_{D_{0}}\right)
\end{equation}
As mentioned before, the functor $F_{\mathcal{E}_{+}}$ is irrelevant for us, because we are concerned with the image of $M_{0}(\mathcal{B}_{D_{0}}^{(0)})$ under $\pi_{\zeta_{+}}$. Indeed, by a careful analysis of \eqref{FEminusB0} we conjecture that $\pi_{\zeta_{+}}\circ M_{0}(\mathcal{B}_{D_{0}}^{(0)})$ corresponds to the following object in $D(X_{\zeta_{+}})$:
\begin{equation}\label{M0D0}
\pi_{\zeta_{+}}\circ M_{0}(\mathcal{B}_{D_{0}}^{(0)})
\simeq 
\left(\mathcal{O}^{\oplus n+1}_{X_{\zeta_{+}}}\rightarrow \mathcal{O}_{D_{0}}\oplus\mathcal{O}_{X_{\zeta_{+}}}(-1,0)\oplus\bigoplus_{I=1}^{n+1}\mathcal{O}_{X_{\zeta_{+}}}(0,\vec{n}_{I})\rightarrow \mathcal{O}_{X_{\zeta_{+}}}(-1,\vec n)\right)[1]\otimes K_{Y}|_X
\end{equation}
We define the object $\calC$ as the element in \eqref{M0D0}, given by:
\begin{equation}\label{calC}
    \calC:=\bigg(\mathcal{O}^{\oplus n+1}_{X_{\zeta_{+}}}\rightarrow \mathcal{O}_{X_{\zeta_{+}}}(-1,0)\oplus\bigoplus_{I=1}^{n+1}\mathcal{O}_{X_{\zeta_{+}}}(0,\vec{n}_{I})\rightarrow \mathcal{O}_{X_{\zeta_{+}}}(-1,\vec n)\bigg).
\end{equation}

\section*{Curve and point classes}

For the case of the curve classes, the matrix factorizations of main interest will be a twist of $\mathcal{B}_{\mu}\in MF_{G}(W)$, namely
\begin{equation}
\mathcal{B}_{\mu}(-\delta_{\mu,0},-\delta_{\mu,\alpha}\mathbf{e}_{\alpha})\in MF_{G}(W),\label{eqn:Bmu}
\end{equation}
moreover, the Chan-Paton vector space $M$, for $\mathcal{B}_{0}(-1,0)$ includes only the representations $\mathbb{C}(-n),\ldots, \mathbb{C}(-1)$, and for $\mathcal{B}_{\mu}(0,-\mathbf{e}_{\mu})$ only the representations $\mathbb{C}(-n),\ldots, \mathbb{C}(0)$. Therefore 
\begin{equation}
\mathcal{B}_{0}(-1,0)\in \mathbb{W}(0)\cap \mathbb{W}(-1),
\end{equation}
so its monodromy is trivial
\begin{equation}\label{trivialC0}
M_{0}(\mathcal{B}_{0}(-1,0))=M_{0}(\mathcal{B}^{(0)}_{0}(-1,0))=\mathcal{B}_{0}(-1,0)\xmapsto{\pi_{\zeta_{+}}}\mathcal{O}_{C_{0}}(-1,0).
\end{equation}
For the case of $\mathcal{B}_{\mu}(0,-\mathbf{e}_{\mu})$ we have that $\mathcal{B}_{\mu}(0,-\mathbf{e}_{\mu})=\mathcal{B}^{(0)}_{\mu}(0,-\mathbf{e}_{\mu})\in \mathbb{W}(0)$ and, in order to grade restrict it to $\mathbb{W}(-1)$ we need to take care of the factor $\mathcal{W}(0,0)\otimes \widetilde{M}_{\mu}$, where $\widetilde{M}_{\mu}$ is a trivial $U(1)\subset G$ representation given by
\begin{equation}
\widetilde{M}_{\mu}:=\wedge^{\bullet}\mathcal{V}_{\mu}(0,-\mathbf{e}_{\mu}),\qquad \mathcal{V}_{\mu}:=\bigoplus_{\alpha=1}^{r}\mathbf{\overline{k}}_{\alpha}^{\oplus (m_{\alpha}-k_{\alpha}-\delta_{\alpha,\mu})}\bigoplus({\det}^{-1}\mathbf{k}_{\mu})^{\oplus (k_\mu-1)}.
\end{equation}
It is straightforward to show that $\widetilde{M}_{\mu}$ corresponds to the Chan-Paton vector space of a matrix factorization $\widetilde{\mathcal{B}}_{\mu}\in MF_{G}(W_{2})$ and that we can then construct:
\begin{equation}
\mathcal{B}_{-}\otimes \widetilde{\mathcal{B}}_{\mu}\in MF_{G}(W),\qquad \pi_{\zeta_{-}}(\mathcal{B}_{-}\otimes \widetilde{\mathcal{B}}_{\mu})\simeq 0
\end{equation}
and so
\begin{equation}\label{M0C}
F_{\mathcal{E}_{-}}(\mathcal{B}_{\mu}(0,-\mathbf{e}_{\mu}))=\mathrm{Cone}(\mathcal{B}_{-}\otimes \widetilde{\mathcal{B}}_{\mu}\rightarrow \mathcal{B}_{\mu}(0,-\mathbf{e}_{\mu})).
\end{equation}
In this case, we do not have a derivation of the resulting object after monodromy, however, results from sect. \ref{sec:Aperiods} suggest that
\begin{equation}
\pi_{\zeta_{+}}\circ M_{0}(\mathcal{B}_{\mu}(0,-\mathbf{e}_{\mu}))=\mathcal{O}_{C_{\mu}}(\mathrm{det}\mathcal{S}^{(\mu)})[2]\otimes (K_{Y}|_X)
\end{equation}
Finally, for the case of $\mathcal{B}_{\mathrm{pt}}$, the representations of $U(1)\subset G$ appearing in its Chan-Paton vector space are $\mathbb{C}(-n),\ldots,\mathbb{C}(0)$, therefore $\mathcal{B}_{\mathrm{pt}}=\mathcal{B}^{(0)}_{\mathrm{pt}}\in\mathbb{W}(0)$ and a reasoning analogous than the one for $\mathcal{B}_{\mu}(0,-\mathbf{e}_{\mu})$ shows that we can grade restrict this object to $\mathbb{W}(-1)$ by taking cone with $\mathcal{B}_{-}\otimes\widetilde{\mathcal{B}}_{\mathrm{pt}}$, where the Chan-Paton vector space of $\widetilde{\mathcal{B}}_{\mathrm{pt}}$  is given by 
\begin{equation}\label{insertionpoint}
\wedge^{\bullet}\mathcal{V}_{\mathrm{pt}},\qquad \mathcal{V}_{\mathrm{pt}}:=\bigoplus_{\alpha=1}^{r}\mathbf{\overline{k}}_{\alpha}^{\oplus (m_{\alpha}-k_{\alpha})}.
\end{equation}
Then, a straightforward analysis of the complex $\pi_{\zeta_{+}}(\widetilde{\mathcal{B}}_{\mathrm{pt}})$ shows that it is generically supported in a subvariety of codimension $n+1+\mathrm{dim}(\vec{\mathbb{P}})$, hence we conjecture $\pi_{\zeta_{+}}(\widetilde{\mathcal{B}}_{\mathrm{pt}})\simeq 0$, which is also supported by the results of sect. \ref{sec:Aperiods}, which shows this object has vanishing central charge. Therefore we conclude
\begin{equation}\label{M0pt}
M_{0}(\mathcal{B}_{\mathrm{pt}})=\mathcal{B}_{\mathrm{pt}}\xmapsto{\pi_{\zeta_{+}}}\mathcal{O}_{\mathrm{pt}}
\end{equation}

\section{Hemisphere partition function and A-periods}\label{sec:Aperiods}

The central charge of B-branes for $\mathcal{N}=(2,2)$ theories is defined in \cite{Cecotti:1991me,hori2000d} by the partition function the A-twisted theory on a disk attached to an infinitely flat cylinder and boundary conditions corresponding to a B-brane. This coupling of an A-twisted theory in the bulk and B-brane boundary conditions is a very natural object to study, for instance in $\mathcal{N}=(2,2)$ SCFTs \cite{Ooguri:1996ck}. Using supersymmetric localization techniques 
\cite{Hori:2013ika,Honda:2013uca,Sugishita:2013jca} the 
central charge for a B-brane $(\mathcal{B},L_{t})$, $\mathcal{B}\in MF_{G}(W)$ was computed in the context of GLSMs and found to be given by a Mellin-Barnes type integral:
\begin{equation}
Z_\mathcal{B}(t):=\int_{L_{t}\subset 
\mathfrak{t}_{\bbC}} \mathrm{d}^{l_{G}}\sigma 
\prod_{\alpha>0}\alpha(\sigma)\sinh(\pi\alpha(\sigma))\prod_{j=1}
^{N}\Gamma\left(iQ_
{j}(\sigma)+\frac{R_{j}}{2}\right)e^{it(\sigma)}f_{\mathcal{B}}(\sigma).
\end{equation}
where $Q_j\in \mathfrak{t}^{\vee}_{\bbC}$ denotes the weights of the representation \eqref{eqn:chiralfields}, while $R_j$ denotes the weights of the $U(1)_R$ symmetry action. The symbol $\prod_{\alpha>0}$ denotes the product over the positive roots of $G$ and $l_{G}:=\mathrm{dim}(\mathfrak{t})$. Denote $\sigma^{(i)}_\alpha$ the coordinates of $(\mathfrak{t}_{U(k_\alpha)})_{\bbC}\cong \mathbb{C}^{k_{\alpha}}$, $i=1,\cdots,k_\alpha$, and define
\begin{equation}
\sigma_\alpha:=\sigma_\alpha^{(1)}+\cdots+\sigma_\alpha^{(k_\alpha)},\qquad,\alpha=0,\ldots, r.
\end{equation}
In particular, $k_0=1$ and $m_0=n+1$. For the GLSM considered here, the hemisphere partition function, with respect to the R-charge integrality \eqref{eqn:RInt}, has an explicit expression as
\begin{eqnarray}
  Z_{\calB}(t) &=& \int_{L_t} \prod_{\alpha=0}^r \mathrm{d}^{k_{\alpha}}\sigma_\alpha\prod_{i_\alpha<j_\alpha}(\sigma_\alpha^{(i_\alpha)}-\sigma_\alpha^{(j_\alpha)})\sinh\pi(\sigma_\alpha^{(i_\alpha)}-\sigma_\alpha^{(j_\alpha)}) e^{it_\alpha \sigma_\alpha}\prod_{i_\alpha=1}^{k_\alpha}\Gamma(i\sigma^{(i_\alpha)}_\alpha)^{m_\alpha}
  \nonumber\\
  &&\times \prod_{I=1}^{n+1}\Gamma\left(-i \vec n_I(\sigma)-i\sigma_0+1\right)\times \prod_{J=n+2}^{K}\Gamma(-i \vec n_J(\sigma)+1) f_{\calB}(\sigma).
  \nonumber\\ \label{eqn:ZBt}
\end{eqnarray}
Where we write:
\begin{eqnarray}
\vec{n}_I(\sigma):=\sum_{\alpha=1}^{r}n_{I}^{(\alpha)}\sigma_{\alpha},\qquad I=1,\ldots,K.
\end{eqnarray}
The contour $L_t$ in \eqref{eqn:ZBt}, satisfying the admissibility conditions of \cite{Hori:2013ika}, for the phases $\zeta_{\pm}$ can be determined in the same way as the so called 'abelian contour' in \cite{Lin:2024fpz}, and is given explicitly by:
\begin{equation}
    L_{\zeta_\pm}= \bigg\{ \Im\sigma_0=\pm\big( \Re\sigma_0 \big)^2-\epsilon \bigg\} \cap\bigcap_{\alpha=1}^r\bigcap_{i_\alpha=1}^{k_\alpha}\bigg\{ \Im\sigma_\alpha^{(i_\alpha)}=\big(\Re\sigma_\alpha^{(i_\alpha)}\big)^2 -\epsilon \bigg\},\quad 0<\epsilon<1.
\end{equation}
The shift $\epsilon$ is introduced to separate the poles of the phases $\zeta_{\pm}$ in the limit that some of the R-charges takes values $0$ or $2$ \footnote{As this is strictly not allowed in the localization computation, but the limit is well defined, in general i.e. after taking residues, for instance, the limit $\epsilon\rightarrow 0$ is well defined.}. Indeed, the contours $L_{\zeta_{\pm}}$ can be deformed so that \eqref{eqn:ZBt} becomes residue integrals around infinite number of poles, specifically these poles are given by
\begin{equation}
    (\sigma_0,\sigma^{(i_1)}_1,\cdots,\sigma^{(i_r)}_r)\in \left\{\begin{array}{cc}
       i\bbZ_{\geq0}\cup \bigcup_{\alpha=1}^r(i\bbZ_{\geq0})^{k_\alpha}  &  \text{for }\zeta_+\text{ \ phase}
         \\
        i\bbZ_{\leq-1}\cup \bigcup_{\alpha=1}^r(i\bbZ_{\geq0})^{k_\alpha}  & \text{for }\zeta_-\text{ \ phase}
    \end{array} \right.\label{eqn:poles}
\end{equation}
Finally, the contribution of the object $\calB$ is contained in the brane factor of $f_{\calB}$ and is given by:
\begin{equation}
f_{\mathcal{B}}(\sigma):=\mathrm{tr}_{M}\left(R_{M}(e^{i
\pi})\rho_{M}(e^{2\pi\sigma
} )\right)
\end{equation}
The function (\ref{eqn:ZBt}) is conjectured to coincide with the IR central charge as defined by \cite{Cecotti:1991me}, for SCFTs i.e. nonanomalous GLSMs. It is important to remark that in (\ref{eqn:ZBt}) the integration variable is dimensionless since it corresponds to $r\sigma$ (where $r$ is the radius of the hemisphere) but we denoted it $\sigma$ in order not to introduce more notation. In addition, (\ref{eqn:ZBt}) has a normalization factor of the form $C(r\Lambda)^{\frac{\hat{c}}{2}}$ where $\Lambda$ corresponds to the UV cut-off and $C$ is just a numerical constant. We will simply set this factor to a convenient numerical constant (specified in each example) in the following, but in general it should be considered for applications involving anomalous models \cite{hori2019notes}. \\
Since we are interested in the monodromies with base point near $\zeta_{+}$, $Z_{\mathcal{B}}(t)$ can be written as the infinite sum of residues:
\begin{eqnarray}\label{resLplus}
  Z_{\calB}(t)|_{\zeta_{+}} = \sum_{il_{\alpha}\in i(\mathbb{Z}_{\geq 0})^{k_{\alpha}}}\oint_{s=0}\prod_{\alpha=0}^r \mathrm{d}^{k_{\alpha}}s_{\alpha} h_{\mathcal{B}}(s^{(i_{\alpha})}_{\alpha}+il^{(i_{\alpha})}_{\alpha}),
\end{eqnarray}
where $h_{\calB}(\sigma_{\alpha}^{(i_{\alpha})})$ denotes the integrand of \eqref{eqn:ZBt}. Since the poles of the gamma functions are simple poles, the integral \eqref{resLplus} can be straightforwardly identified with an integral over $\mathbb{P}^{n}\times (\mathbb{P}^{m_{1}-1})^{k_{1}}\times\ldots (\mathbb{P}^{m_{r}-1})^{k_{r}}$.

Moreover, as it is done in \cite{Lin:2024fpz}, using the results of \cite{Martin:1999ng}, we can write \eqref{eqn:ZBt} as an integral over $\mathbb{P}^{n}\times \vec{\mathbb{P}}$, indeed we can state the following proposition:
\begin{prop}\cite{bertram1996severi}\label{propresgrass}
    Denote $H^{(1)},\cdots,H^{(k)}$ the Chern roots of the dual tautological bundle $\mathcal{S}^{\vee}\rightarrow Gr(k,m)$. Then for any regular totally-symmetric function $h(x)$ of the variables $x^{(1)},\cdots,x^{(k)}$, we have the identity
    \begin{equation}
        \frac{(-1)^{\tiny\begin{pmatrix}
            k\\2
        \end{pmatrix}}}{k!}\oint_0 \frac{\mathrm{d}^k x}{(2\pi i)^k}  \ \frac{\prod_{i<j}(x^{(i)}-x^{(j)})^2}{(x^{(1)}\cdots x^{(k)})^m}  h(x)=\int_{Gr(k,m)} h(H^{(i)}).
    \end{equation}
\end{prop}
Thus, the residues \eqref{resLplus} become an integral over $\mathbb{P}^{n}\times \vec{\mathbb{P}}$ upon the identification
\begin{equation}\label{sigmaident}
   s_\alpha^{(i_\alpha)} \rightarrow \frac{H_\alpha^{(i_\alpha)}}{2\pi}, \qquad c(\mathcal{S}^{\vee}_{\alpha})=\prod_{i_{\alpha}=1}^{k_{\alpha}}(1+H_{\alpha}^{(i_{\alpha})}).
\end{equation}
The integration over $\mathbb{P}^{n}\times \vec{\mathbb{P}}$ can be further reduced to an integration over $X_{\zeta_{+}}$ by use of the adjunction formula. It is then expected that \eqref{resLplus} reduces to the geometric central charge of B-branes or A-periods \cite{Cheung:1997az,Green:1996dd,minasian1997k}, in the $\zeta_{+}$ phase. More precisely we expect that the geometric A-period of a sheaf $\mathcal{E}\in D(X_{\zeta_{+}})$, $Z^{\mathrm{geom}}_{\mathcal{E}}$, coincides with \eqref{resLplus} as
\begin{equation}\label{conjecturegeom}
Z_{\mathcal{B}}|_{\zeta_{+}}=Z^{\mathrm{geom}}_{\pi_{\zeta_{+}}(\mathcal{B})}
\end{equation}
in the following we will simply denote it by $Z_{\mathcal{E}}$, since we will not refer to A-periods in any other phase. Moreover we will work on a particular basis of A-periods, 
\begin{defi}\label{DMdefi}
We define the Doran-Morgan lattice \cite{Doran:2005gu} of B-brane charges by the lattice of A-periods generated by $\langle Z_{\calO_X},\  Z_{\calO_{D_\alpha}},\ Z_{\calO_{\widetilde C_\alpha}},\ Z_{\calO_P}\rangle$, in a geometric phase. Where $\calO_{\widetilde C_\alpha}:=\calO_{C_\alpha}(-\mathbf{e}_\alpha)$ a twisted curve class.
\end{defi}
The lattice described in def. \ref{DMdefi} is expected to coincide with the Grothendieck group $K_{0}(X)$ of $X$ which is spanned by homology classes of holomorphic vector bundles over $X$ \cite{witten1999d,Hosono:2000eb}. The central charge of a general B-brane $\calB$, in the $\zeta_{+}$ phase will take then the form:
\begin{equation}\label{chargeofBB}
Z_{\pi_{\zeta_{+}}(\calB)}=a^0Z_{\calO_X}+a^\alpha Z_{\calO_{D_\alpha}}+a_\alpha 
Z_{\calO_{\widetilde C_\alpha}}+a_0 Z_{\calO_P}=:\left(a^0,a^\alpha,a_\alpha,a_0\right),
\end{equation}
where $a^0,a^\alpha,a_\alpha,a_0\in\mathbb{Z}$.\footnote{The basis in this paper has a mild modification compared to that in \cite{Doran:2005gu,Almkvist:2005qoo}. This also modifies \eqref{LKY} and \eqref{TOX}. However, this modification preserve the lattice of charges (i.e. all autoequivalences are still integrer valued).}
Finally, we make some general statements about $Z_{\mathcal{E}}$, for $\mathcal{E}\in D(X)$ in the zero-instanton sector. The A-period $Z_{\mathcal{E}}$ of $X$, in general, takes the form:
\begin{equation}
\begin{gathered}
Z_{\mathcal{E}}(\kappa)=\int_X 
e^{J}\hat\Gamma_X\mathrm{ch}(\calE)+\text{instantons}=:Z^{0}_{\calE}(\kappa)+\text{instantons } ,
\end{gathered}\label{LVZ}
\end{equation}
where $J:=B+i\frac{\omega}{2\pi}\in H^2(X,\mathbb{C})$, $B\in 
H^{2}(X,\mathbb{R}/\mathbb{Z})$ is the $B$-field and $\omega\in 
\mathcal{K}_{X}\subset H^{2}(X,\mathbb{R})$. $\mathcal{K}_{X}$ denotes the 
K\"ahler cone of $X$ and the instantons are weighted by
\begin{equation}\label{instantonexp}
\exp\left(2\pi i\int_{\beta}J\right),\quad \beta\in H_{2}(X,\mathbb{Z})
\text{ an effective curve class.}
\end{equation}
Fix a basis $\{J_{\alpha}\}$ of $H^2(X,\bbC)$, then $J=\kappa^\alpha J_\alpha$, and $\kappa^{\alpha}$ corresponds to the so-called flat coordinates. The Gamma class\footnote{ For $X$ CY, the $\hat{A}$-genus 
equals the Todd class $\mathrm{Td}_{X}$. Because of (\ref{gammaroot}), we can 
regard $\hat\Gamma_X$ as a root of $\hat{A}_{X}$. Moreover, this root is not unique and $\hat\Gamma_X$ is just a particular choice \cite{Halverson:2013qca}, however this choice is different from the one corresponding to the 
Ramond-Ramond (RR) charge computed in \cite{Cheung:1997az,Green:1996dd,minasian1997k}. The choice $\hat\Gamma_X$ encodes the perturbative corrections to the central charge. The Gamma class appears implicitly in the works 
\cite{libgober1999chern,Hosono:2000eb} and then it was further defined in a mathematical context in \cite{iritani2009integral,katzarkov2008hodge}.}
$\hat\Gamma_X$ is a multiplicative characteristic class, given by 
\begin{equation}
\hat\Gamma_X:=\prod_{j}\Gamma\left(1-\frac{\lambda_{j}}{2\pi i}\right)
\end{equation}
where $\lambda_{j}$ are the Chern roots of the holomorphic tangent bundle $TX$ 
of 
$X$. It satisfies the important property
\begin{equation}\label{gammaroot}
\hat\Gamma_X\hat{\Gamma}^{*}_{X}=\hat{A}_{X},\qquad \hat\Gamma^{*}_X:=\prod_{j}\Gamma\left(1+\frac{\lambda_{j}}{2\pi i}\right)
\end{equation}
with $\hat{A}_{X}$ the $\hat{A}$-genus of $X$. When $X$ is a CY threefold, we have the explicit expansion:
\begin{equation}\label{gammaCY3}
\hat\Gamma_X=1+\frac{1}{24}c_2(X)+\frac{\zeta(3)}{(2\pi i)^3}c_3(X).
\end{equation}
Then, we have the explicit formulae for $Z^{0}_{\calE}(\kappa)$ in the Doran-Morgan basis:
\begin{eqnarray}\label{classAperiods}
Z^{0}_{\calO_X}(\kappa)&=&\frac{1}{3!}c_{\alpha\beta\gamma}\kappa^\alpha \kappa^\beta\kappa^\gamma +c_\alpha \kappa^\alpha+\frac{\zeta(3)}{(2\pi i)^3}\chi(X),\nonumber\\
Z^{0}_{\calO_{D_\alpha}}(\kappa) &=&\frac{1}{2}c_{\alpha\beta\gamma}\kappa^\beta 
\kappa^\gamma-\frac{1}{2}c_{\alpha\alpha\beta}\kappa^\beta+\frac{1}{6}c_{
\alpha\alpha\alpha}+c_\alpha,\nonumber\\
Z^{0}_{\calO_{\widetilde C_\alpha}}(\kappa) &=&\kappa^\alpha,\nonumber\\
Z^{0}_{\calO_P}(\kappa)&=& 1,
\end{eqnarray}
where
\begin{equation}
    c_{\alpha\beta\gamma}:=\int_X H_\alpha H_\beta H_\gamma,\quad 
c_\alpha:=\frac{1}{24}\int_X c_2(X)H_\alpha,\quad \chi(X)=\int_X c_3(X).
\end{equation}

\subsection{Monodromy of A-periods from window monodromy}

In this subsection we will derive the monodromy action on A-periods. Using the notation in sect. \ref{sec:wincatMon}, we are going to compute explicit expressions for 
\begin{equation}
Z_{\pi_{\zeta_{+}}(M_{0}(\mathcal{B}))}(t),
\end{equation}
in the Doran-Morgan basis \ref{DMdefi}. For this purpose it is useful to define the rank $2r+2$ row vector:
\begin{equation}
\Pi(t):=\left(Z_{\widehat{\calE}^{(0)}_{-}}(t),Z_{\calB^{(0)}_{D_{\alpha}}}(t),Z_{\calB^{(0)}_{{\alpha}}(-\mathbf{e}_{\alpha})}(t),Z_{\calB^{(0)}_{\mathrm{pt}}}(t)\right)
\end{equation}
where all the B-branes in $\Pi(t)$ are assumed to be all in $\mathbb{W}(0)$.
\begin{rmk}\label{zeroinstrmk}
The vector $\Pi(t)$ satisfies the following properties. By the conjecture \eqref{conjecturegeom} we have:
\begin{equation}\label{eqgeomplus}
\Pi_{+}(t):=\pi_{\zeta_{+}}(\Pi(t)):=\left(Z_{\pi_{\zeta_{+}}(\mathcal{B}_{s})}(t)\right)_{s=1}^{2r+2}=\left(Z_{\calO_{X_{\zeta_{+}}}}(t),Z_{\calO_{D_{\alpha}}}(t),Z_{\calO_{\widetilde C_{\alpha}}}(t),Z_{\calO_{P}}(t)\right)
\end{equation}
where the RHS of \eqref{eqgeomplus} denotes the geometric central charge. However, in general $t\neq \kappa$, but for the zero-instanton sector $t$ and $\kappa$ are related by some linear transformation, which we will derive explicitly below. Then we denote the zero-instanton sector vector as
\begin{equation}
\Pi^{0}_{+}(t):=\left(Z^{0}_{\pi_{\zeta_{+}}(\mathcal{B}_{s})}(t)\right)_{s=1}^{2r+2}
\end{equation}
where $Z^{0}_{\calB}(t)$ denotes the residue in \eqref{resLplus} around the origin in $\sigma$-space.
\end{rmk}
Using proposition \ref{propresgrass} in the residue around the origin on \eqref{resLplus}, and the identification \eqref{sigmaident} we obtain:
\begin{eqnarray}\label{Zzeroint}
Z_{\pi_{\zeta_{+}}(\calB)}^{0}(t)&=&C\int_{\mathbb{P}^{n}\times\vec{\mathbb{P}}}\prod_{\alpha=1}^{r}\left(e^{it_{\alpha}\frac{H_{\alpha}}{2\pi}}\frac{\prod_{i_\alpha=1}^{k_\alpha}\Gamma\left(i\frac{H^{(i_{\alpha})}}{2\pi}+1\right)^{m_{\alpha}}}{\prod_{i_\alpha<j_\alpha}\Gamma\left(1+H^{(i_{\alpha})}_{\alpha}-H^{(j_{\alpha})}_{\alpha}\right)\Gamma\left(1-H^{(i_{\alpha})}_{\alpha}+H^{(j_{\alpha})}_{\alpha}\right)}\right)\nonumber\\
&\times & \frac{e^{-\sum_{I=1}^{K}\frac{n_{I}(H)}{2}-(n+1)\frac{L}{2}}}{\prod_{I=1}^{n+1}\Gamma\left(i\frac{n_{I}(H)}{2\pi}+i\frac{L}{2\pi}\right)\prod_{I=n+2}^{K}\Gamma\left(i\frac{n_{I}(H)}{2\pi}\right)}\frac{f_{\calB}\left(\frac{H}{2\pi}\right)}{f_{\widehat{\calE}_{-}}\left(\frac{H}{2\pi}\right)}
\end{eqnarray}
where
\begin{eqnarray}
f_{\widehat{\calE}_{-}}\left(\frac{H}{2\pi}\right)=\prod_{I=1}^{n+1}\left(1-e^{-n_{I}(H)-L}\right)\prod_{I=n+2}^{K}\left(1-e^{-n_{I}(H)}\right)
\end{eqnarray}
and $C$ is a numerical constant, which we fix once and for all by the condition
\begin{eqnarray}
Z_{\pi_{\zeta_{+}}(\calB_{\mathrm{pt}})}^{0}=1,
\end{eqnarray}
then, using the adjunction formula for $X_{\zeta_{+}}\subset \mathbb{P}^{n}\times\vec{\mathbb{P}}$:
\begin{equation}
    \int_{X_{\zeta_{+}}}\alpha =\int_{\mathbb{P}^{n}\times \vec{\mathbb{P}}}e(N_{X/\mathbb{P}^{n}\times \vec{\mathbb{P}}})\alpha,\qquad e(N_{X/\mathbb{P}^{n}\times \vec{\mathbb{P}}})=\prod_{I=1}^{n+1}(\vec{n}_{I}(H)+L)\prod_{I=n+2}^K\vec n_I(H)
\end{equation}
where $\alpha$ denotes an element of $H^{*}(\mathbb{P}^{n}\times\vec{\mathbb{P}})$ as well as its restriction to $H^{*}(X_{\zeta_{+}})$, equation \eqref{Zzeroint} can be written as 
\begin{equation}
Z_{\pi_{\zeta_{+}}(\calB)}^{0}=\int_{X_{\zeta_{+}}}e^{J}\hat{\Gamma}_{X_{\zeta_{+}}}\mathrm{ch}(\pi_{\zeta_{+}}(\calB))
\end{equation}
where
\begin{eqnarray}\label{geomeqsZ}
J&=&\kappa^{\alpha}J_{\alpha}=it_{\alpha}\frac{H_{\alpha}}{2\pi}-\sum_{I=1}^{K}\frac{n_{I}(H)}{2}-(n+1)\frac{L}{2},\nonumber\\ \hat{\Gamma}_{X_{\zeta_{+}}}&=& \frac{1}{\prod_{I=1}^{n+1}\Gamma\left(i\frac{n_{I}(H)}{2\pi}+i\frac{L}{2\pi}+1\right)\prod_{I=n+2}^{K}\Gamma\left(i\frac{n_{I}(H)}{2\pi}+1\right)}\nonumber\\
&\times&\prod_{\alpha=1}^{r}\frac{\prod_{i_\alpha=1}^{k_\alpha}\Gamma\left(i\frac{H^{(i_{\alpha})}}{2\pi}+1\right)^{m_{\alpha}}}{\prod_{i_\alpha<j_\alpha}\Gamma\left(1+H^{(i_{\alpha})}_{\alpha}-H^{(j_{\alpha})}_{\alpha}\right)\Gamma\left(1-H^{(i_{\alpha})}_{\alpha}+H^{(j_{\alpha})}_{\alpha}\right)},\nonumber\\
\mathrm{ch}(\pi_{\zeta_{+}}(\calB))&=&\frac{f_{\calB}\left(\frac{H}{2\pi}\right)}{f_{\widehat{\calE}_{-}}\left(\frac{H}{2\pi}\right)}
\end{eqnarray}
by an explicit computation, we can show
\begin{prop} \label{PropClassicalA}
Conjecture \eqref{eqgeomplus} holds at the zero-instanton sector.
\end{prop}
\begin{proof}
First, notice that since $Z_{\pi_{\zeta_{+}}(\calB^{(0)})}=Z_{\pi_{\zeta_{+}}(\calB)}$ for any $\calB\in MF_{G}(W)$, as we only used objects whose central charge vanishes under $\pi_{\zeta_{+}}$ projection, therefore it is enough to compute using the brane factors $f_{\mathcal{B}}$, for objects before we grade restrict them to $\mathbb{W}(0)$. The constant $C$ in \eqref{Zzeroint}, is fixed by noticing that
\begin{eqnarray}
    f_{\calB_{\mathrm{pt}}}\left(\frac{H}{2\pi}\right) 
    &=& \left( 1-e^{-L}\right)^{n}\times\prod_{\alpha=1}^r\prod_{i_\alpha=1}^{k_\alpha}\left( 1-e^{-H^{(i_\alpha)}_\alpha} \right)^{m_\alpha-k_\alpha}, \nonumber
    \\
    &=& L^n\times\prod_{\alpha=1}^r \widetilde{H}_\alpha^{m_\alpha-k_\alpha}
\end{eqnarray}
as a class in $\mathbb{P}^{n}\times \vec{\mathbb{P}}$, then set, once and for all, $C=1$, and choose the normalization
\begin{eqnarray}
   \int_{\mathbb{P}^{n}\times \vec{\mathbb{P}}}L^n\times\prod_{\alpha=1}^r \widetilde{H}_\alpha^{m_\alpha-k_\alpha}=1.
\end{eqnarray}
then,
\begin{equation}
 Z^0_{\pi_{\zeta_{+}}(\calB_{\text{pt}})}(t)= \int_{\bbP^n\times\vec\bbP} (1+\cdots) L^n \times\prod_{\alpha=1}^r \widetilde{H}_\alpha^{m_\alpha-k_\alpha}=1=Z_P(\kappa). 
\end{equation}
The A-periods for the structure sheaf and divisor classes are straightforwardly computed by noticing that (using the expression for the Chern character in \eqref{geomeqsZ}):
\begin{equation}
\mathrm{ch}(\pi_{\zeta_{+}}(\widehat{\mathcal{E}}_{-}))=1,\qquad \mathrm{ch}(\pi_{\zeta_{+}}(\calB_{D_{\alpha}}))=1-e^{-H_{\alpha}},
\end{equation}
therefore using \eqref{gammaCY3}:
\begin{eqnarray}
Z^{0}_{\pi_{\zeta_{+}}(\widehat{\mathcal{E}}_{-})}(t)&=& \int_{X_{\zeta_+}} e^J\hat\Gamma_{X_{\zeta_+}} 1
\nonumber\\
&=& \int_{X_{\zeta_+}}\left( \frac{1}{6} J^3+\frac{1}{24} c_2(X_{\zeta_+})J+\frac{\zeta(3)}{(2\pi i)^3}c_3(X_{\zeta_{+}}) \right)
\nonumber\\
&=&Z_{\mathcal{O}_{X_{\zeta_{+}}}}(\kappa),
\\
Z^{0}_{\pi_{\zeta_{+}}(\calB_{D_{\alpha}})}(t) &=& \int_{X_{\zeta_+}} e^J\hat\Gamma_{X_{\zeta_+}} \left( J_\alpha-\frac{1}{2}J_\alpha^2+\frac{1}{6}J_\alpha^3 \right)
\nonumber\\
&=&\int_{X_{\zeta_+}} \left( \frac{1}{2}J_\alpha J^2-\frac{1}{2}J_\alpha^2J+\frac{1}{6}J_\alpha^3+\frac{1}{24}c_2(X_{\zeta_+})J_\alpha \right)
\nonumber\\
&=&Z_{\mathcal{O}_{D_{\alpha}}}(\kappa).
\end{eqnarray}
where $\kappa$ is defined as in \eqref{geomeqsZ}. Finally, using the expansion
\begin{eqnarray}
    e^J\hat{\Gamma}_{X_{\zeta_{+}}} \frac{e\left(N_{X/\mathbb{P}^{n}\times \vec{\mathbb{P}}}\right)}{f_{\widehat{\calE}_{-}}\left(\frac{H}{2\pi}\right)}&=&\left(1+ \kappa_0 L+\vec \kappa( H)+\cdots\right)\left( 1+\cdots \right)
    \nonumber
    \\
    &\times&\prod_{I=1}^{n+1}\left(1+\frac{L+\vec n_I(H)}{2}\right)\prod_{I=n+2}^K\left( 1+\frac{\vec n_I(H)}{2} \right)
    \nonumber
    \\
    &=&1+\left(\kappa_0+\frac{n+1}{2}\right)L+\left(\vec \kappa(H)+\frac{\vec m(H)}{2}\right)+\cdots, 
\end{eqnarray}
it is straightforward to compute the curve central charges/A-periods in the zero-instanton sector:
\begin{eqnarray}
   Z^0_{\pi_{\zeta_{+}}(\calB_0(-1,0))}(t)&=& \int_{\bbP^n\times\vec\bbP}\left( 1+\left(\kappa_0+\frac{n+1}{2}\right)L+\cdots \right) \left(\frac{-n-1}{2}L^n+ L^{n-1} \right) \times\prod_{\alpha=1}^r \widetilde{H}_\alpha^{m_\alpha-k_\alpha}\nonumber
    \\
    &=& \left(\kappa_0+\frac{n+1}{2}\right)+\frac{-n-1}{2}=\kappa_0=Z^{0}_{\widetilde C_0}(\kappa) ,
    \\
    Z^0_{\pi_{\zeta_{+}}(\calB_{\alpha}(0,-\mathbf{e}_\alpha))}(t)&=& \int_{Gr(k_\alpha,m_\alpha)}\left( 1+ (\kappa_\alpha +\frac{m_\alpha}{2})H_\alpha \right)\times \widetilde{H}_\alpha^{m_\alpha-k_\alpha-1}\left( -\frac{m_\alpha}{2}H_\alpha^{k_\alpha}+H_\alpha^{k_\alpha-1} \right)
    \nonumber
    \\
    &=& \kappa_\alpha \int_{Gr(k_\alpha,m_\alpha)}\widetilde{H}_\alpha^{m_\alpha-k_\alpha-1}H_\alpha^{k_\alpha}=\kappa_\alpha=Z^{0}_{\widetilde C_\alpha}(\kappa).
\end{eqnarray}

\end{proof}

\begin{thm}\label{thm:TPN}
Under the monodromy $M_{0}$, we have 
\begin{equation}
    \Pi_{+}(t)|_{M_{0}(\mathcal{B}^{(0)})}=\Pi_{+}(t)\cdot \mathrm{M}_{0},\qquad \mathrm{M}_{0}:=T_N\cdot L_{K_{Y}}
\end{equation}
where 
\begin{equation}
   T_N:=(T_{\calO_{\widetilde C_0}})^N=\begin{pmatrix}
       1 & & & 
       \\
        & \ \ \delta_{\alpha\beta} & &
       \\
        & -N\delta_{\alpha\beta,00}\   & \delta_{\alpha\beta} &
       \\
       & & & \ \ \  1
   \end{pmatrix},\label{eqn:MD0}
\end{equation}
corresponds to a matrix of $N$ times of the spherical twist by $\calO_{C_0}(-\mathbf{e}_\alpha)$, acting trivially on all components of $\Pi_{+}(t)$, except $Z_{\mathcal{O}_{D_{0}}}(t)\rightarrow Z_{\mathcal{O}_{D_{0}}}(t)-NZ_{\mathcal{O}_{\widetilde C_{0}}}(t)$. The coefficient $N$ is defined as:
\begin{equation}
    N:=\int_{Y}(A_2(\vec{n})^2-A_1(\vec{n})A_3(\vec{n})),\label{eqn:N}
\end{equation}
and \footnote{$\delta_{\mu\alpha}$ and $c_{\mu\mu\alpha}$ must be read as column vectors of rank $r+1$. Likewise, $b_{\mu\beta}$ and $\delta_{\mu\beta}$ must be read as rank $r+1$ row vectors.}
\begin{equation}\label{LKY}
   L_{K_{Y}}:= \prod_{\mu=1}^r L_{\mu}^{-n^{(\alpha)}},\qquad L_\mu=\begin{pmatrix}
       \ 1 &  & & 
        \\
        \  \delta_{\mu\alpha} & \ \delta_{\alpha\beta} &  & 
        \\
        c_{\mu\mu\alpha} & c_{\mu\alpha\beta} & \delta_{\alpha\beta} &
        \\
       \ 0 & \ b_{\mu\beta} & \delta_{\mu\beta} & \ 1
   \end{pmatrix},\quad  b_{\alpha\beta}:=\frac{c_{\alpha\alpha\beta}-c_{\alpha\beta\beta}}{2}
   \end{equation}
Where $L_\mu$ is given by 
\begin{eqnarray}
    \Pi(\kappa_\alpha+\delta_{\mu\alpha})=\Pi(\kappa_\alpha)\cdot L_\mu.
\end{eqnarray}
\end{thm}
\begin{proof}
First, note that since $\Pi(t)$ forms a complete set of solutions to a system of partial differential equations, it is enough to check the action of the monodromy for the zero-instanton sector, which becomes an exercise on the computation of characteristic classes, after proposition \ref{PropClassicalA}. For the case of $Z_{\pi_{\zeta_{+}}(\widehat{\calE}_{-})}(t)$, and $Z_{\pi_{\zeta_{+}}(\calB_{D_{\alpha}})}$, for $\alpha=1,\ldots,r$, it is clear from the equations \eqref{complexesOX} and
\eqref{strshfquasi}, respectively, that the effect of the mondodromy on the brane factor is
\begin{equation}
f_{\mathcal{B}}(\sigma)\rightarrow e^{2\pi\vec{n}(\sigma)}f_{\mathcal{B}}(\sigma),\qquad \vec{n}(\sigma):= \sum_{I=1}^{n+1}n^{(\alpha)}_{I}\sigma_{\alpha}
\end{equation}
then, using the reasoning as in the proof of proposition \ref{PropClassicalA}, with the extra insertion of $\exp(\vec{n(H)})$, the action of $L_{K_{Y}}$ defined in \eqref{LKY}, on these components of $\Pi(t)$, follows.

Secondly, for $\calB_{\text{pt}}$, the monodromy action adds a brane factor, to the original one, with Chan-Paton space \eqref{insertionpoint}, which leads to the insertion of the class
\begin{equation}\label{extrapt}
    -f_{\calE_-}\times \prod_{\alpha=1}^r \widetilde{H}_\alpha^{m_\alpha-k_\alpha}
\end{equation}
into the integral over $\mathbb{P}^{n}\times \vec{\mathbb{P}}$. By a simple degree argument, the insertion of \eqref{extrapt} identically vanishes on $\bbP^n\times\vec\bbP$ since $f_{\hat\calE_-}$ has no constant term. The monodromy action on $Z_{\calB_{\mathrm{pt}}}(t)$ is trivial, consistent with the matrix \eqref{LKY}.\\
For  $\calB_{\mu}$, $\mu\neq0$, the monodromy action in \eqref{M0C} adds the following term as
\begin{equation}
    -f_{\calE_-}\times \left(\prod_{\alpha\neq\mu} \widetilde{H}_\alpha^{m_\alpha-k_\alpha} \right) \times\widetilde{H}_\mu^{m-k-1}\left( -\frac{m_\mu}{2}H^k_\mu + H_\mu^{k-1}\right).
\end{equation}
By expanding $f_{\calB_-}$ to the leading term
\begin{eqnarray}
    f_{\calE_-}&=&\prod_{I=1}^{n+1}\left(1-e^{-L-\vec n_I(H)}\right)=L^n\ \vec n(H)+\cdots
\end{eqnarray}
the only contribution to the integral over $\mathbb{P}^{n}\times \vec{\mathbb{P}}$ is $L^n\times n^{(\mu)}H_\mu$, resulting in
\begin{eqnarray}
    &&\int_{\bbP^n\times\vec\bbP} -f_{\widehat\calE_-}\times \left(\prod_{\alpha\neq\mu} \widetilde{H}_\alpha^{m_\alpha-k_\alpha} \right) \times \widetilde{H}_\mu^{m-k-1}\left( -\frac{m_\mu}{2} H^k_\mu + H_\mu^{k-1}\right) \nonumber
    \\
    &=&-n^{(\mu)}\int_{\bbP^n\times Gr(k_\mu,m_\mu)} L^n \widetilde{H}_\mu^{m-k-1}H_\mu^{k}=-n^{(\mu)},\quad \vec{n}^{(\mu)}:=\sum_{I=1}^{n+1}n_{I}^{(\mu)}.
\end{eqnarray}
Thus 
\begin{equation}
    Z^0_{\pi_{\zeta_{+}}(M_{0}(\calB_{\mu}(0,-\mathbf{e}_\mu)))}=\kappa_\mu-n^{(\mu)}= Z^0_{\pi_{\zeta_{+}}(M_{0}(\calB_{\mu}(0,-\mathbf{e}_\mu)))}-n^{(\mu)}Z^{0}_{\pi_{\zeta_{+}}(\calB_{\mathrm{pt}})}=Z_{\calO_{\widetilde C_\mu}}\cdot L_\mu.
\end{equation}
The monodromy action on $\calB_0$ is trivial \eqref{trivialC0}, consistent with \eqref{LKY}\footnote{It is straightforward to show that the change $f_{\calB_{\mathrm{pt}}}(\sigma)\rightarrow \exp(2\pi\vec{n}(\sigma))f_{\calB_{\mathrm{pt}}}(\sigma)$ has no effect in $Z_{\pi_{\zeta_{+}}(\calB_{\mathrm{pt}})}$.}.
Finally, according to \eqref{FEminusB0} the monodromy action on the A-period $Z_{\calB_{D_{0}}}$ boils down to two effects: adding the central charge $Z_{\mathcal{C}}$ of the object $\calC$ in \eqref{calC} and twisting by $K_{Y}$. The effect of twisting by $K_{Y}$ on  $Z_{\pi_{\zeta_{+}}(\calB_{D_{0}})}$ is completely analogous to the computation for $Z_{\pi_{\zeta_{+}}(\calB_{D_{\alpha}})}$, so the only nontrivial effect is adding $Z_{\mathcal{C}}$ and in order to compute this effect we need the Chern character 
\begin{equation}\label{chernCC}
\begin{aligned}
    \mathrm{ch}( \calC)=&\frac{f_{\calC}\left(\frac{H}{2\pi}\right)}{f_{\widehat\calE_-}\left(\frac{H}{2\pi}\right)}=e^{-L+\vec n(H)}+(n+1)-e^{-L}-\sum_{I=1}^{n+1} e^{\vec n_I(H)}
    \\
    =&\sum_{I=1}^{n+1}\left(1-e^{\vec n_I(H)}\right)-e^{-L}\left(1-e^{\vec n(H)}\right)
    \\
    =&\frac{A_1(\vec n)}{2}L^2-\left( A_1(\vec n)+\frac{1}{2}A_1(\vec n)^2 \right)L+A_2(\vec n)-\frac{1}{2}\bigg(A_3(\vec n)-A_1(\vec n)A_2(\vec n)\bigg)
\end{aligned}
\end{equation}
Note that the top degree summand in \eqref{chernCC} identically vanishes, upon using \eqref{plusnatform}: 
\begin{equation}
\begin{aligned}
   &\frac{1}{2}\int_X A_1L^2+A_1A_2-A_1^2L-A_3
    \\
    =&\frac{1}{2}\int_Y A_1A_3+A_1^2A_2-A_1^2A_2-A_1A_3
    \\
    =&0
\end{aligned}
\end{equation}
Then, by direct computation, we get the general form of the Gambelli-Thom-Porteous formula \cite{Candelas:1987kf} in terms of the vectors $\vec n_I$, time $Z_{\calO_{C_0}}$: 
\begin{equation}
\begin{aligned}
    Z_{\calC}=&\int_{X} (\kappa_0 L+\vec \kappa(H))(A_2-A_1L)
    \\
    =&\kappa_0\int_X \left(A_2 L- A_1 L^2\right)+ \int_X (A_1 L-A_2)\vec\kappa(H)
    \\
    =&\kappa_0\int_Y \left(A_2^2-A_1A_3 \right)+\int_Y (A_1A_2-A_2A_1)\vec\kappa(H)
    \\
    =& N\kappa_0=N Z_{\calO_{\widetilde C_0}}.
\end{aligned}
\end{equation}
This completes the proof.
\end{proof}

Theorem \ref{thm:TPN} is reminiscent of the results in \cite{donovan2024derived}, it will be interesting to clarify this relation.

\begin{rmk}\label{rmkN}
    As pointed out in rmk. \ref{rmkbundles}, we can replace the nef partition of $-K_{Y}$ by a vector bundle $\calN$. Then the derivation of $N$ in theorem \ref{thm:TPN} can be carried on analogously, if $\calN$ is constructed from an associated bundle $\mathcal{V}\times_{h}G$ with $h$ having positive weights. Then, $A_k(\vec n)$ is identified with $c_k(\calN)$. We will analyze several examples of this type in sect. \ref{sec:NAexamples}
\end{rmk}

\section{Examples}\label{sec:examples}

In this section we will present some examples of explicit computation of the matrix $M_{0}$, obtained by the window shift monodromy and show how the quantum nature of $\mathcal{M}_{K}$, allows us say more about the structure of $M_{0}$ i.e., we will show that we can write $M_{0}$ as a product of matrices associated to simpler autoequivalences, associated to different components of nested torus links, whose fundamental group properties are analyzed in appendix \ref{sec:appknot}. There are some general facts we need to recall that will be useful along this section.

On the Doran-Morgan basis, the monodromy matrices $L_\alpha$ and $T_{\calE}$ take explicit forms with some coefficients given by the topological data of $X$ \cite{Doran:2005gu,Almkvist:2005qoo}. $L_{\alpha}$ is given in \eqref{LKY}. The spherical twist \cite{Seidel:2000ia} $T_{\calE}$ by a spherical object $\calE \in D^bCoh(X)$\footnote{For a more general case known as the EZ-twist, see \cite{Horja:2001cp}. The matrix form can be found in \cite{Cota:2019cjx}.} corresponds to the action on the central charges
\begin{equation} \label{TE}
Z_{T_\calE(\calB)}=Z_\calB-\chi(\calE,\calB)Z_\calE, \qquad \calB \in D^bCoh(X)
\end{equation}
with $\chi(\calE,\calB)$ the Euler characteristic (which can also be computed using the GLSM annulus partition function as in \cite{Lin:2024fpz}), explicitly:
\begin{equation}
    \chi_{\alpha\beta}=\begin{pmatrix}
        0 & \ \ \  l_\beta & 0 & 1
        \\
        \ -^tl_\alpha & -b_{\alpha\beta} & -\delta_{\alpha\beta} & 0 
        \\
        0 & \ \ \ \delta_{\alpha\beta} & 0 & 0
        \\
        -1 & \ 0 & 0 & 0 & 
    \end{pmatrix}, \quad l_\alpha=\frac{1}{6}c_{\alpha\alpha\alpha}+2c_{\alpha}.
\label{eqn:chiab}
\end{equation}
For instance, for $\calE=\calO_X$, we get:
\begin{equation}
    T_{\calO_X}=\begin{pmatrix}
        1 & \ -l_\beta & 0 & -1
        \\
        & \ \ \ \ \delta_{\alpha\beta} & &
        \\
         &  & \delta_{\alpha\beta} & 
        \\
         &  &  & 1
    \end{pmatrix},\label{TOX}
\end{equation}
and for $\calE=\calO_{\widetilde C_\mu}$, we have
\begin{equation}
   T_{\calO_{\widetilde C_\mu}}=\begin{pmatrix}
       1 & & & 
       \\
        & \ \delta_{\alpha\beta} & &
       \\
        & -\delta_{\alpha\beta,\mu\mu}\   & \delta_{\alpha\beta} &
       \\
       & & & \ \ \  1
   \end{pmatrix}.
\end{equation}

\subsection{Abelian examples in "Five Guys\textsuperscript{{\textregistered}}"}\label{fiveguys}

In this section, we will focus on the following CICY3 hypersurfaces $X^\natural$. In all these cases $Y=\vec\bbP$ as
\begin{equation}
   X^\natural= \bbP^4[5],
    \quad 
    \left[\begin{array}{c|c}
     \bbP^2& 3  \\
     \bbP^2& 3
\end{array} \right],
\quad
\left[\begin{array}{c|c}
     \bbP^3& 4  \\
     \bbP^1& 2
\end{array} \right],
\quad
\left[\begin{array}{c|c}
     \bbP^2& 3  
     \\
     (\bbP^1)^2& 2
\end{array} \right],
\quad
(\bbP^1)^4[2].\label{eqn:Xnatural}
\end{equation}
Denote them by $X^\natural(\vec n)$ for
\begin{equation}
    \vec n=(n^{(1)},\cdots,n^{(r)})=(5),\ (3,3),\ (4,2),\ (3,2,2),\ (2,2,2,2).
\end{equation}
Then, by the argument in \eqref{extrtransition}, given any nontrivial multi-partition of $\vec{n}$ (i.e. a partition for each $n^{(i)}$) we can write a splitting configuration of the type \eqref{eqn: CICYI} (and consequently a GLSM). By a straightforward matrix computation carried in appendix \ref{sec:appcomputation}, we can state the following theorem:
\begin{thm}\label{thm:CICY}
Given a $X^\natural(\vec n)\subset\vec{\bbP}$ as in \eqref{eqn:Xnatural}, the following matrix identity holds on a splitting configuration $X$ (as in \eqref{eqn: CICYI}) of $X^\natural$: 
\begin{eqnarray}
    M_0\cdot L_1^{n^{(1)}}\cdots L_r^{n^{(r)}}=T_N,
\end{eqnarray}
where the matrices ${L}_{\alpha}$ and $T_{N}$ are as in theorem \ref{thm:TPN} and moreover, $M_{0}$ can be further decomposed into the following recursive formula:
\begin{equation}\label{eqfiveguysM}
     M_r:=T_{\mathcal{O}_{X_{\zeta_{+}}}},\qquad M_\alpha=(M_{\alpha+1}L_{\alpha+1})^{n^{(\alpha+1)}}L_{\alpha+1}^{-n^{(\alpha+1)}},\qquad \alpha=0,\cdots,r-1,
\end{equation}
where the matrices in \eqref{eqfiveguysM} satisfies the $\mathfrak{L}_{n^{(\alpha)}}^{(1)}$ link relation:
\begin{eqnarray}
    (M_\alpha L_\alpha)^{n^{(\alpha)}}=(L_\alpha M_\alpha)^{n^{(\alpha)}},\qquad \alpha=1,\cdots,r.
\end{eqnarray}
In particular, the above result holds under shuffling the order of $n^{(1)},\cdots,n^{(r)}$. This result is compatible with conjecture \ref{conjecturenested}.

\end{thm}
Finally, to illustrate the possible extension of our results to the cases where weighted projective spaces are included in $\vec{\mathbb{P}}$ we show by a direct computation, in sect. \ref{sec:weighted}, that similar results than in theorem \ref{thm:CICY} holds for $X^{\natural}=\bbW\bbP_{96111}[18]$.

\subsection{Nonabelian examples}\label{sec:NAexamples}

In this section, we discuss two families of nonabelian two-parameter examples of splitting configurations of $X^{\natural}=Gr(2,4)[4]$ and $X^{\natural}=Gr(2,5)[3,1,1]$, respectively. In each family, the links related to the discriminant $\Delta$ have three  components, in contrast with the abelian examples of sect. \ref{fiveguys} where each relevant link have only two components. We conjecture that the extra component always corresponds a spherical twist by $\calS_X$: the pull back of the tautological bundle on $Gr(k,n)$, to $X$. The appearance of $T_{\calS_{X}}$ is a phenomena that has been observed already in GLSMs related to CYs in Grassmannians \cite{eager2017beijing,Lin:2024fpz,EHKR}. This object has geometric central charge (in the notation of \eqref{chargeofBB})\footnote{The central charge of any object $\calE$ in our basis is written as $Z_{\calE}=\int_X \Ch_0(\calE)(\frac{1}{6}J^3+\frac{c_2(X)}{24}J+\frac{\zeta(3)}{(2\pi i)^3}c_3(X))+\Ch_1(\calE)(\frac{1}{2}J^2+\frac{c_2(X)}{24})+\Ch_2(\calE) J+\Ch_3(\calE)$.}
\begin{equation}
    Z_{\calS^{}_X}=\left(\operatorname{rk}\calS_X^{},-\int_{C_\alpha}c_1(\calS_X),\ -\int_{D_\alpha} c_2(\calS_X),\ \int_X\Ch_3\calS^{}_X\right)
\end{equation}
whose spherical twist matrix $T_{\calS_X}$ is given by \eqref{TE}.\\
The way we will classify the splitting configurations in these examples will be by specifying a partition $ n_{1}+\ldots+n_{\ell}=k_{1}$, where $k_{1}=4$ for $X^{\natural}=Gr(2,4)[4]$ and $k_{1}=3$ for $X^{\natural}=Gr(2,5)[3,1,1]$. Then, we will consider a (reducible) vector bundle
\begin{equation}\label{sumNbundles}
\mathcal{N}=\oplus_{i=1}^{\ell}\mathcal{N}_{i}\rightarrow Y
\end{equation}
whose summands $\mathcal{N}_{i}$ satisfy $c_{1}(\mathcal{N}_{i})=n_{i}$, $i=1,\ldots,\ell$. Then, the choices for $\calN_{i}$, in general, are not unique. Since we restrict to bundles $\mathcal{N}_{i}$ whose weights under $U(2)$ are positive, the possibilities, given a partition, are finite and 
\begin{equation}
\sum_{i=1}^{\ell}c_{0}(\mathcal{N}_{i})=n+1.
\end{equation}
We will explicitly list them in the following. Finally, as mentioned in rmk. \ref{rmkN}, $A_k(\vec n)$ is essentially the Chern class $c_k(\calN)$ for a generalized bundle $\calN$. Thus, we denote them by $A_k(\calN)$ for consistency for non-abelian cases. Moreover, the same multi-partition $\vec n_I$ in a non-abelian $U(k)$ case is defined similarly in a $U(1)^k$ case, which is given by the weight of the Cartan subalgebra $\prod_{\alpha=1}^rU(1)^{k_\alpha}\subset\prod_{\alpha=1}^rU(k_\alpha)$ on the corresponding representation of $\calN$. Thus, the similar result in the non-abelian case can be written as
\begin{equation}
    c(\calN)=:\prod_{I=1}^{n+1}\left( 1+\vec n_I(H) \right),\quad \vec n_I(H)=\sum_{\alpha=1}^r\sum_{i_\alpha=1}^{k_\alpha} n_I^{(\alpha,\ i_\alpha)}H_\alpha^{(i_\alpha)}.
\end{equation}

\subsubsection{$X^{\natural}=Gr(2,4)[4]$}

In this case, the splitting configuration $X$ corresponding to $1+1+1+1=4$\footnote{In terms of multi-partition $\vec n_I=(n_I^{(1,1)},n_I^{(1,2)})$, it is written as $4(1,0)+4(0,1)=(4,4)$.}, where $\mathcal{V}_{I}=\calS^{\vee}$ for all $i=1,\ldots, 4$, was analyzed in detail in \cite{Lin:2024fpz} and it is found that $\Delta\cap S^{3}\cong \mathfrak{L}_{4,2}^{(2)}$ with $S^{3}$ centered at $\{z_{1}=0\}\cap \Delta$ (where $\zeta_{+}$ phase is located near $(z_{0},z_{1})=(0,0)$). Moreover, in \cite{Lin:2024fpz} it was shown that the loops around the components of $\mathfrak{L}_{4,2}^{(2)}$  are given by (using the notation of appendix \ref{sec:appknot}):
\begin{equation}
a_{1}\rightarrow T_{\mathcal{O}_{X}},\qquad a_{2}\rightarrow T_{\calS_{X}} .
\end{equation}
We conjecture that this holds for all splitting configurations of $Gr(2,4)[4]$. The following proposition can be proven by direct computation, and is a strong evidence for the conjecture:
\begin{prop}
    The following matrix identities, from the fundamental group of $\mathfrak{L}_{4,2}^{(2)}$, holds on any splitting configuration $Gr(2,4)[4]$:
    \begin{eqnarray}
        &&(T_{\calS_X}T_{\calO_X}L_{1})^2(T_{\calO_X}L_{1})^2 a=a(T_{\calS_X}T_{\calO_X}L_{1})^2(T_{\calO_X}L_{1})^2,\quad a\in\{L_{1},T_{\calO_X}\},
        \nonumber\\
        &&(T_{\calS_X}T_{\calO_X}L_{1})^2T_{\calS_X}=T_{\calS_X}(T_{\calS_X}T_{\calO_X}L_{1})^2.\\
        \label{eqn:linkG24}
    \end{eqnarray}
   Recall that $L_{1}$ corresponds to the matrix associated to the twist $-\otimes\mathrm{det}^{-1}\calS_{X}$ Moreover, 
    \begin{equation}
         M_0\cdot L_{1}^{4}= (T_{\calS_X}T_{\calO_X}L_{1})^2(T_{\calO_X}L_{1})^2=T_N. \label{eqn:TPNG24}
    \end{equation}
    for $N$ given by \eqref{eqn:N} from a partition $4$.
\end{prop}

\begin{proof}
Consider the general splitting associated to a partition $n_{1}+\ldots n_{\ell}=4$ as in rmk. \ref{rmkN}:
\begin{equation}
    X=\left[\begin{array}{c|ccc}
        \bbP^n &  1 & \cdots & 1
        \\
        Gr(2,4) & n_1 & \cdots & n_{\ell} 
    \end{array}\right].
\end{equation}
Where $n_{i}=c_{1}(\mathcal{N}_{i})$ as in \eqref{sumNbundles}. In this case $Y=Gr(2,4)$. Next, we evaluate $N$ in $T_N$ using \eqref{eqn:c00a}-\eqref{eqn:c0ab}. Since $A_2(\calN)$ is a class of degree two on $Y$, it can always be expanded in the basis $H^2=c_1(Y)^2$ and $\widetilde{H}=c_2(Y)$ as
\begin{equation}
    A_2(\calN)=h_1H^2+h_2 \widetilde{H},\qquad h_{1},h_{2}\in \mathbb{Z}
\end{equation}
then using Littlewood-Richardson rule on $Gr(2,4)$\footnote{$\int_YH^4=2\int_YH^2\widetilde{H}=2\int_Y\widetilde{H}^2=2$}, one has
\begin{eqnarray}
    c_{011}&=&\int_{Y} A_2(\calN)H^2=2h_1+h_2,
    \\
    g&=&\int_{Y}A_2(\calN)\widetilde{H}=h_1+h_2.
\end{eqnarray}
It can be rewritten that
\begin{eqnarray}
    \int_{Y}A_2(\calN)^2&=&\int_{Y}h_1^2H^4+2h_1h_2H^2\widetilde{H}+h_2^2\widetilde{H}^2
    \nonumber\\
   &=&2h_1^2+2h_1h_2+h_2^2
   \nonumber
   \\
    &=&2g^2+c_{011}^2-2c_{011}g 
   \nonumber
   \\
\end{eqnarray}
and
\begin{equation}
    \int_{Y}A_1(\calN)A_3(\calN)=\int_{Y}4H A_3(\calN)=4c_{001}.
\end{equation}
Thus for any $X$ and choice of partition $\{n_i\}_{i=1}^{\ell}$, $N$ is always given by
\begin{eqnarray}
    N&=&\int_{Y} A_2(\calN)^2-A_1(\calN)A_3(\calN)
    \nonumber\\
    &\equiv&2g^2+c_{011}^2-2c_{011}g-4c_{001}.\label{eqn:NG24}
\end{eqnarray}
Where $c_{011}$ and $g$ depends on the choice of partition and bundles. To show the link relations \eqref{eqn:linkG24} and \eqref{eqn:TPNG24} we use the explicit form for the matrices:
\begin{eqnarray}
     L_{1}&=&\begin{pmatrix}
 1 &  &  & \  & \  &  \\
 0 & 1 &  & \  & \  &  \\
 1 & 0 & 1 & \  & \  &  \\
 c_{011} & c_{001} & c_{011} & \ \ 1 & \  &  \\
 8 & c_{011} & 8 & \ \ 0 & \ \ 1 &  \\
 0 & b & 0 & \ \ 0 & \ \ 1 & \ \ 1
    \end{pmatrix},\qquad b=\frac{c_{011}-c_{001}}{2},
    \\
    T_{\calO_X}&=&\begin{pmatrix}
        1 & -l & -6 & 0 & \ 0 & -1
        \\
         & \ 1 & &&&
         \\
         && 1 &&&
         \\
         &&& 1 &&
         \\
         &&&& \ 1 &
         \\
         &&&&&1
    \end{pmatrix},\qquad l=\frac{1}{6}c_{000}+2c_0
    \\
    T_{\calS_X}&=&\left(
\begin{array}{cccccc}
 -7 & 2g-2b-4 l & -16 & 0 & -2 & -4 \\
 0 & 1 & 0 & 0 & 0 & 0 \\
 4 & b-g+2 l & 9 & 0 & 1 & 2 \\
 4 g & bg-g^2+2 l g & 8 g & 1 &
   g & 2 g \\
 16 & 4b-4 g+8 l & 32 & 0 & 5 & 8 \\
 -8 & 2g-2b-4 l & -16 & 0 & -2 & -3 \\
\end{array}
\right),
\end{eqnarray}
where, the central charge of $\calS_X$ is given by
\begin{equation}
    Z_{\calS_X}=(2,0,-1,-g,-4,2).
\end{equation}
Then by direct matrix multiplication we verify \eqref{eqn:linkG24} and \eqref{eqn:TPNG24}, where $N$ is precisely given by \eqref{eqn:NG24}.
    
\end{proof}

\begin{rmk}
   We list all general splitting configurations $\calN$ as the following. Denote homogeneous bundles on $Gr(2,4)$ by $S(p,q)=\calO(p)\otimes\Sym^q S$ and $S=\calS^\vee$ denotes the dual tautological bundle. The Chern character $\Ch S(p,q)$ is given by
   \begin{equation}
       \Ch S(p,q)=e^{p(H^{(1)}+H^{(2)})}\sum_{\substack{l_1,l_2\geq0\\l_1+l_2=q}} e^{l_1H^{(1)}+l_2H^{(2)}}.
   \end{equation}
   The possible irreducible bundles in the splitting configuration must satisfy $c_1(S(p,q))\leq4c_1(S)$ and they are given by
\begin{equation}
\begin{array}{c|c|cc}
   S(p,q)  &   c_0=q+1 & c_1=\frac{(2p+q)(q+1)}{2} & 
     \\\hline
   \calO(p) & 1 & p & 1\leq p\leq4
   \\
   \Sym^qS & q+1 & q(q+1)/2 & q=1,2
   \\
   S(1) & 2 & 3
\end{array}\label{eqn:Spq}
\end{equation}
And all admissible $\calN$ and their corresponding $N$ are listed below:
\begin{itemize}
    \item $c_0=2$
    \begin{eqnarray}
        \calO(1)\oplus\calO(3) &\quad& N=18
        \\
        \calO(2)^{\oplus2} &\quad& N=32
    \end{eqnarray}

    \item $c_0=3$
    \begin{eqnarray}
        S(1)\oplus \calO(1) &\quad& N=41
        \\
        S\oplus \calO(3) &\quad& N=13
        \\
        \calO(1)^{\oplus2}\oplus \calO(2) &\quad& N=34
    \end{eqnarray}

    \item $c_0=4$
    \begin{eqnarray}
        \Sym^2S\oplus \calO(1) &\quad& N=58
        \\
        S(1)\oplus \calS &\quad& N=42
        \\
        S\oplus \calO(1)\oplus \calO(2) &\quad& N=33
        \\
        \calO(1)^{\oplus4} &\quad& N=40
    \end{eqnarray}

    \item $c_0=5$
    \begin{eqnarray}
        \Sym^2S\oplus S &\quad& N=65
        \\
        S\oplus \calO(1)^{\oplus3} &\quad& N=41
        \\
        S^{\oplus2}\oplus \calO(2) &\quad& N=34
    \end{eqnarray}

    \item $c_0=6$
    \begin{eqnarray}
        S^{\oplus2}\oplus \calO(1)^{\oplus2} \quad N=44
    \end{eqnarray}

    \item $c_0=7$
    \begin{equation}
        S^{\oplus3}\oplus \calO(1) \quad N=49
    \end{equation}

    \item $c_0=8$
    \begin{equation}
        S^{\oplus4} \quad N=56 \quad\text{(GN model \cite{Jockers:2012zr})}
    \end{equation}
\end{itemize}
\end{rmk}

\subsubsection{$X^{\natural}=Gr(2,5)[3,1,1]$}

\begin{figure}[h]
\centering
\includegraphics[width=.5\textwidth]{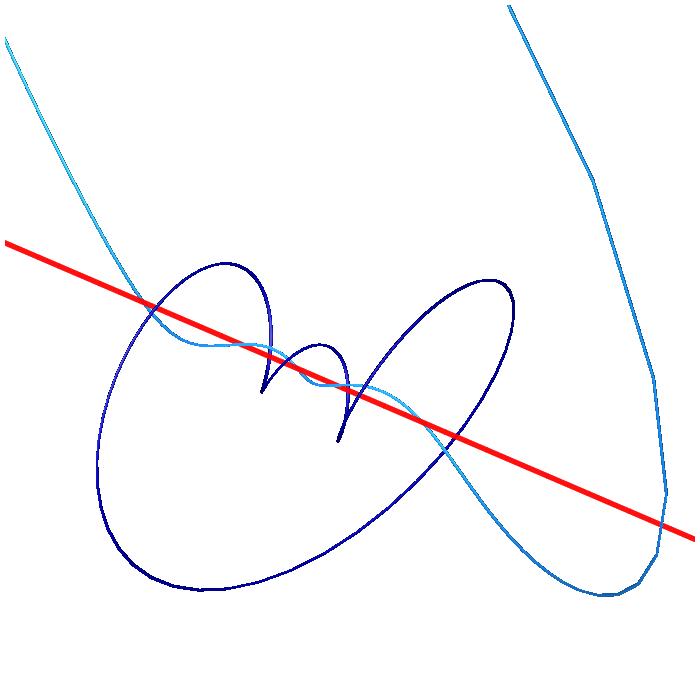}
\caption{The $\mathfrak{L}^{(2)}_{3,3}$ link of $\Delta$ in 5-intersection model. The inner component (light blue) is conjecturally the spherical twist $T_{\calO_X}$ and the outer component (dark blue) is $T_{\calS_X}$}\label{fig:knotGr25}
\end{figure}

The second family of splitting configurations are characterized by partitions of $n_{1}+\ldots+n_{\ell}=3$. Before stating the general result we will start with a particular case of the partition $1+1+1=3$\footnote{In terms of multi-partition $\vec n_I$ it is written as $(1,1)+(1,1)+(1,1)=(3,3)$.}, corresponding to the 5-intersection genus one fibration model discussed in \cite{Knapp:2021vkm}:
\begin{equation}
    X=\left[ \begin{array}{c|ccccc}
       \bbP^2  & 1 & 1 & 1 & 0 & 0  
         \\
       Gr(2,5)  & 1 & 1 & 1 & 1 & 1 
    \end{array} \right]
\end{equation}
where $Gr(2,5)[1,1,1,1,1]$ is a non-abelian GLSM realized elliptic curve \cite{Knapp:2023izn}. The GLSM discriminant is given by \cite{Knapp:2021vkm}
\begin{eqnarray}
    \Delta &=& 1 - 33z_1 + 360z_1^2 - 1265z_1^3 - 360z_1^4 - 33z_1^5 - z_1^6 - 6z_0 - 132z_1z_0 - 1602z_1^2z_0 
\nonumber\\
&+& 4686z_1^3z_0 - 2601z_1^4z_0 + 33z_1^5z_0 + 15z_0^2 + 561z_1z_0^2 + 1755z_1^2z_0^2 - 4686z_1^3z_0^2 \nonumber\\
&-& 360z_1^4z_0^2 - 20z_0^3 - 561z_1z_0^3 - 1602z_1^2z_0^3 + 1265z_1^3z_0^3 + 15z_0^4 + 132z_1z_0^4 
\nonumber\\
&+& 360z_1^2z_0^4 - 6z_0^5 + 33z_1z_0^5 + z_0^6
\end{eqnarray}
which intersects with the $Gr(2,5)$ large volume limit $z_1=0$ at $(z_0,z_1)=(1,0)$. Around the intersection point, the local form of discriminant at $\tilde z_0=z_0-1$, $\widetilde z_1=z_1$ is given by
\begin{eqnarray}
  \Delta &=& \widetilde{z}_0^6+33 \left(\widetilde{z}_0 \left(\widetilde{z}_0+9\right)+9\right) \widetilde{z}_1 \widetilde{z}_0^3+33 \widetilde{z}_1^5 \widetilde{z}_0+11
   \left(\widetilde{z}_0 \left(115 \widetilde{z}_0-81\right)-81\right) \widetilde{z}_1^3 \widetilde{z}_0-\widetilde{z}_1^6
   \nonumber\\
   &-&9 \left(\widetilde{z}_0
   \left(40 \widetilde{z}_0+369\right)+369\right) \widetilde{z}_1^4+9 \left(\widetilde{z}_0 \left(\widetilde{z}_0 \left(2 \widetilde{z}_0
   \left(20 \widetilde{z}_0-9\right)-99\right)-162\right)-81\right) \widetilde{z}_1^2.
   \nonumber
\end{eqnarray}
Then $\Delta\cap S^3\cong\mathfrak{L}_{3,3}^{(2)}$, with $S^3$ centered at $(1,0)$, as visualized in figure \ref{fig:knotGr25}. 
We conjecture that the loops around the components of $\mathfrak{L}^{(2)}_{3,3}$ are given by (using the notation of appendix \ref{sec:appknot}):
\begin{equation}
a_{1}\rightarrow T_{\mathcal{O}_{X}},\qquad a_{2}\rightarrow T_{\calS_{X}} .
\end{equation}
which can be tested they satisfy the expected link relations. 

We conjecture this is the case for any splitting configuration $X$ of $Gr(2,5)[3,1,1]$. The following proposition, proved by a direct computation, serves as a strong evidence:
\begin{prop}
The following matrix identities, from the fundamental group of $\mathfrak{L}_{3,3}^{(2)}$, holds for any splitting configuration $X$ of $Gr(2,5)[3,1,1]$:
    \begin{eqnarray}
        (T_{\calS_X}T_{\calO_X}L_{1})^3a=a(T_{\calS_X}T_{\calO_X}L_{1})^3,\quad a\in\{L,T_{\calO_X},T_{\calS_X}\}.\label{eqn:linkG25}
    \end{eqnarray}
    Moreover,
    \begin{equation}
         M_1\circ L_{1}^3=(T_{\calS_X}T_{\calO_X}L_{1})^3=T_N.\label{eqn:TPNG25}
    \end{equation}
\end{prop}
\begin{proof}
    Consider a splitting configuration of the degree 3 surface in $Y=Gr(2,5)[1,1]$, denoted as
\begin{equation}
     X=\left[\begin{array}{c|ccccc}
        \bbP^n &  1 & \cdots & 1 & 0 & 0
        \\
        Gr(2,5) & n_1 & \cdots & n_{\ell} & 1 & 1
    \end{array}\right].
\end{equation}
Firstly, we compute the value of $N$ in $T_N$. Since $A_2(\calN)$ is a class of degree two on $Y$, it can always be expanded in the basis $H^2=c_1(Y)^2$ and $\widetilde{H}=c_2(Y)$ as $A_2(\calN)=h_1 H^2+h_2\widetilde H$ again. Using the Littlewood-Richardson rule on $Y$\footnote{Using $ H^6=5\widetilde H^3,\ H^4\widetilde H=2H^2\widetilde{H}^2=2\widetilde{H}^3$, $\int_{Gr(2,5)}\widetilde H^3=1$, one has $\int_Y H^4=5,\ \int_Y H^2\widetilde H=2 \int_Y\widetilde H^2=2$.}, one has
\begin{eqnarray}
    c_{011}=&\int_{Gr(2,5)} A_2(\calN)c_1^4=5h_1+2h_2,
    \\
    g=&\int_{Gr(2,5)}A_2(\calN)c_1^2c_2=2h_1+h_2.
\label{eqn:liftG25}
\end{eqnarray}
It can be rewritten that
\begin{eqnarray}
    \int_Y A_2(\calN)^2&=&\int_{Y} h_1^2H^4+2h_1h_2H^2\widetilde H+h_2^2\widetilde{H}^2
    \nonumber
    \\
    &=& 5h_1^2+4h_1h_2+h_2^2
    \nonumber
    \\
    &=&  5g^2+c_{011}^2-4c_{011}g
\end{eqnarray}
and
\begin{equation}
     \int_YA_1(\calN)A_3(\calN)=\int_Y 3H A_3(\calN)=3c_{001}.
\end{equation}
Thus for any $X$ and choice of partition $\{n_I\}_{I=1}^{\ell}$, $N$ is always given by
\begin{eqnarray}
    N&=&\int_{Y} A_2(\calN)^2-A_1(\calN)A_3(\calN)
    \nonumber\\
    &\equiv&5g^2+c_{011}^2-4c_{011}g-3c_{001}.\label{eqn:NG25}
\end{eqnarray}
Where $c_{011}$ and $g$ depends on the choice of partition and bundles. To show the link relations \eqref{eqn:linkG25} and \eqref{eqn:TPNG25} we use the explicit form for the matrices:
\begin{eqnarray}
     L&=&\begin{pmatrix}
 1 &  &  &  &  &  \\
 0 & 1 &  &  &  &  \\
 1 & 0 & 1 &  &  &  \\
 c_{011} & c_{001} & c_{011} & \  1 &  &  \\
 15 & c_{011} & 15 & \ 0 & \ \ 1 &  \\
 0 & b & 0 & \ 0 & \ \ 1 & \ \ 1
    \end{pmatrix},\qquad b=\frac{c_{011}-c_{001}}{2}
    \\
    T_{\calO_X}&=&\begin{pmatrix}
        1 & -l & -8 & \ 0 & \ \ 0 & \ -1
        \\
         & \ 1 & &&&
         \\
         && 1 &&&
         \\
         &&& \  1 &&
         \\
         &&&& \ \ 1 &
         \\
         &&&&&\ 1
    \end{pmatrix},\qquad l=\frac{1}{6}c_{000}+2c_0,
    \\
    T_{\calS_X}&=&\left(
\begin{array}{cccccc}
 -9 & 2g-b-4 l & -20 & 0 & -2 & -4 \\
 0 & 1 & 0 & 0 & 0 & 0 \\
 5 & b-g+2 l & 11 & 0 & 1 & 2 \\
 5 g & bg-g^2+2 l g & 10 g & 1 &
   g & 2 g \\
 30 & 6b-6 g+12 l & 60 & 0 & 7 & 12 \\
 -15 & 3 g-3b-6 l & -30 & 0 & -3
   & -5 \\
\end{array}
\right)
\end{eqnarray}
where, the central charge of $\calS_X$ is given by
\begin{equation}
    Z_{\calS_X}=(2,0,-1,-g,-6,3).
\end{equation}
Then by direct computation, we verify  \eqref{eqn:linkG25} and \eqref{eqn:TPNG25}, in where the last result is a simple $T_N$ matrix in the form of \eqref{eqn:MD0} with $N$ written in \eqref{eqn:NG25}.

\end{proof}
\begin{rmk}
   The bundles $\calN$ involved in the splitting configurations, are given by sums of the homogeneous bundles in \eqref{eqn:Spq} satisfying $c_1(S(p,q))\leq 3c_1(S)$. All admissible $\calN$ and their $N$'s are listed below:
\begin{itemize}
    \item $c_0=2$
    \begin{eqnarray}
        S(1) &\quad& N=29
        \\
        \calO(1)\oplus \calO(2) &\quad& N=20
    \end{eqnarray}

     \item $c_0=3$
    \begin{eqnarray}
        \Sym^2S &\quad& N=44
        \\
        S\oplus \calO(2) &\quad& N=17
        \\
        \calO(1)^{\oplus3} &\quad& N=30\quad \text{(genus one 5-intersection \cite{Knapp:2021vkm})}\label{TPN30}
    \end{eqnarray}
    
    \item $c_0=4$
    \begin{eqnarray}
        S\oplus \calO(1)^{\oplus 2} &\quad& N=31
    \end{eqnarray}
    
   \item $c_0=5$
   \begin{eqnarray}
       S^{\oplus2}\oplus \calO(1) &\quad& N= 34
   \end{eqnarray}

   \item $c_0=6$
   \begin{eqnarray}
       S^{\oplus3} &\quad& N=39
   \end{eqnarray}
\end{itemize}
\end{rmk}

\appendix

\section{Fundamental group of nested torus links}\label{sec:appknot}

\begin{figure}[h]
\centering
\includegraphics[width=.6\textwidth]{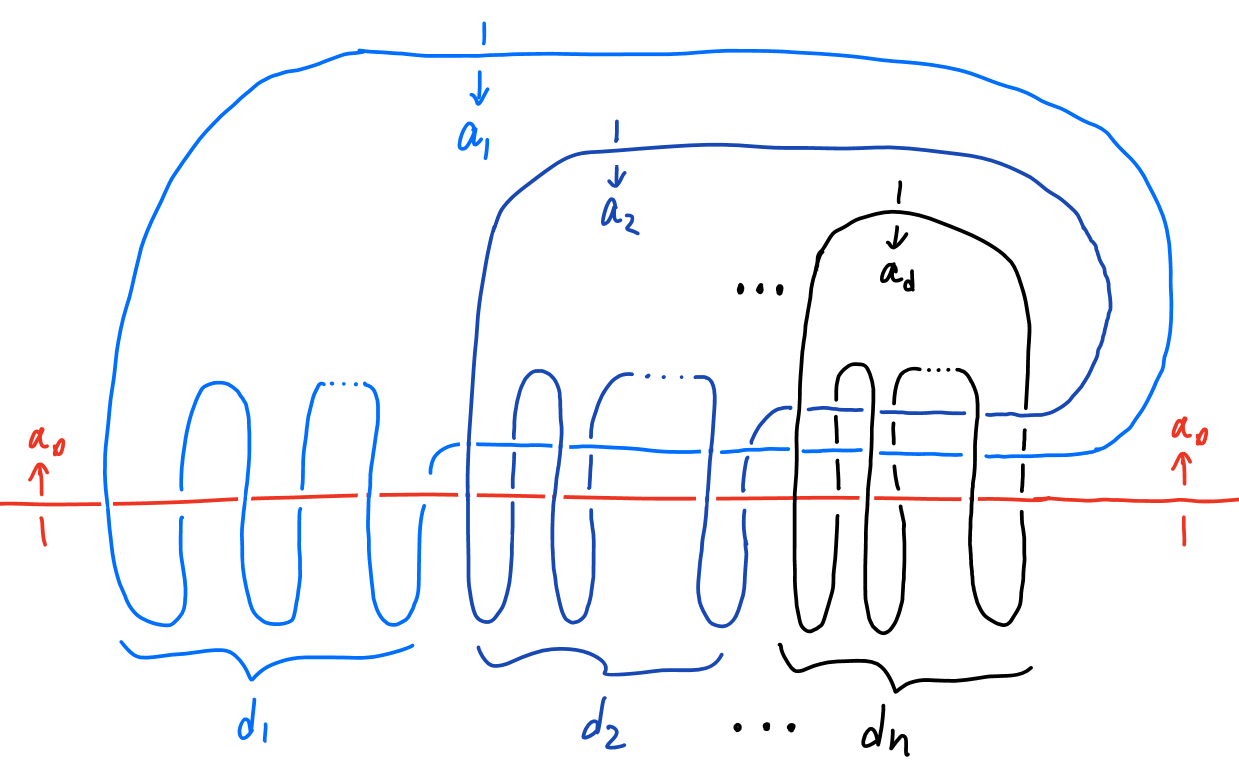}
\caption{Illustration of the link of type $\mathfrak{L}^{(n)}_{\{d_k\}}$ and generator assignment}
\label{link}
\end{figure}

\begin{defi}[Simple nested torus link]\label{simplenested}
Consider a $n+1$ component link in $S^3$ and denote its component as $\mathcal C_k$, indexed by $k=0,1,\cdots,n$. We call this a link of of type $\mathfrak{L}^{(n)}_{\{d_k\}}:=\mathfrak{L}_{d_1,\cdots,d_n}$, a simple nested  if $\mathcal C_0$ is an unknot and $\mathcal C_i$ forms a $(2,2d_{i})$-torus link with $\mathcal C_0$ and winds around all the components $\mathcal C_l$ of $l<i$, $d_{i}$-times. For instance, $\mathfrak{L}^{(1)}_{d}$ is a $(2,2d)$-torus link. The link $L^{(n)}_{\{d_k\}}$ is illustrated in figure~\ref{link} (with $\mathcal C_0$ sterographically projected as an infinite line in $\bbR^3$).
\end{defi}

The fundamental group $\pi_{1}(S^{3}\setminus \mathfrak{L}^{(n)}_{\{d_k\}})$ can be computed using Wirtinger presentation. The generators of $\pi_{1}(S^{3}\setminus \mathfrak{L}^{(n)}_{\{d_k\}})$ are given by loops with a base point above the plane where the link diagram is drawn and winds around a component of the link from behind, as illustrated by small arrows labelled by $a_i$ in figure \ref{link}. Then, we assign a generator to each $a_{k}$, $k=0,\ldots,n$. 
Then relations between $a_{k}$'s can be determined using classical techniques \cite{stillwell1993classical}. The fundamental group of a nested torus link (actually, a more general case than the simple nested torus link from definition \ref{simplenested}) was derived in \cite{Argyres:2019kpy}, but here we provide a simpler form more suitable for our purposes: define the following elements in $\pi_{1}(S^{3}\setminus \mathfrak{L}^{(n)}_{\{d_k\}})$:  \begin{equation}
\begin{gathered}
\Pi_k=a_ka_{k-1}\cdots a_0,\quad \delta_k=d_k-d_{k+1},\quad d_0=d_{n+1}:=0,\qquad \text{for \ }k=0,\cdots,n.
\end{gathered}
\end{equation}
\begin{thm}\label{thmnested}
The link group $\pi_{1}(S^{3}\setminus \mathfrak{L}^{(n)}_{\{d_k\}})$ is generated by $n+1$ generators $a_k$ subjecting to the following $n+1$ relations
\begin{equation}
\begin{gathered}
\left(\Pi_n^{\delta_n}\Pi_{n-1}^{\delta_{n-1}}\cdots\Pi_{k}^{\delta_k}\right)a_k=a_k\left(\Pi_n^{\delta_n}\Pi_{n-1}^{\delta_{n-1}}\cdots\Pi_{k}^{\delta_k}\right),
\\
k=1,\cdots,n.
\end{gathered}\label{relation2}
\end{equation}
together with the equation for $a_0$
\begin{equation}
\left(\Pi_n^{\delta_n}\Pi_{n-1}^{\delta_{n-1}}\cdots\Pi_{1}^{\delta_1}\right)a_0=a_0\left(\Pi_n^{\delta_n}\Pi_{n-1}^{\delta_{n-1}}\cdots\Pi_{1}^{\delta_1}\right).
\end{equation}
Moreover, the element that corresponds to the longitude cycle along $\mathcal{C}_{0}$ i.e. the loop homotopic to $\mathcal{C}_{0}$ is given by
\begin{equation}
    \Pi_n^{\delta_n}\Pi_{n-1}^{\delta_{n-1}}\cdots\Pi_{1}^{\delta_1}\Pi_0^{\delta_0}. \label{eqn:linkC0}
\end{equation}
\end{thm}
\begin{proof}

Using Reidemeister moves, we computed the fundamental group in a straightforward manner with fixed projection and generators $a_i$ in the following inductive way: Start from the left side of the zeroth component $\mathcal C_0$ and the first component $\mathcal C_1$ with generators $a_0$ and $a_1$. Define the generators for the $i$-th segment of $\mathcal C_1$ and $\mathcal C_0$ as $e^{(1)}_i$ and $\tilde{e}^{(1)}_i$ respectively in Fig.\ref{pf1}. Note that $e_0^{(1)}=a_1$ and $\tilde{e}_0^{(1)}=a_0$. 
\begin{figure}[h]
\centering
\includegraphics[scale=0.5]{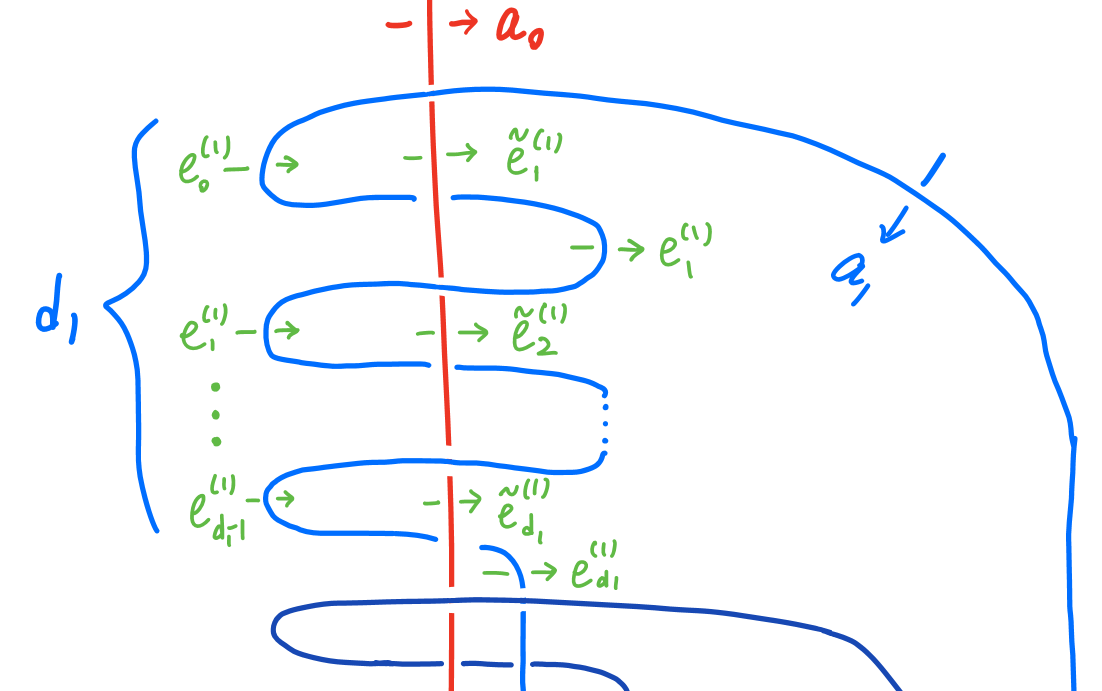}
\caption{Labelling the first component}
\label{pf1}
\end{figure}

By the Reidemeister move on the top intersection, we have relation
\begin{equation}
a_1a_0=\widetilde{e}_1^{(1)}e_0^{(1)}\quad \Rightarrow\quad \widetilde{e}_1^{(1)}=a_1a_0a_1^{-1}
\end{equation}
Similarly, for the $i$-th intersection of segments, we have relations
\begin{equation}
\begin{gathered}
\widetilde{e}^{(1)}_i\circ e_{i-1}^{(1)}=e_{i-1}^{(1)}\circ \widetilde{e}^{(1)}_{i-1}\quad\Rightarrow\quad \widetilde{e}^{(1)}_i=e_{i-1}^{(1)}\circ \widetilde{e}^{(1)}_{i-1}\circ(e_{i-1}^{(1)})^{-1}
\\
e_i^{(1)}\circ \widetilde{e}_i^{(1)}=\widetilde{e}_i^{(1)}\circ e_{i-1}^{(1)}\quad\Rightarrow\quad e_i^{(1)} = \widetilde{e}_i^{(1)}\circ e_{i-1}^{(1)}\circ (\widetilde{e}_i^{(1)})^{-1}
\end{gathered}
\end{equation}
which by induction gives generators for the $i$-th segment as
\begin{equation}
\begin{gathered}
e_i^{(1)}=\left(e_0^{(1)}\widetilde{e}_0^{(1)}\right)^ie_0^{(1)}\left(e_0^{(1)}\widetilde{e}_0^{(1)}\right)^{-i}=(a_1a_0)^ia_1(a_1a_0)^{-i},
\\
\widetilde{e}_i^{(1)}=
\left(e_0^{(1)}\widetilde{e}_0^{(1)}\right)^{i-1}\left({{e}}_0^{(1)} \widetilde{e}_0^{(1)}({{e}^{(1)}_0})^{-1}\right)
\left(e_0^{(1)}\widetilde{e}_0^{(1)}\right)^{-(i-1)}=(a_1a_0)^{i-1}(a_1a_0a_1^{-1})(a_1a_0)^{-(i-1)},
\\
i=1,\cdots,d_1.
\end{gathered}\label{rel}
\end{equation}

If $\mathfrak{L}^{(n)}_{\{d_i\}}=\mathfrak{L}^{(1)}_{d}$ has only one discriminant, $e_{d}^{(1)}$ will be identified to $a_1$ at the botton and give the relation:
\begin{equation}
(a_1a_0)^{d}a_1(a_1a_0)^{-d}=a_1\quad \Rightarrow\quad (a_0a_1)^{d}=(a_1a_0)^{d}.
\end{equation}
And a path homotopic to $\calC_0$ from deforming $\calC_0$ to the left is given by connecting the path of arcs from the left hand side:
\begin{eqnarray}
    e^{(1)}_{d-1}\circ\cdots\circ e_1^{(1)}\circ e^{(1)}_{0}&=&(a_1a_0)^{d-1}a_1 (a_1a_0)^{-(d-1)}\cdots (a_1a_0)a_1(a_1a_0)^{-1}a_1
    \nonumber\\
    &=&(a_1a_0)^{d-1}\big(a_1(a_1a_0)^{-1}\big)^{d-1}a_1
    \nonumber\\
    &=&(a_1a_0)^{d-1}a_1(a_0)^{-(d-1)}
    \nonumber\\
    &=&(a_1a_0)^d(a_0)^{-d}.
\end{eqnarray}

This analysis can be generalized to the component $\mathcal C_k$ without burden. For instance, assign the generators of $\mathcal C_2$ as in Fig.\ref{pf2}. Note that $\widetilde{e}_0^{(2)}$ winds around both segments of $\mathcal C_0$ and $\mathcal C_1$.
\begin{figure}[h]
\centering
\includegraphics[scale=0.5]{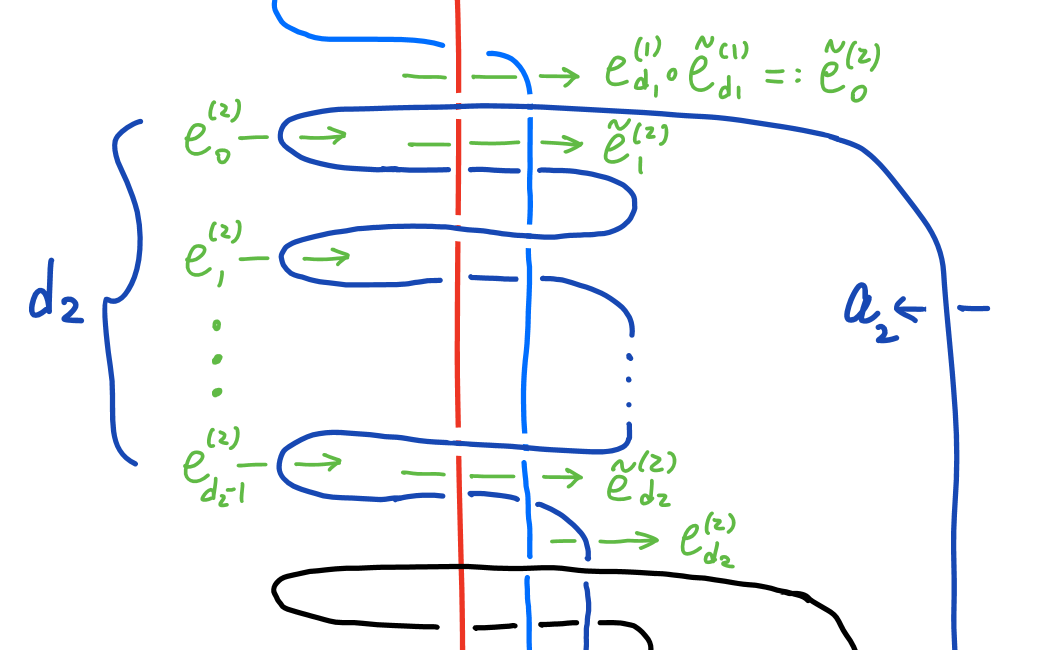}
\caption{Labelling the second component}
\label{pf2}
\end{figure}
Then the induction will be the same as in the first component, by requiring any loop crossing the central segment to wind around both segments from $\mathcal C_0$ and $\mathcal C_1$. Thus \eqref{rel} is a straightforward formula to be generalized to the $k$-th component:
\begin{eqnarray}
e_i^{(k)}&=&\left(e_0^{(k)}\widetilde{e}_0^{(k)}\right)^ie_0^{(k)}\left(e_0^{(k)}\widetilde{e}_0^{(k)}\right)^{-i}
\nonumber\\
&=&\left(a^{}_k\widetilde{e}_0^{(k)}\right)^ia^{}_k\left(a^{}_k\widetilde{e}_0^{(k)}\right)^{-i},
\\
\widetilde{e}_i^{(k)}&=&\left(e_0^{(k)}\widetilde{e}_0^{(k)}\right)^{i-1}
\left({{e}}_0^{(k)} \widetilde{e}_0^{(k)}({{e}^{(k)}_0})^{-1}\right)\left(e_0^{(k)}\widetilde{e}_0^{(k)}\right)^{-(i-1)}
\nonumber\\
&=&\left(a^{}_k\widetilde{e}_0^{(k)}\right)^{i-1}\left(a^{}_k\widetilde{e}_0^{(k)}a_k^{-1}\right)\left(a^{}_k\widetilde{e}_0^{(k)}\right)^{-(i-1)},
\\
\widetilde{e}_0^{(k+1)}&=&e_{d_{k}}^{(k)}\widetilde{e}_{d_{k}}^{(k)},\qquad i=1,\cdots,d_k,\qquad k=1,\cdots,n.
\end{eqnarray}
In fact, $\widetilde{e}_0^{(k+1)}$ is noting but the underneath loop from outside to the interior of $\mathcal C_k$, induced from all components $\mathcal C_{k-1},\cdots\mathcal C_1$. That is, one has inductively a simple result as
\begin{eqnarray}
\widetilde{e}_0^{(k+1)}&=&\left(e_0^{(k)}\widetilde{e}_0^{(k)}\right)^{d_k}e_0^{(k)}\left(e_0^{(k)}\widetilde{e}_0^{(k)}\right)^{-d_k}\left(e_0^{(k)}\widetilde{e}_0^{(k)}\right)^{d_k-1}
\left({{e}}_0^{(k)} \widetilde{e}_0^{(k)}({{e}^{(k)}_0})^{-1}\right)\left(e_0^{(k)}\widetilde{e}_0^{(k)}\right)^{-(d_k-1)}
\nonumber\\
&=&e_0^{(k)}\widetilde{e}_0^{(k)}=a_ka_{k-1}\cdots a_0.
\end{eqnarray}
Denote this product by notation
\begin{equation}
\Pi_k:=a_k\cdots a_0.
\end{equation}

Proceedingly, we can analyzing the $k$ vertical segments labelled by $[i]^{(k)}_a$ from $\mathcal C_a=\mathcal C_0,\mathcal C_1,\cdots,\mathcal C_{k-1}$ before the $i$-th segment $e_i^{(k)}$ in $\mathcal C_k$, as illustrated in Fig.\ref{pf3}. Note that 
\begin{equation}
[0]_a^{(k+1)}=\left\lbrace\begin{array}{cl}
[d_{k}]^{(k)}_a & a=1,\cdots,k-1
\\
e_{d_{k}}^{(k)} & a=k
\end{array}\right.,\quad k=1,\cdots,n
\end{equation}

\begin{figure}[h]
\centering
\includegraphics[scale=0.5]{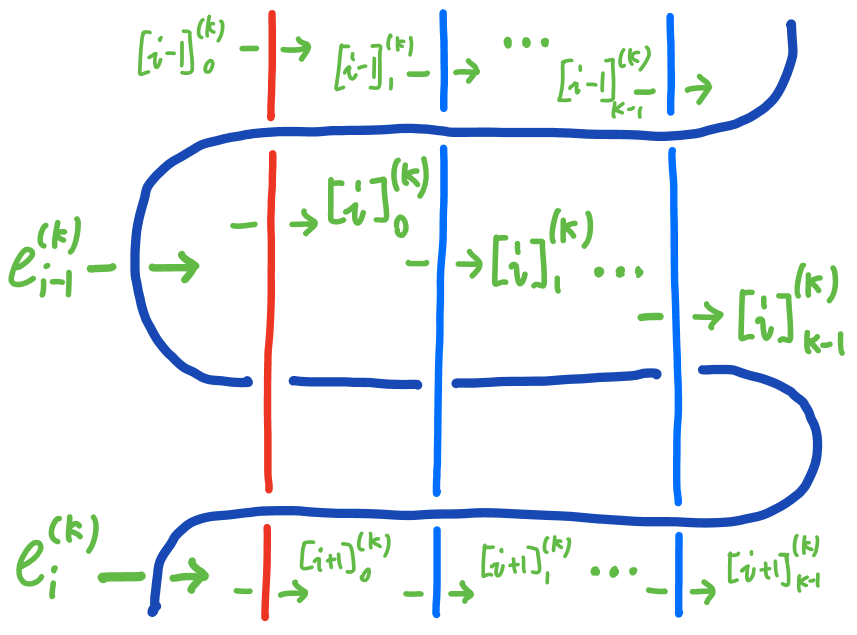}
\caption{Labelling the $k$ segments between $e_{i-1}^{(k)}$ and $e_{i}^{(k)}$ in $\mathcal C_k$.}
\label{pf3}
\end{figure}

The induction rule is simple:
\begin{equation}
\begin{gathered}
[i]_a^{(k)}
\\
=e_{i-1}^{(k)}[i-1]_a^{(k)}(e_{i-1}^{(k)})^{-1}
\\
\vdots
\\
=\left(e_{i-1}^{(k)}e_{i-2}^{(k)}\cdots e_{0}^{(k)}\right)[0]^{(k)}_a\left(e_{i-1}^{(k)}e_{i-2}^{(k)}\cdots e_{0}^{(k)}\right)^{-1}
\end{gathered}
\end{equation}
such that
\begin{eqnarray}
[i]_{k-1}^{(k)}\cdots[i]_{0}^{(k)}&=&\left(e_{i-1}^{(k)}e_{i-2}^{(k)}\cdots e_{0}^{(k)}\right)[0]_{k-1}^{(k)}\cdots[0]_0^{(k)}\left(e_{i-1}^{(k)}e_{i-2}^{(k)}\cdots e_{0}^{(k)}\right)^{-1}
\nonumber\\
&=&\left(e_{i-1}^{(k)}e_{i-2}^{(k)}\cdots e_{0}^{(k)}\right)\widetilde{e}_0^{(k)}\left(e_{i-1}^{(k)}e_{i-2}^{(k)}\cdots e_{0}^{(k)}\right)^{-1}
\nonumber\\
&=&\widetilde{e}_i^{(k)}
\end{eqnarray}
as expected. Define the notation for the connected path of arcs on the left hand side as
\begin{equation}
\Sigma^{(k)}:=\left(e_{d_k-1}^{(k)}e_{d_k-2}^{(k)}\cdots e_{0}^{(k)}\right),
\end{equation}
which can be simplified as
\begin{eqnarray}
\Sigma^{(k)}&=&e_{d_k-1}^{(k)}e_{d_k-2}^{(k)}\cdots e_{0}^{(k)}
\nonumber\\
&=&(e_0^{(k)}\widetilde{e}_0^{(k)})^{d_k-1}e_0^{(k)}(e_0^{(k)}\widetilde{e}_0^{(k)})^{1-d_k}\cdot (e_0^{(k)}\widetilde{e}_0^{(k)})^{d_k-2}e_0^{(k)}(e_0^{(k)}\widetilde{e}_0^{(k)})^{2-d_k}
\nonumber\\&&\cdots (e_0^{(k)}\widetilde{e}_0^{(k)})^{1}e_0^{(k)}(e_0^{(k)}\widetilde{e}_0^{(k)})^{-1}e_0^{(k)}
\nonumber\\
&=&(e_0^{(k)}\widetilde{e}_0^{(k)})^{d_k-1}\left( e_0^{(k)}(e_0^{(k)}\widetilde{e}_0^{(k)})^{-1} \right)^{d_k-1}e_0^{(k)}
\nonumber\\
&=&(e_0^{(k)}\widetilde{e}_0^{(k)})^{d_k}(\widetilde{e}_0^{(k)})^{-d_k}
\nonumber\\
&=&\Pi_k^{d_k}\Pi_{k-1}^{-d_k}.
\end{eqnarray}
The $j$-th segment $[d_{k}]_j^{(k)}$ at the bottom of $\mathcal C_k$ is inductively given by that from $\mathcal C_{k-1}$, $\mathcal C_{k-2}$, etc. as
\begin{eqnarray}
[d_k]_j^{(k)}&=&\Sigma^{(k)}\ [0]_j^{(k)}( \Sigma^{(k)} )^{-1}
\nonumber\\
&=&\left( \Sigma^{(k)}\cdots\Sigma^{{(j+2)}}\Sigma^{(j+1)} \right) e_{d_j}^{(j)}\left( \Sigma^{(k)}\cdots\Sigma^{{(j+2)}}\Sigma^{(j+1)} \right)^{-1}
\nonumber\\
&=&\left( \Sigma^{(k)}\cdots\Sigma^{{(j+2)}}\Sigma^{(j+1)} \right) (e_0^{(j)}\widetilde{e}_0^{(j)})^{d_j}e_0^{(j)}(e_0^{(j)}\widetilde{e}_0^{(j)})^{-d_j}\left( \Sigma^{(k)}\cdots\Sigma^{{(j+2)}}\Sigma^{(j+1)} \right)^{-1}
\nonumber\\
&=&\bigg(\left( \Pi_k^{d_k}\Pi_{k-1}^{-d_k}\Pi_{k-1}^{d_{k-1}}\Pi_{k-2}^{-d_{k-2}}\cdots\Pi_{j+1}^{d_{j+1}}\Pi_{j}^{-d_{j+1}} \right)\Pi_j^{d_j}\bigg) a_j(\cdots)^{-1}
\nonumber\\
&=&\left(\Pi_k^{d_k}\Pi_{k-1}^{(d_{k-1}-d_k)}\cdots\Pi_{j}^{(d_j-d_{j+1})}\right)a_j(\cdots)^{-1}
\end{eqnarray}
which is identified to $e_0^{(j)}$ at the bottom of the final component $\mathcal C_n$ and gives $n$ relations in generators of fundamental group:
\begin{eqnarray}
a_j&=&e_{0}^{(j)}=[d_n]_j^{(n)}
\nonumber\\
&=&\left(\Pi_n^{d_n}\Pi_{n-1}^{(d_{n-1}-d_n)}\cdots\Pi_{j}^{(d_j-d_{j+1})}\right)a_j(\cdots)^{-1}
\nonumber\\
\Leftrightarrow\quad \left(\Pi_n^{d_n}\Pi_{n-1}^{(d_{n-1}-d_n)}\cdots\Pi_{j}^{(d_j-d_{j+1})}\right)a_j&=&a_j\left(\Pi_n^{d_n}\Pi_{n-1}^{(d_{n-1}-d_n)}\cdots\Pi_{j}^{(d_j-d_{j+1})}\right),
\nonumber\\
j&=&0,\cdots, n-1.
\end{eqnarray}
together with the relation from component $\mathcal C_n$, which can also be regarded as the case $j=n$ that has no $\Sigma$ adjoint:
\begin{equation}
e_0^{(n)}=e^{(n)}_{d_n}=(e_0^{(n)}\widetilde{e}_0^{(n)})^{d_n}e_0^{(n)}(e_0^{(n)}\widetilde{e}_0^{(n)})^{-d_n}=\Pi_n^{d_n}e_0^{(n)}\Pi_n^{-d_n}.
\end{equation} 
Notice that the $j=0$ case whose induction on $\Sigma$ ends on $\widetilde{e}_{d_1}^{(1)}$ gives the same equation as $j=1$:
\begin{equation}
\begin{gathered}
a_0=e_0^{(0)}=\left( \Sigma^{(k)}\cdots\Sigma^{{(j+2)}}\Sigma^{(2)} \right) 
\widetilde{e}_{d_1}^{(1)}
\left( \Sigma^{(k)}\cdots\Sigma^{{(j+2)}}\Sigma^{(2)} \right)^{-1}
\\
=\left( \Sigma^{(k)}\cdots\Sigma^{{(j+2)}}\Sigma^{(2)} \right) 
(e_0^{(1)}\widetilde{e}_0^{(1)})^{d_1-1}(e_0^{(1)}\widetilde{e}_0^{(1)}(e_{0}^{(1)})^{-1})(e_0^{(1)}\widetilde{e}_0^{(1)})^{1-d_1}
\left( \Sigma^{(k)}\cdots\Sigma^{{(j+2)}}\Sigma^{(2)} \right)^{-1}
\\
=\left(\Pi_n^{d_n}\Pi_{n-1}^{(d_{n-1}-d_n)}\cdots\Pi_{1}^{(d_1-d_{2})}\right)a_0\left(\Pi_n^{d_n}\Pi_{n-1}^{(d_{n-1}-d_n)}\cdots\Pi_{1}^{(d_1-d_{2})}\right)^{-1}.
\end{gathered}
\end{equation}

Finally, the central component that is homotopic to $\calC_0$ is simply given by the product of each arc on the left hand side:
\begin{eqnarray}
    \Sigma^{(n)}\circ\Sigma^{(n-1)}\circ\cdots \circ\Sigma^{(1)} &=& \Pi_n^{d_n}\Pi_{n-1}^{-d_{n}}\Pi_{n-1}^{d_{n-1}}\Pi_{n-2}^{-d_{n-1}}\cdots \Pi_1^{d_1}\Pi_0^{-d_1}
    \nonumber\\
    &=& \Pi_n^{d_n}\Pi_{n-1}^{d_{n-1}-d_n}\cdots\Pi_1^{d_1-d_2}\Pi_0^{-d_1}.
\end{eqnarray}

\end{proof}

The following examples of relations, from theorem \ref{thmnested}, are relevant to several of our models:
\begin{itemize}
\item  $\mathfrak{L}^{(1)}_{n}$:
\begin{equation}
(a_1a_0)^n=(a_0a_1)^n \label{eqn:L1n}
\end{equation}
Central component:
\begin{equation}
    (a_1a_0)^n(a_0)^{-n}.
\end{equation}

\item $\mathfrak{L}^{(2)}_{4,2}$:
\begin{equation}
(a_2a_1a_0)^2(a_1a_0)^2a_{0,1}=a_{0,1}(a_2a_1a_0)^2(a_1a_0)^2 \label{eqn:L24_1}
\end{equation}
\begin{equation}
(a_2a_1a_0)^2a_2=a_2(a_2a_1a_0)^2 \label{eqn:L24_2}
\end{equation}
Central component:
\begin{equation}
    (a_2a_1a_0)^2(a_1a_0)^2(a_0)^{-4}.
\end{equation}

\item $\mathfrak{L}^{(2)}_{3,3}$:
\begin{equation}
(a_2a_1a_0)^3a_i=a_{i}(a_2a_1a_0)^3,\quad i=0,1,2 \label{eqn:GOF}
\end{equation}
Central component:
\begin{equation}
    (a_2a_1a_0)^3(a_0)^{-3}.
\end{equation}

\end{itemize}

We remark that the two link groups $L_5^{(1)}$ and $L_{4,2}^{(2)}$ have already appear in \cite{Lin:2024fpz}, but with a different presentation.

Finally, we present a conjecture for computing the class of the loop presented on sect. \ref{sec:wincatMon} in terms of loops surrounding components of $\Delta$ for the case $\mathrm{dim}(\mathcal{M}_{K})\geq 3$, in certain models. Denote the coordinates of $\mathcal{M}_{K}$ by $z_{\alpha}=\exp(-t_{\alpha})$, $\alpha=0,\ldots,r$. Therefore the geometric phase $X_{\zeta_{+}}$ is at a neighborhood of the origin of $z$ coordinates, and our base point is located there, where define the two-dimensional subspaces: 
\begin{equation}
\widehat{D}_{\alpha,\beta}=\{z_{\gamma}=0:\gamma\neq\alpha,\gamma\neq\beta\},
\end{equation}
so for instance, the loop we are computing the monodromy around can be fully contained inside $\{z_{1}=0\}\cap\widehat{D}_{0,1}$. Then, consider in particular 
\begin{equation}
\widehat{D}_{0,1} ,
\end{equation}
and assume $\Delta\cap\widehat{D}_{0,1}$ is irreducible\footnote{This, assumption fails to be true in most nonabelian models, for example.}, then an analogous analysis than the one in sect. \ref{sec:comment}, restricted to $\widehat{D}_{0,1}$ shows that the relevant link\footnote{We mean that $S^{3}\cap \Delta$, where $S^{3}$ is centered at $\Delta\cap\{z_1=0\}\cap \widehat{D}_{0,1}$, is a link of type $\mathfrak{L}^{(1)}_{n^{(1)}}$ and the monodromy loop runs parallel to $\{z_1=0\}$.} is $\mathfrak{L}^{(1)}_{n^{(1)}}$, with $n^{(1)}:=\sum_{I=1}^{n+1}n^{(1)}_{I}$. By theorem \ref{thmnested}, the monodromy $M_{0}$ is given by
\begin{equation}
    M_0=(M_1L_1)^{n^{(1)}}L_1^{-n^{(1)}}.
\end{equation}
where $L_{\alpha}$ is defined in \eqref{LKY} and $M_{1}$ is the monodromy around the component of $\Delta$. So, next we need to compute $M_{1}$, for this purpose we consider the two-dimensional subspace
\begin{equation}
\widehat{D}_{1,2},
\end{equation}
then, $M_{1}$ is represented by a loop running parallel to $\{z_{2}=0\}$. Again, assuming irreducibility of $\Delta\cap\widehat{D}_{1,2}$, a similar analysis than sect. \ref{sec:comment} shows that the relevant link is $\mathfrak{L}^{(1)}_{n^{(2)}}$ and therefore 
\begin{equation}
    M_1=(M_2L_2)^{n^{(2)}}L_2^{-n^{(2)}}.
\end{equation}
where $M_{2}$ represents a loop surrounding the component of $\Delta$. We repeat this computation, always keeping the assumption $\Delta\cap \widehat{D}_{\alpha-1,\alpha}$ until we reach the subspace $\widehat{D}_{r-1,r}$ and we propose that
$M_{r}=T_{\mathcal{O}_{X_{\zeta_{+}}}}$. This then gives the following recursive expression for $M_{0}$:
\begin{equation}\label{iteratform}
    M_r:=T_{\mathcal{O}_{X_{\zeta_{+}}}},\qquad M_\alpha=(M_{\alpha+1}L_{\alpha+1})^{n^{(\alpha+1)}}L_{\alpha+1}^{-n^{(\alpha+1)}},\qquad \alpha=0,\cdots,r-1.
\end{equation}
In particular, at each step of iteration, the generators $a_0=L_{\alpha}$ and $a_1=M_\alpha$ of the $\mathfrak{L}_{n^{(\alpha)}}^{(1)}$ link also satisfy the link relation in \eqref{eqn:L1n}:
\begin{equation}
    (M_\alpha L_\alpha)^{n^{(\alpha)}}=(L_\alpha M_\alpha)^{n^{(\alpha)}},\qquad \alpha=1,\cdots,r.\label{iteratlink}
\end{equation}
Then, we propose
\begin{conj}\label{conjecturenested}
The form \eqref{iteratform} and \eqref{iteratlink} of the monodromy $M_{0}$, hold for every $X_{\zeta_{+}}$ constructed as in sect. \ref{sec:flopsGLSM}, whenever $\Delta\cap \widehat{D}_{\alpha-1,\alpha}$ is irreducble for all $\alpha$. Moreover we conjecture that  \eqref{iteratform} and \eqref{iteratlink} are independent of the labeling of $z_{1},\ldots,z_{r}$.
\end{conj}

\section{Topological numbers of $X_{\zeta_{+}}$ and $X^{\natural}$}\label{sec:appcharclass}

In this subsection we collect some useful formulae for the computations of A-periods of sheaves over $X_{\zeta_{+}}$ and $X^{\natural}$, in particular we compute explicit expressions for their topological numbers. These invariants will enter into the computation of monodromy actions over the Doran-Morgan basis of A-periods.\\

Denote by $L$ the hyperplane class of $\bbP^n$ and 
\begin{equation}
H_\alpha:=c_1(\calS^\vee_\alpha),\qquad \alpha=1,\ldots,r.
\end{equation}
In term of the chern roots of $\calS^\vee_\alpha$:
\begin{equation}
    c_{1}(\calS^\vee_\alpha)=H_{\alpha}=\sum_{i=1}^{k_{\alpha}}H_{\alpha}^{(i)}.
\end{equation}
we will also define for convenience: 
\begin{equation}
     \widetilde{H}_\alpha:=c_{k_\alpha}(\calS^\vee_\alpha)=H_\alpha^{(1)}\cdots H_\alpha^{(k_\alpha)}.
\end{equation}
Thus, the total Chern class of $\calN$, defined by the line bundles in \eqref{eqn:nefp}, is written in terms of $\vec n_I(H)=c_1(\calL_I)$ as 
\begin{equation}
    c(\calN)=\prod_{I=1}^{n+1}(1+\vec n_I(H)),\quad \vec n_I(H)=n_I^{(1)}H_1+\cdots+ n_I^{(r)}H_r.
\end{equation}
And one has
\begin{equation}
    c(-K_Y)=1+c_1(Y)=1+\sum_{I=1}^{n+1}\vec n_I(H)=:1+\vec n(H).
\end{equation}
With these results at hand we can then compute the Chern classes of $X_{\zeta_{+}}$ by expanding its total Chern class:
\begin{eqnarray}
    c(X)&=&\frac{c(Y)c(\bbP^n)}{c(\calN\boxtimes\calO_{\bbP^n}(1))} \nonumber\\
    &=&\frac{c(Y)(1+L)^{n+1}}{\prod_{I=1}^{n+1}(1+L+\vec n_I(H))}\nonumber\\
    &=&\bigg( 1+c_1(Y)+c_2(Y)+c_3(Y)\bigg) \times\bigg(1+c_1(\bbP^n)+c_2(\bbP^n)+c_3(\bbP^n)\bigg) \nonumber
    \\\nonumber
    &\times&\bigg(1-A_1(L+\vec n)+A_1(L+\vec n)^2-A_2(L+\vec n)
    \\\nonumber
    &&-A_1(L+\vec n)^3+2A_1(L+\vec n)A_2(L+\vec n)-A_3(L+\vec n)\bigg)
    \\\nonumber
    &=&c(X^\natural)-A_2(\vec n)+A_1(\vec n)A_2(\vec n)-A_3(\vec n)
    \\
    &+&L\bigg( A_1(\vec n)+2A_2(\vec n)-A_1(\vec n)^2 \bigg)-L^2 A_1(\vec n).
\end{eqnarray}
where the total Chern class of the hypersurface $X^{\natural}$ is given by:
\begin{equation}
    c(X^\natural)=\frac{c(Y)}{c(-K_Y)}=1+0+c_2(Y)+c_3(Y)-c_1(Y)c_2(Y),
\end{equation}
and we defined the characteristic class
\begin{equation}
     A_k(\vec n):=\sum_{1\leq I_1<\cdots<I_k\leq n+1} \vec n_{I_1}\cdots \vec n_{I_k}
\end{equation}
which satisfies the following expansion (in where $(L+\vec n_I):=(L+\vec n_1,\cdots,L+\vec n_{n+1})$ for short)
\begin{equation}
    A_{k}(L+\vec n)=\sum_{j=0}^k \begin{pmatrix}
        n+1-j\\k-j
    \end{pmatrix} L^{k-j}A_j(\vec n).
\end{equation}
In order to get explicit expressions for the topological numbers, is useful to use the adjunction formula for the following Euler classes of normal bundles:
\begin{equation}
   e(N_{X^{\natural}/Y})=\vec{n}(H),\qquad e(N_{X_{\zeta_{+}}/Y})=\prod_{I=1}^{n+1}(L+\vec{n}_{I}(H))
\end{equation}
Then, for any class $\delta\in H^{*}(X_{\zeta_{+}})$ that we can expand as
\begin{equation}
    \delta=\sum_{k=0}^{3}L^{k}\delta_{k}(H)
\end{equation}
we have
\begin{eqnarray}\label{plusnatform}
    \int_X\delta&=&\int_{\bbP^n\times Y}\delta \prod_{I=1}^{n+1}\left( L+\vec n_I (H)\right) \nonumber
    \\
    &=&\int_{\bbP^n\times Y}\delta\left( L^n A_1(\vec n)+L^{n-1}A_2(\vec n)+\cdots\right) \nonumber
    \\
        &=&\int_Y\sum_{k=1}^{4} \delta_{k-1} A_k(\vec n).
\end{eqnarray}
Then, from \eqref{plusnatform} is clear that
\begin{equation}
    \int_Y \delta_0 A_1(\vec n)=\int_{X^\natural} \delta_0
\end{equation}
thus the topological numbers that do not involve the restriction from the $\mathbb{P}^{n}$ class $L$ are the same for $X_{\zeta_{+}}$ and $X^{\natural}$:
\begin{equation}
    c_{\alpha\beta\gamma}(X_{\zeta_{+}})=c_{\alpha\beta\gamma}(X^\natural),\quad c_\alpha(X_{\zeta_{+}})=c_\alpha(X^\natural),\qquad\alpha,\beta,\gamma=1,\ldots, r
\end{equation}
while the rest of the topological numbers for $X_{\zeta_{+}}$ are given by
\begin{eqnarray}
    c_{000}(X_{\zeta_{+}})&=&\int_Y A_4(\vec n) \label{eqn:c000}
    \\
    c_{00\alpha}(X_{\zeta_{+}})&=&\int_YA_3(\vec n)H_\alpha \label{eqn:c00a}
    \\
    c_{0\alpha\beta}(X_{\zeta_{+}})&=&\int_Y A_2(\vec n)H_\alpha H_\beta \label{eqn:c0ab}
    \\
    24c_0(X_{\zeta_{+}})&=&\int_Y c_2(Y)A_2(\vec n)-N\\
    \chi(X_{\zeta_{+}})&=&\int_{X^\natural} c_3(X^\natural) +\int_XA_1A_2-A_3+2LA_2-LA_1^2-L^2A_1 \nonumber
    \\
    &=&\chi(X^\natural)+\int_Y( A_1^2A_2-A_1A_3+2A_2^2-A_1^2A_2-A_1A_3 )\nonumber
    \\
    &=&\chi(X^\natural)+2N.
\end{eqnarray}

\section{Computation of "Five Guys\textsuperscript{{\textregistered}}"}\label{sec:appcomputation}

Here we list the matrices for the five families of the splitting configurations of CICY3 hypersurfaces \eqref{eqn:Xnatural}. We will show explicitly that, in the cases $\dim\mathcal{M}_{K}(X_{\zeta_{+}})\geq 3$, conjecture \ref{conjecturenested} holds.

The monodromy matrices we compute in this appendix are very conveniently expressed in terms of topological invariants of $X^{\natural}$. Therefore, to make our results easier to read, in this appendix we will reshuffle the vector $\Pi_{+}$, w.r.t \eqref{eqgeomplus}, to 
\begin{equation}
   \Pi_{+}\rightarrow (Z_{{D_0}}, Z_{\widetilde{C_0}}\vert Z_{\calO_X},Z_{{D_\alpha}},Z_{\widetilde{C_\alpha}},Z_{P}),
\end{equation}
therefore the matrices that we need, under this change of basis, become
\begin{eqnarray}\label{matshuffleb}
    L_\mu&=&\begin{pmatrix}
        1 & & & & & 
        \\
        0 & 1 & & & &
        \\
        \delta_{\mu\alpha} & 0 & \delta_{\alpha\beta}
        \\
        c_{\mu\mu0} & c_{\mu00} & c_{\mu0\beta} & 1
        \\
        c_{\mu\mu\alpha} & c_{\mu0\alpha} & c_{\mu\alpha\beta} & 0 & \ \ \delta_{\alpha\beta}
        \\
        0 & b_{\mu0} & b_{\mu\beta} & 0 & \ \ \delta_{\mu\beta} & \ \ 1
    \end{pmatrix}
    \rightarrow
    \left(\begin{array}{cc|cccc}
        1 &   & 0 & 0 & 0 & 0
        \\
        c_{\mu00} & 1 & c_{\mu\mu0} & c_{\mu0\beta} & 0 & 0
        \\\hline
        0 & 0 & 1 & & &
        \\
        0 & 0 & \delta_{\mu\alpha} & \delta_{\alpha\beta} 
        \\
         c_{\mu0\alpha} & 0 & c_{\mu\mu\alpha} & c_{\mu\alpha\beta} & \delta_{\alpha\beta}
        \\
         b_{\mu0} & 0 & 0 & b_{\mu\beta} & \delta_{\mu\beta} &\  1
    \end{array}\right),
    \nonumber\\
    \\
     T_{\calO_X}&=&\begin{pmatrix}
        1 & -l_0 & -l_\beta & 0 & 0 & -1
        \\
        & 1 &
        \\ 
        &&\delta_{\alpha\beta}
        \\
        &&&1
        \\
        &&&&\  \delta_{\alpha\beta}
        \\
        &&&&&1
    \end{pmatrix}
    \rightarrow
    \left(\begin{array}{cc|cccc}
        1 &  
        \\
         & 1
        \\\hline
        -l_0 & 0 & 1 & -l_\beta & 0 & -1
        \\
        &&&\ \ \delta_{\alpha\beta}
        \\
        &&&&\delta_{\alpha\beta}
        \\
        &&&&&1
    \end{array}\right),
    \\
    T_N&\rightarrow&\left(\begin{array}{cc|cccc}
         \ \ 1 & &
         \\
         -N & \ 1
         \\\hline
         &&1 
         \\
         &&&\ \delta_{\alpha\beta}
         \\
         &&&&\delta_{\alpha\beta}
         \\
         &&&&&1
    \end{array}\right).
\end{eqnarray}

\subsection{$X^\natural(5)$}

In this class of examples we have the following:
\begin{prop}
For any nontrivial partition of $5$: $n_1+\cdots+n_{n+1}=5$, we have a splitting configuration
\begin{equation}
    X=\left[\begin{array}{c|ccc}
         \bbP^n& 1 & \cdots & 1  
         \\
        \bbP^4 & n_1 & \cdots & n_{n+1}
    \end{array}\right].
\end{equation}
For all $X$ we can identify the monodromy $M_{0}$ with the fundamental group of $\mathfrak{L}_{5}^{(1)}$ in the following way:
\begin{eqnarray}
    &M_0=(T_{\calO_X}L_{1})^5L_{1}^{-5}=T_NL_{1}^{-5},\label{eqn:TPN5}
    \\
    &(T_{\calO_X}L_{1})^5=(L_{1}T_{\calO_X})^5.\label{eqn:link5}
\end{eqnarray}
\end{prop}
\begin{proof}

Note that $Y=\bbP^4$. The $T_N$ number can be evaluated as (in this case $A_k(\vec n(H))=A_k(\vec n)H^k$. $H:=H_{1}$)
\begin{eqnarray}
    N&=&\int_{Y}A_2(\vec n)^2-A_1(\vec n)A_3(\vec n)\equiv c_{011}^2-5c_{001}.
\end{eqnarray}
Then, the corresponding matrices ( the numbers $c_{011},c_{001}$ depend on $\vec n_I$) are given by
\begin{equation}
\begin{aligned}
L=\left(
\begin{array}{cccccc}
 1 & 0 & 0 & 0 & 0 & 0 \\
 c_{001} & 1 & c_{011} & c_{011} & 0 & 0 \\
 0 & 0 & 1 & 0 & 0 & 0 \\
 0 & 0 & 1 & 1 & 0 & 0 \\
 c_{011} & 0 & 5 & 5 & 1 & 0 \\
 b_{01} & 0 & 0 & 0 & 1 & 1 \\
\end{array}
\right),
\quad
T_{\calO_X}=\left(
\begin{array}{cccccc}
 1 & 0 & 0 & 0 & 0 & 0 \\
 0 & 1 & 0 & 0 & 0 & 0 \\
 -l_{0} & 0 & 1 & -5 & 0 & -1 \\
 0 & 0 & 0 & 1 & 0 & 0 \\
 0 & 0 & 0 & 0 & 1 & 0 \\
 0 & 0 & 0 & 0 & 0 & 1 \\
\end{array}
\right).
\end{aligned} 
\end{equation}
and direct computation gives \eqref{eqn:TPN5} and \eqref{eqn:link5}.
\end{proof}
The monodromy of the case $1+\ldots+1=5$ was analyzed in detail in \cite{Lin:2024fpz}.

\subsection{$X^\natural(3,3)$}
Consider the topological data of $X^\natural(3,3)$, the bicubic hypersurface in $\bbP^2\times\bbP^2$:
\begin{equation}
    c_{\alpha\beta\gamma}=\left\{\begin{array}{cl}3 & \alpha=\beta\neq\gamma\\ 0 & \text{otherwise}\end{array}\right.,\quad  c_\alpha=\frac{3}{2},\qquad \alpha,\beta,\gamma=1,2.
\end{equation}
Then, the submatrices, corresponding to $X^{\natural}$, in \eqref{matshuffleb} are thus
\begin{equation}
    L^\natural_1=\begin{pmatrix}
        1 & 0 & 0 & 0 & 0 & 0 \\
 1 & 1 & 0 & 0 & 0 & 0 \\
 0 & 0 & 1 & 0 & 0 & 0 \\
 0 & 0 & 3 & 1 & 0 & 0 \\
 3 & 3 & 3 & 0 & 1 & 0 \\
 0 & 0 & 0 & 1 & 0 & 1 \\
    \end{pmatrix},
    \quad
    L^\natural_2=\begin{pmatrix}
        1 & 0 & 0 & 0 & 0 & 0 \\
 0 & 1 & 0 & 0 & 0 & 0 \\
 1 & 0 & 1 & 0 & 0 & 0 \\
 3 & 3 & 3 & 1 & 0 & 0 \\
 0 & 3 & 0 & 0 & 1 & 0 \\
 0 & 0 & 0 & 0 & 1 & 1 \\
    \end{pmatrix},
    \quad 
    T^\natural_{\calO_X}=\begin{pmatrix}
         1 & -3 & -3 & 0 & 0 & -1 \\
 0 & 1 & 0 & 0 & 0 & 0 \\
 0 & 0 & 1 & 0 & 0 & 0 \\
 0 & 0 & 0 & 1 & 0 & 0 \\
 0 & 0 & 0 & 0 & 1 & 0 \\
 0 & 0 & 0 & 0 & 0 & 1 \\
    \end{pmatrix}.
\end{equation}

\begin{prop}\label{prop:33}
For any splitting configuration $X$, constructed from $X^{\natural}(3,3)$ the relevant links, in the sense of conjecture \ref{conjecturenested}, appearing in the subspaces $\widehat{D}_{0,1}$ and $\widehat{D}_{1,2}$, both corresponds to $\mathfrak{L}_{3}^{(1)}$. The monodromy matrix $M_{0}$ is given by:
\begin{eqnarray}
   & M_1=(T_{\calO_X}L_2)^3L_2^{-3},\quad M_0=(M_1L_1)^3L_1^{-3}=T_N(L_1L_2)^{-3},\label{eqn:TPN33}
 \end{eqnarray}
 where $M_{\alpha}$, $L_{\alpha}$ satisfy the $\mathfrak{L}^{(1)}_{3}$ relations:
\begin{eqnarray} 
    &(M_\alpha L_\alpha)^3=(L_\alpha M_\alpha)^3,\quad \alpha=1,2. \label{eqn:link33}
\end{eqnarray}
\end{prop}
\begin{proof}
A splitting configuration $X$ with $Y=\bbP^2\times\bbP^2$ is given by a multi-partition of $(3,3)$ as $\vec n=n_1^{(\alpha)}+\cdots+n_{n+1}^{(\alpha)}=(3,3) $, $\alpha=1,2$. By expanding $\vec n_I(H)=n_I^{(1)}H_1+n_I^{(2)}H_2$ and using $c_{0\alpha\beta}$, one has
\begin{equation}
    \int_{Y}A_2(\vec n(H))^2=(\sum_{I\neq J}n_I^{(1)}n_J^{(2)})^2+2(\sum_{I<J}n_I^{(1)}n_{J}^{(1)})(\sum_{K<L}n_K^{(2)}n_{L}^{(2)})=c_{012}^2+2c_{011}c_{022}
\end{equation}
and
\begin{equation}
    \int_{Y} A_1(\vec n(H))A_3(\vec n(H))=\int_{\vec\bbP}(3H_1+3H_2) A_3(\vec n)=3(c_{001}+c_{002}).
\end{equation}
Thus
\begin{equation}
    N\equiv {c_{012}}^2 + 2 c_{011} c_{022}-3(c_{001} +c_{002}).
\end{equation}
where the topological numbers depend on the choice of $\vec{n}$. By a direct computation to verify that eqns.~\eqref{eqn:TPN33} and \eqref{eqn:link33} are satisfied by the following matrices:
\begin{eqnarray}
L_1&=&\left(
\begin{array}{cccccccc}
 1 & 0 & 0 & 0 & 0 & 0 & 0 & 0 \\
 c_{001} & 1 & c_{011} & c_{011} & c_{012} &
   0 & 0 & 0 \\
 0 & 0 & 1 & 0 & 0 & 0 & 0 & 0 \\
 0 & 0 & 1 & 1 & 0 & 0 & 0 & 0 \\
 0 & 0 & 0 & 0 & 1 & 0 & 0 & 0 \\
 c_{011} & 0 & 0 & 0 & 3 & 1 & 0 & 0 \\
 c_{012} & 0 & 3 & 3 & 3 & 0 & 1 & 0 \\
 b_{10} & 0 & 0 & 0 & 0 & 1 & 0 & 1 \\
\end{array}
\right),
\quad
L_2=\left(
\begin{array}{cccccccc}
 1 & 0 & 0 & 0 & 0 & 0 & 0 & 0 \\
 c_{002} & 1 & c_{022} & c_{012} & c_{022} &
   0 & 0 & 0 \\
 0 & 0 & 1 & 0 & 0 & 0 & 0 & 0 \\
 0 & 0 & 0 & 1 & 0 & 0 & 0 & 0 \\
 0 & 0 & 1 & 0 & 1 & 0 & 0 & 0 \\
 c_{012} & 0 & 3 & 3 & 3 & 1 & 0 & 0 \\
 c_{022} & 0 & 0 & 3 & 0 & 0 & 1 & 0 \\
 b_{20} & 0 & 0 & 0 & 0 & 0 & 1 & 1 \\
\end{array}
\right)
\nonumber\\
T_{\calO_X}&=&\left(
\begin{array}{cccccccc}
 1 & 0 & 0 & 0 & 0 & 0 & 0 & 0 \\
 0 & 1 & 0 & 0 & 0 & 0 & 0 & 0 \\
 -l_0 & 0 & 1 & -3 & -3 & 0 & 0 & -1 \\
 0 & 0 & 0 & 1 & 0 & 0 & 0 & 0 \\
 0 & 0 & 0 & 0 & 1 & 0 & 0 & 0 \\
 0 & 0 & 0 & 0 & 0 & 1 & 0 & 0 \\
 0 & 0 & 0 & 0 & 0 & 0 & 1 & 0 \\
 0 & 0 & 0 & 0 & 0 & 0 & 0 & 1 \\
\end{array}
\right)
\end{eqnarray}
    
\end{proof}

\begin{rmk}
    The following splitting configuration of bi-cubic
    \begin{eqnarray}
        X=\left[\begin{array}{c|cc}
    \bbP^1 & 1 & 1
     \\
     \bbP^2& 3 & 0
     \\
     \bbP^2& 0 & 3
\end{array}\right]
    \end{eqnarray}
    corresponding to the multi-partition $(n^{(1)}_{1},n_{2}^{(1)})=(3,0)$, $(n^{(1)}_{1},n_{2}^{(1)})=(0,3)$
    has $\chi(X)=0$ and $h^{1,1}=h^{2,1}=19\neq3$, so our GLSM do not capture all K\"ahler moduli of this CY, i.e. is not K\"ahler-favorable. However, this does not affect the result of theorem \ref{thm:TPN} and prop. \ref{prop:33}. In particular, the $T_N$ number is correctly given by
\begin{equation}
    N=\int_{\bbP^2\times\bbP^2}(3H_1\cdot 3H_2)^2=81=\frac{0-(-162)}{2}=\frac{\chi(X)-\chi(X^\natural)}{2}.
\end{equation}
\end{rmk}

\subsection{$X^\natural(4,2)$}
Denote the hyperplane classes for $\bbP^1$ and $\bbP^3$ by $H_1$ and $H_2$ respectively. The topological numbers of $X^\natural$ are given by
\begin{eqnarray}
    c_{122}=4,\quad c_{222}=2
\end{eqnarray}
and the submatrices, corresponding to $X^{\natural}$, in \eqref{matshuffleb} are
\begin{equation}
    L^\natural_1=\begin{pmatrix}
        1 & 0 & 0 & 0 & 0 & 0 \\
 1 & 1 & 0 & 0 & 0 & 0 \\
 0 & 0 & 1 & 0 & 0 & 0 \\
 0 & 0 & 0 & 1 & 0 & 0 \\
 0 & 0 & 4 & 0 & 1 & 0 \\
 0 & 0 & -2 & 1 & 0 & 1 \\
    \end{pmatrix},
    \quad
    L^\natural_2=\begin{pmatrix}
        1 & 0 & 0 & 0 & 0 & 0 \\
 0 & 1 & 0 & 0 & 0 & 0 \\
 1 & 0 & 1 & 0 & 0 & 0 \\
 4 & 0 & 4 & 1 & 0 & 0 \\
 2 & 4 & 2 & 0 & 1 & 0 \\
 0 & 2 & 0 & 0 & 1 & 1 \\
    \end{pmatrix},
    \quad
    T^\natural_{\calO_X}=\begin{pmatrix}
          1 & -2 & -4 & 0 & 0 & -1 \\
 0 & 1 & 0 & 0 & 0 & 0 \\
 0 & 0 & 1 & 0 & 0 & 0 \\
 0 & 0 & 0 & 1 & 0 & 0 \\
 0 & 0 & 0 & 0 & 1 & 0 \\
 0 & 0 & 0 & 0 & 0 & 1 \\
    \end{pmatrix}
\end{equation}
Similarly, it is conjectured that $\Delta\cap D_1$ forms a $\mathfrak{L}_2^{(1)}$ link while $\Delta\cap D_2$ is a $\mathfrak{L}_4^{(1)}$ link. Then it can be tested that indeed,
\begin{equation}
    (L^\natural_1T^\natural_{\calO_X})^2=(T^\natural_{\calO_X}L^\natural_1)^2,\quad (L^\natural_2T^\natural_{\calO_X})^4=(T^\natural_{\calO_X}L^\natural_2)^4.
\end{equation}

\begin{prop}
For any splitting configuration of $X^{\natural}(4,2)$, the relevant links, in the sense of conjecture \ref{conjecturenested}, appearing in the subspaces $\widehat{D}_{0,1}$ and $\widehat{D}_{1,2}$, corresponds to $\mathfrak{L}_2^{(1)}$ and $\mathfrak{L}_4^{(1)}$, respectively. The monodromy matrix $M_{0}$ is given by:
\begin{eqnarray}
   M_1:=(T_{\calO_X}L_2)^4L_2^{-4},\quad M_0=(M_1L_1)^2L_1^{-2}=T_N L_1^{-2}L_2^{-4}.\label{eqn:TPN42}
\end{eqnarray}
where 
\begin{eqnarray}
(M_1L_1)^2=(L_1 M_1)^2,\quad (T_{\calO_X}L_2)^4=(L_2 T_{\calO_X})^4.\label{eqn:link42}
\end{eqnarray}
If we consider the subspaces $\widehat{D}_{0,2}$ and $\widehat{D}_{2,1}$, instead, the result is unchanged.
\end{prop}
\begin{proof}
A splitting configuration $X$ with $Y=\bbP^1\times\bbP^3$ is given by a multi-partition of $(2,4)$ as $\vec n=n_1^{(\alpha)}+\cdots+ n_{n+1}^{(\alpha)}=(2,4)$. Similarly one has
\begin{equation}
    \int_{Y}A_2(\vec n(H))^2=2\sum_{I<J}\sum_{K<L} n_K^{(2)}n_L^{(2)}\{n_I^{(1)},n_J^{(2)}\}=2c_{012}c_{022}.
\end{equation}
where 
\begin{equation}
    \{n_I^{(\alpha)},n_J^{(\beta)}\}:=n_I^{(\alpha)}n_J^{(\beta)}+n_I^{(\beta)}n_J^{(\alpha)}.
\end{equation}
And
\begin{equation}
    \int_{Y}A_1(\vec n(H))A_3(\vec n(H))=\int_Y(2H_1+4H_2)A_3(\vec n)=2c_{001}+4c_{002}.
\end{equation}
Thus
\begin{equation}
    N=2c_{012}c_{022}-2c_{001}-4c_{002}.
\end{equation}
Then, it is a direct computation to verify \eqref{eqn:TPN42} and \eqref{eqn:link42} by the following matrices:
\begin{eqnarray}
        L_1&=&\left(
\begin{array}{cccccccc}
 1 & 0 & 0 & 0 & 0 & 0 & 0 & 0 \\
 c_{001} & 1 & 0 & 0 & c_{012} & 0 & 0 & 0 \\
 0 & 0 & 1 & 0 & 0 & 0 & 0 & 0 \\
 0 & 0 & 1 & 1 & 0 & 0 & 0 & 0 \\
 0 & 0 & 0 & 0 & 1 & 0 & 0 & 0 \\
 0 & 0 & 0 & 0 & 0 & 1 & 0 & 0 \\
 c_{012} & 0 & 0 & 0 & 4 & 0 & 1 & 0 \\
 b_{10} & 0 & 0 & 0 & -2 & 1 & 0 & 1 \\
\end{array}
\right),
\quad 
L_2=\left(
\begin{array}{cccccccc}
 1 & 0 & 0 & 0 & 0 & 0 & 0 & 0 \\
 c_{002} & 1 & c_{022} & c_{012} & c_{022} &
   0 & 0 & 0 \\
 0 & 0 & 1 & 0 & 0 & 0 & 0 & 0 \\
 0 & 0 & 0 & 1 & 0 & 0 & 0 & 0 \\
 0 & 0 & 1 & 0 & 1 & 0 & 0 & 0 \\
 c_{012} & 0 & 4 & 0 & 4 & 1 & 0 & 0 \\
 c_{022} & 0 & 2 & 4 & 2 & 0 & 1 & 0 \\
 b_{20} & 0 & 0 & 2 & 0 & 0 & 1 & 1 \\
\end{array}
\right),
\nonumber\\
T_{\calO_X}&=&\left(
\begin{array}{cccccccc}
 1 & 0 & 0 & 0 & 0 & 0 & 0 & 0 \\
 0 & 1 & 0 & 0 & 0 & 0 & 0 & 0 \\
 -l_0 & 0 & 1 & -2 & -4 & 0 & 0 & -1 \\
 0 & 0 & 0 & 1 & 0 & 0 & 0 & 0 \\
 0 & 0 & 0 & 0 & 1 & 0 & 0 & 0 \\
 0 & 0 & 0 & 0 & 0 & 1 & 0 & 0 \\
 0 & 0 & 0 & 0 & 0 & 0 & 1 & 0 \\
 0 & 0 & 0 & 0 & 0 & 0 & 0 & 1 \\
\end{array}
\right).
\end{eqnarray}  

\end{proof}

\subsection{$X^\natural(3,2,2)$}
Denote the hyperplane class for $\bbP^2$ by $H_1$ and other two $\bbP^1$'s by $H_2$, $H_3$. The topological numbers of $X^\natural$ are given by
\begin{equation}
    c_{112}=c_{113}=2,\quad c_{123}=3
\end{equation}
and the submatrices, corresponding to $X^{\natural}$, in \eqref{matshuffleb} are
\begin{eqnarray}
    L^\natural_1&=&\begin{pmatrix}
        1 & 0 & 0 & 0 & 0 & 0 & 0 & 0 \\
 1 & 1 & 0 & 0 & 0 & 0 & 0 & 0 \\
 0 & 0 & 1 & 0 & 0 & 0 & 0 & 0 \\
 0 & 0 & 0 & 1 & 0 & 0 & 0 & 0 \\
 0 & 0 & 2 & 2 & 1 & 0 & 0 & 0 \\
 2 & 2 & 0 & 3 & 0 & 1 & 0 & 0 \\
 2 & 2 & 3 & 0 & 0 & 0 & 1 & 0 \\
 0 & 0 & 1 & 1 & 1 & 0 & 0 & 1 \\
    \end{pmatrix},
    \quad 
    L^\natural_2=\begin{pmatrix}
        1 & 0 & 0 & 0 & 0 & 0 & 0 & 0 \\
 0 & 1 & 0 & 0 & 0 & 0 & 0 & 0 \\
 1 & 0 & 1 & 0 & 0 & 0 & 0 & 0 \\
 0 & 0 & 0 & 1 & 0 & 0 & 0 & 0 \\
 0 & 2 & 0 & 3 & 1 & 0 & 0 & 0 \\
 0 & 0 & 0 & 0 & 0 & 1 & 0 & 0 \\
 0 & 3 & 0 & 0 & 0 & 0 & 1 & 0 \\
 0 & -1 & 0 & 0 & 0 & 1 & 0 & 1 \\
    \end{pmatrix},
    \nonumber\\
    L^\natural_3&=&\begin{pmatrix}
        1 & 0 & 0 & 0 & 0 & 0 & 0 & 0 \\
 0 & 1 & 0 & 0 & 0 & 0 & 0 & 0 \\
 0 & 0 & 1 & 0 & 0 & 0 & 0 & 0 \\
 1 & 0 & 0 & 1 & 0 & 0 & 0 & 0 \\
 0 & 2 & 3 & 0 & 1 & 0 & 0 & 0 \\
 0 & 3 & 0 & 0 & 0 & 1 & 0 & 0 \\
 0 & 0 & 0 & 0 & 0 & 0 & 1 & 0 \\
 0 & -1 & 0 & 0 & 0 & 0 & 1 & 1 \\
    \end{pmatrix},\quad
    T^\natural_{\calO_X}=\begin{pmatrix}
        1 & -l_\alpha & 0 & -1
        \\
         & \bbI_3 & &
         \\
         &&\bbI_3&
         \\
         &&&1
    \end{pmatrix},\quad l_\alpha=(3,2,2).
    \nonumber\\
\end{eqnarray}

\begin{prop}
For any splitting configuration of $X^{\natural}(3,2,2)$, the relevant links, in the sense of conjecture \ref{conjecturenested}, appearing in the subspaces $\widehat{D}_{0,1}$, $\widehat{D}_{1,2}$ and $\widehat{D}_{2,3}$ corresponds to $\mathfrak{L}_3^{(1)}$, $\mathfrak{L}_2^{(1)}$ and $\mathfrak{L}_2^{(1)}$, respectively. The monodromy matrix $M_{0}$ is given by:
\begin{eqnarray}
    M_2=(T_{\calO_X}L_3)^2L_3^{-2},\quad M_1=(M_2L_2)^2L_2^{-2},\quad M_0=(M_1L_1)^3L_1^{-3}=T_NL_1^{-3}L_2^{-2}L_3^{-2}.
    \nonumber\\\label{eqn:TPN322}
\end{eqnarray}
where
\begin{eqnarray}
    (M_1L_1)^3=(L_1 M_1)^3,\quad (M_\alpha L_\alpha)^2=(L_\alpha M_\alpha)^2,\quad \alpha=2,3.\label{eqn:link322}
\end{eqnarray}
We can choose the subspaces $\widehat{D}_{0,\alpha}$, etc, in any other order and the result for $M_{0}$ is unchanged.
\end{prop}
\begin{proof}
A splitting configuration $X$ with $Y=\bbP^2\times\bbP^1\times\bbP^1$ is given by a multi-partition of $(3,2,2)$ as $\vec n=n_1^{(\alpha)}+\cdots+ n_{n+1}^{(\alpha)}=(3,2,2)$, $\alpha=1,2,3$. One has
\begin{eqnarray}
\int_{Y}A_2(\vec n(H))^2&=&\sum_{I<J}\sum_{K<L}\left( n_I^{(1)}n_J^{(1)}\{n_K^{(2)},n_L^{(3)}\}+n_K^{(1)}n_L^{(1)}\{n_I^{(2)},n_J^{(3)}\}\right.
\\
&&\left.\quad\quad\quad+\{n_I^{(1)},n_J^{(3)}\}\{n_K^{(1)},n_L^{(2)}\}+\{n_I^{(1)},n_J^{(2)}\}\{n_K^{(1)},n_L^{(3)}\}\right)
\\
&=&2(c_{012}c_{013}+c_{011}c_{023}).
\end{eqnarray}
and
\begin{eqnarray}
    \int_Y A_1(\vec n(H))A_3(\vec n(H))=\int_Y(3H_1+2H_2+2H_3)A_3(\vec n)=3c_{001}+2c_{002}+2c_{003}.
\end{eqnarray}
Thus
\begin{eqnarray}
    N\equiv2(c_{012}c_{013}+c_{011}c_{023})-3c_{001}-2c_{002}-2c_{003}.
\end{eqnarray}

Then, it is a direct computation to verify \eqref{eqn:TPN322} and ~\eqref{eqn:link322} by the following matrices:
{\tiny 
\begin{eqnarray}
  L_1=&\left(
\begin{array}{cccccccccc}
 1 & 0 & 0 & 0 & 0 & 0 & 0 & 0 & 0 & 0 \\
 c_{001} & 1 & c_{001} & c_{001} & c_{012} &
   c_{013} & 0 & 0 & 0 & 0 \\
 0 & 0 & 1 & 0 & 0 & 0 & 0 & 0 & 0 & 0 \\
 0 & 0 & 1 & 1 & 0 & 0 & 0 & 0 & 0 & 0 \\
 0 & 0 & 0 & 0 & 1 & 0 & 0 & 0 & 0 & 0 \\
 0 & 0 & 0 & 0 & 0 & 1 & 0 & 0 & 0 & 0 \\
 c_{001} & 0 & 0 & 0 & 2 & 2 & 1 & 0 & 0 & 0 \\
 c_{012} & 0 & 2 & 2 & 0 & 3 & 0 & 1 & 0 & 0 \\
 c_{013} & 0 & 2 & 2 & 3 & 0 & 0 & 0 & 1 & 0 \\
 b_{10} & 0 & 0 & 0 & 1 & 1 & 1 & 0 & 0 & 1 \\
\end{array}
\right),
\quad
L_2=\left(
\begin{array}{cccccccccc}
 1 & 0 & 0 & 0 & 0 & 0 & 0 & 0 & 0 & 0 \\
 c_{002} & 1 & c_{012} & c_{012} & 0 &
   c_{023} & 0 & 0 & 0 & 0 \\
 0 & 0 & 1 & 0 & 0 & 0 & 0 & 0 & 0 & 0 \\
 0 & 0 & 0 & 1 & 0 & 0 & 0 & 0 & 0 & 0 \\
 0 & 0 & 1 & 0 & 1 & 0 & 0 & 0 & 0 & 0 \\
 0 & 0 & 0 & 0 & 0 & 1 & 0 & 0 & 0 & 0 \\
 c_{012} & 0 & 0 & 2 & 0 & 3 & 1 & 0 & 0 & 0 \\
 0 & 0 & 0 & 0 & 0 & 0 & 0 & 1 & 0 & 0 \\
 c_{023} & 0 & 0 & 3 & 0 & 0 & 0 & 0 & 1 & 0 \\
 b_{20} & 0 & 0 & -1 & 0 & 0 & 0 & 1 & 0 & 1 \\
\end{array}
\right),
\nonumber\\
L_3=&\left(
\begin{array}{cccccccccc}
 1 & 0 & 0 & 0 & 0 & 0 & 0 & 0 & 0 & 0 \\
 c_{003} & 1 & c_{013} & c_{013} & c_{023} &
   0 & 0 & 0 & 0 & 0 \\
 0 & 0 & 1 & 0 & 0 & 0 & 0 & 0 & 0 & 0 \\
 0 & 0 & 0 & 1 & 0 & 0 & 0 & 0 & 0 & 0 \\
 0 & 0 & 0 & 0 & 1 & 0 & 0 & 0 & 0 & 0 \\
 0 & 0 & 1 & 0 & 0 & 1 & 0 & 0 & 0 & 0 \\
 c_{013} & 0 & 0 & 2 & 3 & 0 & 1 & 0 & 0 & 0 \\
 c_{023} & 0 & 0 & 3 & 0 & 0 & 0 & 1 & 0 & 0 \\
 0 & 0 & 0 & 0 & 0 & 0 & 0 & 0 & 1 & 0 \\
 b_{30} & 0 & 0 & -1 & 0 & 0 & 0 & 0 & 1 & 1 \\
\end{array}
\right),
\quad
T_{\calO_X}=\left(
\begin{array}{cccccccccc}
 1 & 0 & 0 & 0 & 0 & 0 & 0 & 0 & 0 & 0 \\
 0 & 1 & 0 & 0 & 0 & 0 & 0 & 0 & 0 & 0 \\
 -l_0 & 0 & 1 & -3 & -2 & -2 & 0 & 0 & 0 & -1 \\
 0 & 0 & 0 & 1 & 0 & 0 & 0 & 0 & 0 & 0 \\
 0 & 0 & 0 & 0 & 1 & 0 & 0 & 0 & 0 & 0 \\
 0 & 0 & 0 & 0 & 0 & 1 & 0 & 0 & 0 & 0 \\
 0 & 0 & 0 & 0 & 0 & 0 & 1 & 0 & 0 & 0 \\
 0 & 0 & 0 & 0 & 0 & 0 & 0 & 1 & 0 & 0 \\
 0 & 0 & 0 & 0 & 0 & 0 & 0 & 0 & 1 & 0 \\
 0 & 0 & 0 & 0 & 0 & 0 & 0 & 0 & 0 & 1 \\
\end{array}
\right).
\nonumber\\
\end{eqnarray}  
}
    
\end{proof}

\subsection{$X^\natural(2,2,2,2)$}
The topological numbers of $X^\natural(2,2,2,2)$ are given by
\begin{equation}
    c_{\alpha\beta\gamma}=2,\quad\alpha\neq\beta\neq\gamma=1,\cdots,4.
\end{equation}
Then, the submatrices, corresponding to $X^{\natural}$, in \eqref{matshuffleb} are thus
\begin{equation}
    L^\natural_\alpha=\begin{pmatrix}
        1 & 0 & 0 & 0
        \\
        e_\alpha & \bbI_4 & 0 & 0
        \\
        0 & h_\alpha & \bbI_4 & 0
        \\
        0 & 0 & \;^te_\alpha & 1
    \end{pmatrix},
    \quad
    T^\natural_{\calO_X}=\begin{pmatrix}
        1 & -\vec{2} & 0 & -1
        \\
         & \bbI_4 & & &
         \\
         & & \bbI_4 & &
         \\
         & & & 1
    \end{pmatrix}
\end{equation}
where $\vec{2}:=(2,2,2,2)$, $e_\alpha$ is the unit column vector, $h_\alpha$ is a $4\times4$ matrix for each $\alpha$, given by
\begin{equation}
   (h_\alpha)_{ij}=\left\{\begin{array}{cl}
    2 & i\neq j\neq\alpha  
        \\ 
    0 & \text{otherwise}
   \end{array}\right.
\end{equation}
\begin{equation}
    (L^\natural_\alpha T^\natural_{\calO_X})^2=(T^\natural_{\calO_X}L^\natural_\alpha)^2,\quad \alpha=1,\cdots,4.
\end{equation}

\begin{prop}
For any splitting configuration $X$ of $X^{\natural}(2,2,2,2)$, the relevant links, in the sense of conjecture \ref{conjecturenested}, appearing in the subspaces $\widehat{D}_{0,1}$, $\widehat{D}_{1,2}$, $\widehat{D}_{2,3}$ and $\widehat{D}_{3,4}$ corresponds to $\mathfrak{L}_2^{(1)}$, for all of them. The monodromy matrix $M_{0}$ is given by:
\begin{eqnarray}
    M_3&=&(T_{\calO_X}L_4)^2L_4^{-2},\  M_2=(M_3L_3)^2L_3^{-2},\   M_1=(M_2L_2)^2L_2^{-2},  
    \nonumber\\
    M_0&=&(M_1L_1)^2L_1^{-2}=T_N(L_1L_2L_3L_4)^{-2} \label{eqn:TPN2222}
    \\
    &&(T_{\calO_X}M_\alpha)^2=(M_\alpha T_{\calO_X})^2,\quad \alpha=1,\cdots,4.\label{eqn:link2222}  
\end{eqnarray}
\end{prop}
\begin{proof}
A splitting configuration $X$ with $Y=\bbP^1\times\bbP^1\times\bbP^1\times\bbP^1$ is given by a multi-partition of $(2,2,2,2)$ as $\vec n=n_1^{(\alpha)}+\cdots+ n_{n+1}^{(\alpha)}=(2,2,2,2)$, $\alpha=1,\cdots,4$. One has
\begin{eqnarray}
    \int_{Y}A_2(\vec n(H))^2&=&\sum_{I<J}\sum_{K<L}\sum_{\alpha_1\neq\cdots\neq\alpha_4}\{n_I^{(\alpha_1)},n_J^{(\alpha_2)} \}\{n_K^{(\alpha_3)},n_L^{(\alpha_4)} \}
    \nonumber\\
    &=&2(c_{012}c_{034}+c_{013}c_{024}+c_{014}c_{023}).
\end{eqnarray}
and
\begin{eqnarray}
    \int_Y A_1(\vec n(H))A_3(\vec n(H))&=&\int_Y2(H_1+H_2+H_3+H_4)A_3(\vec n)
    \nonumber\\
    &=&2(c_{001}+c_{001}+c_{002}+c_{003}+c_{004}).
\end{eqnarray}
Thus
\begin{eqnarray}
    N\equiv2(c_{012}c_{034}+c_{013}c_{024}+c_{014}c_{023})-2(c_{001}+c_{001}+c_{002}+c_{003}+c_{004}).
\end{eqnarray}
We omit their gigantic matrices in this case, which can be written down and tested similarly for \eqref{eqn:TPN2222} and ~\eqref{eqn:link2222}.
\end{proof}

\subsection{$X^{\natural}=\bbW\bbP_{96111}[18]$}\label{sec:weighted}

Here we illustrate, with a single example, that in principle, weighted projective spaces can also be included in $\vec\bbP$. The hypersurface $X^{\natural}=\bbW\bbP_{96111}[18]$ is generically singular, but it enjoys a toric crepant resolution and the corresponding two parameter GLSM can be found in \cite{Candelas:1994hw}. We consider a splitting configuration $X$ of $X^{\natural}$ characterized by the partition $6+6+6=18$\footnote{Opposite to its ominous meaning in Western culture, in China $666$ is a slang expression meaning “awesome,” “skillful,” or “smooth,” and is used to praise someone’s performance or talent.}
The GLSM for splitting configuration is
\begin{eqnarray}
    \begin{array}{c|cccccc}
         &  x_{3}^{(2)} & x_2^{(2)} & x_{1}^{(2)} & x_{1,2,3}^{(1)}  & y_{1,2,3} & p_{1,2,3}
         \\\hline
       U(1)_0  & 0 & 0 & 0 & 0 & 1 & -1
       \\
       U(1)_1 & 3 & 2 & 1 & 0 & 0 & -2
       \\
       U(1)_2 & 0 & 0 & -3 & 1 & 0 & 0
       \\\hline
       U(1)_R & \varepsilon_1 & \varepsilon_2 & \varepsilon_3 & \varepsilon & \varepsilon_0 & \epsilon_{1,2,3}
    \end{array}
\end{eqnarray}
where the superpotential is 
\begin{eqnarray}
    W=\sum_{I,j=1}^3p_I\ y_J F_{IJ}(x_2^{(2)},x_1^{(2)},x_{1,2,3}^{(1)}).
\end{eqnarray}
The R-charge can be assigned as in \eqref{eqn:RInt} for the $\zeta_\pm$ phases.\\
The hemisphere partition function can be written down as in \eqref{eqn:ZBt} and evaluated at the zeroth pole i.e. the zero-instanton sector. We list the B-brane factor that corresponds to the Doran-Morgan basis of A-periods\footnote{Note that in this basis $\chi(D_1,C_2)=-3\neq0$, in contrast with the cases we consider in the rest of this work.}:
\begin{eqnarray}
    f_{\hat\calE_-}&=& \left( 1-e^{-2\pi\sigma_0-4\pi\sigma_1} \right)^3
    \\
    f_{\calB_{D_0}}&=&f_{\hat\calE_-}\left( 1-e^{-2\pi\sigma_0} \right)
    \\
    f_{\calB_{D_1}}&=&f_{\hat\calE_-}\left( 1-e^{-2\pi\sigma_1+6\pi\sigma_2} \right)
    \\
    f_{\calB_{D_2}}&=&f_{\hat\calE_-}\left( 1-e^{-2\pi\sigma_2} \right)
    \\
    f_{\calB_0}&=&\left( 1-e^{-12\pi\sigma_1} \right)\left( 1-e^{-2\pi\sigma_1+6\pi\sigma_2} \right)\left( 1-e^{-2\pi\sigma_2} \right)^2\left( 1-e^{-2\pi\sigma_0} \right)e^{-2\pi\sigma_0}
    \\
    f_{\calB_1}&=&\left( 1-e^{-12\pi\sigma_1} \right)\left( 1-e^{-2\pi\sigma_2} \right)^2\left( 1-e^{-2\pi\sigma_0} \right)^2
    \\
    f_{\calB_2}&=&\left( 1-e^{-12\pi\sigma_1} \right)\left( 1-e^{-2\pi\sigma_1+6\pi\sigma_2} \right)\left( 1-e^{-2\pi\sigma_2} \right)\left( 1-e^{-2\pi\sigma_0} \right)^2e^{-2\pi\sigma_2}
    \\
    f_{\calB_{\text{pt}}}&=&\left( 1-e^{-12\pi\sigma_1} \right)\left( 1-e^{-2\pi\sigma_1+6\pi\sigma_2} \right)\left( 1-e^{-2\pi\sigma_2} \right)^2\left( 1-e^{-2\pi\sigma_0} \right)^2
\end{eqnarray}
with $\kappa$ parameters:
\begin{eqnarray}
    t_0&=&-2\pi i\left(\kappa_0+\frac{3}{2}\right)
    \\
    t_1&=&-2\pi i(\kappa_1+3)
    \\
    t_2&=&-2\pi i\kappa_2.
\end{eqnarray}
Then, the topological numbers can be read directly from the hemisphere partition function:
\begin{equation}
\begin{array}{cccccc}
    c_{000}=0 & c_{001}=12 & c_{002}=4 & c_{011}=18 & c_{012}=6 & c_{022}=2 
     \\
    c_{111}=9 & c_{112}=3 & c_{122}=1 & c_{222}=0, & &
    \\
    c_0=5/2 & c_1=17/4 & c_2=3/2, & \chi=-252. & &
\end{array}
\end{equation}
Then we have (where $X^{\natural}$ should be understood as its crepant resolution)
\begin{equation}
N=\frac{\chi(X)-\chi(X^{\natural})}{2}=(-252-(-540))/2=144.
\end{equation}
The moduli space $\calM_{K}(X^{\natural})$ was analyzed in \cite{Candelas:1994hw} and the monodromies, as autoequivalences, in \cite{Karp:2008qz}. It the case at hand, the monodromy $M_{0}$, associated to $X$ can be shown to satisfy
\begin{eqnarray}\label{weightedmono}
   M_2=T_{\calO_X},\quad  M_1=(T_{\calO_X}L_2)^3L_2^{-3},\quad M_0=(M_1L_1)^6L_1^{-6}=T_{144}L_{K_Y}.
\end{eqnarray}
which suggests that the link associated to $\widehat{D}_{0,1}$ and $\widehat{D}_{1,2}$ are $\mathfrak{L}^{(1)}_{6}$ and $\mathfrak{L}^{(1)}_{3}$, respectively. Is also important to note that since we are working with a crepant resolution $K_{Y}=K_{\bbW\bbP_{96111}}$, in agreement with \eqref{weightedmono}. For completeness, we register here the explicit form of the matrices involved in \eqref{weightedmono}
\begin{eqnarray}
    &L_0=\left(
\begin{array}{cccccccc}
 1 & 0 & 0 & 0 & 0 & 0 & 0 & 0 \\
 0 & 1 & 0 & 0 & 0 & 0 & 0 & 0 \\
 1 & 0 & 1 & 0 & 0 & 0 & 0 & 0 \\
 3 & 0 & 0 & 1 & 0 & 0 & 0 & 0 \\
 12 & 12 & 0 & 6 & 1 & 0 & 0 & 0 \\
 6 & 18 & 0 & 3 & 0 & 1 & 0 & 0 \\
 0 & 6 & 0 & 1 & 0 & 0 & 1 & 0 \\
 0 & 3 & 0 & 1 & 0 & 1 & 0 & 1 \\
\end{array}
\right),\quad L_1=\left(
\begin{array}{cccccccc}
 1 & 0 & 0 & 0 & 0 & 0 & 0 & 0 \\
 0 & 1 & 0 & 0 & 0 & 0 & 0 & 0 \\
 1 & 0 & 1 & 0 & 0 & 0 & 0 & 0 \\
 3 & 0 & 0 & 1 & 0 & 0 & 0 & 0 \\
 12 & 12 & 0 & 6 & 1 & 0 & 0 & 0 \\
 6 & 18 & 0 & 3 & 0 & 1 & 0 & 0 \\
 0 & 6 & 0 & 1 & 0 & 0 & 1 & 0 \\
 0 & 3 & 0 & 1 & 0 & 1 & 0 & 1 \\
\end{array}
\right),
\nonumber\\&L_2=\left(
\begin{array}{cccccccc}
 1 & 0 & 0 & 0 & 0 & 0 & 0 & 0 \\
 0 & 1 & 0 & 0 & 0 & 0 & 0 & 0 \\
 0 & 0 & 1 & 0 & 0 & 0 & 0 & 0 \\
 1 & 0 & 0 & 1 & 0 & 0 & 0 & 0 \\
 2 & 4 & 0 & 2 & 1 & 0 & 0 & 0 \\
 1 & 6 & 0 & 1 & 0 & 1 & 0 & 0 \\
 0 & 2 & 1 & 0 & 0 & 0 & 1 & 0 \\
 0 & -1 & 2 & 0 & 0 & 0 & 1 & 1 \\
\end{array}
\right), \quad T_{\calO_X}=\left(
\begin{array}{cccccccc}
 1 & -5 & -1 & -3 & 0 & 0 & 0 & -1 \\
 0 & 1 & 0 & 0 & 0 & 0 & 0 & 0 \\
 0 & 0 & 1 & 0 & 0 & 0 & 0 & 0 \\
 0 & 0 & 0 & 1 & 0 & 0 & 0 & 0 \\
 0 & 0 & 0 & 0 & 1 & 0 & 0 & 0 \\
 0 & 0 & 0 & 0 & 0 & 1 & 0 & 0 \\
 0 & 0 & 0 & 0 & 0 & 0 & 1 & 0 \\
 0 & 0 & 0 & 0 & 0 & 0 & 0 & 1 \\
\end{array}
\right).
\nonumber\\
\end{eqnarray}

\bibliographystyle{fullsort}
\bibliography{biblio.bib}

\end{document}